\documentclass[runningheads]{llncs}

\usepackage{makeidx}  
\usepackage{graphicx}
\usepackage{subfigure}
\usepackage{multirow}
\usepackage{color,soul} 
\usepackage{ dsfont }
\usepackage{amssymb,amsfonts}
\usepackage{physics} 
\usepackage{cases}
\usepackage{hyperref}
\usepackage{multirow}

\usepackage{graphicx}

\usepackage{amsmath}

\usepackage{subfigure}
\usepackage{algorithm}
\usepackage{algorithmic}

\newcommand{\be}{\begin{equation}}
\newcommand{\ee}{\end{equation}}

\makeatletter
 \def\SOUL@hlpreamble{%
 \setul{}{2.4ex}
 \let\SOUL@stcolor\SOUL@hlcolor
 \SOUL@stpreamble
 }
\makeatother

\begin{document}
\mainmatter              

\title{FoldIt: Haustral Folds Detection and Segmentation in Colonoscopy Videos}
\titlerunning{FoldIt}  
%

\author{Shawn Mathew*\inst{1} \and Saad Nadeem*\inst{2}{\let\thefootnote\relax\footnote{{\hspace{-4mm}*Equal contribution. Email: nadeems@mskcc.org}}} \and Arie Kaufman\inst{1}}
\index{Mathew, Shawn}
\index{Nadeem, Saad}
\index{Kaufman, Arie}

\authorrunning{Mathew \emph{et al}.} 

\tocauthor{Shawn Mathew, Saad Nadeem, and Arie Kaufman}

\institute{Department of Computer Science, Stony Brook University\\
\and
Department of Medical Physics, Memorial Sloan Kettering Cancer Center\\
}

\maketitle              

\begin{abstract}
Haustral folds are colon wall protrusions implicated for high polyp miss rate during optical colonoscopy procedures. If segmented accurately, haustral folds can allow for better estimation of missed surface and can also serve as valuable landmarks for registering pre-treatment virtual (CT) and optical colonoscopies, to guide navigation towards the anomalies found in pre-treatment scans. We present a novel generative adversarial network, FoldIt, for feature-consistent image translation of optical colonoscopy videos to virtual colonoscopy renderings with haustral fold overlays. A new transitive loss is introduced in order to leverage ground truth information between haustral fold annotations and virtual colonoscopy renderings. We demonstrate the effectiveness of our model on real challenging optical colonoscopy videos as well as on textured virtual colonoscopy videos with clinician-verified haustral fold annotations. All code and scripts to reproduce the experiments of this paper will be made available via our Computational Endoscopy Platform at \url{https://github.com/nadeemlab/CEP}.

\end{abstract}
\section{Introduction}
High polyp miss rates during colonoscopy procedure are mainly attributed to colon wall protrusions, formed by circumferential contraction of the colon inner muscular layer, also known as haustral folds. These folds are extremely difficult to segment in optical colonoscopy (OC) videos due to texture and lighting variations along with specular reflections, fluid motion, and organ movements. If segmented, however, these folds can guide endoscope navigation towards high-occlusion areas, potentially reducing the polyp miss rate. Moreover, even though the colon can stretch and distort shape considerably in OC versus the pre-treatment CT virtual colonoscopy (VC), the haustral folds remain intact across the two modalities and hence can serve as useful landmarks for registration.

Mathew et al. \cite{mathew2020augmenting} recently introduced a new unsupervised model, XDCycleGAN, for inferring scale-consistent depth maps from OC video frames using geometric information from 3D VC data, extracted from abdominal CT scans. The unsupervised model was shown to handle variations in texture, lighting and specular reflections much more effectively than previous supervised approaches that were trained completely on OC or on VC datasets but not both simultaneously. Xu et al. \cite{xu2020ofgan} also showed superior performance by using cycle-consistency and optical flow for spatially- and temporally-consistent translation of simulated VC flythroughs to real OC videos.

In this work, we present FoldIt, a new generative adversarial network that can accurately detect and segment haustral folds in OC videos using unpaired image-to-image translation of OC frames to VC haustral fold renderings. \textit{We show that the haustral fold segmentation via our model leads to feature-consistent domain translation for OC video sequences; the feature-consistency refers to consistency of haustral fold annotations between consecutive frames}. FoldIt is available on GitHub via our \href{https://github.com/nadeemlab/CEP}{\textcolor{blue}{Computational Endoscopy Platform}}. The contributions of this work are as follows:
\begin{enumerate}
    \item A method for haustral fold detection and segmentation in real OC images.
    \item A semi-supervised approach to perform image-to-image domain translation via a common domain.
    \item A transitive loss to drive the domain translation while retaining/preserving haustral fold features in the common VC domain.
\end{enumerate}

\section{Related Works}


Deep learning approaches have recently shown promising results in various endoscopic intervention tasks such as depth estimation, 3D reconstruction, and surface coverage \cite{bae2020deep,freedman2020detecting,liu2020reconstructing,mathew2021visualizing}. Deep learning models are data driven and the supervised category requires ground truth information. The issue with the supervised approaches, specifically for colonoscopy, is the need for realistic training data that models the specular reflections, fluid motion, and erratic camera and spasm movements. The ground truth data creation is extremely time consuming and even then the network can easily fail since the input data is not representative of the real domain. To overcome this, the supervised methods require additional measures to handle real data. Rau et al. \cite{rau2019implicit} trained a pix2pix variant on textured VC data which needed extra training on OC images to handle real OC data. Mahmood et al. \cite{mahmood2018unsupervised} required an additional network to transform their data into a synthetic-like (VC) domain before being able to estimate depth maps. Chen et al. \cite{chen2019slam} trained on VC input and tested on porcine phantom models, so it is unclear how well their approach performs on real OC data.

In contrast, unsupervised approaches already work on the input domain, so no additional modules are needed to handle real OC images. They, however, require more complicated network design and careful assessment to avoid overfitting. In FoldIt, we present a semi-supervised approach which leverages real OC images. Ground truth haustral fold annotations on 3D VC triangular mesh models were computed using Fiedler vector representation \cite{nadeem2016corresponding} and verified by a clinician. We capture the best aspects of both the unsupervised and supervised approaches in our FoldIt model.

Generative Adversarial Networks (GAN) \cite{goodfellow2014generative} have shown promising results for image-to-image translation. GANs help the generator output match the training data distribution. Approaches such as pix2pix \cite{isola2017image}, CycleGAN \cite{zhu2017unpaired}, and StarGAN \cite{choi2018stargan} all utilize adversarial learning in their approaches. Pix2pix does paired image-to-image translation while CycleGAN can handle unpaired data. Both pix2pix and CycleGAN assume a one-to-one mapping between domains. Mathew et al. \cite{mathew2020augmenting} introduced a one-to-many mapping to handle OC images since there are large variations in texture, lighting, and specular reflections. 

Haustral fold annotation requires many-to-many mapping since the depth of the folds can be unclear from a single frame. Travel-GAN \cite{amodio2019travelgan} introduced a many-to-many approach, however, their translation is not constrained, making it less suitable for our task. To address this, we present a model that performs many-to-many translation via a common domain. This common domain has a one-to-many mapping with the other two domains to help constrain the output. Fang et al. \cite{fang2020triple} have recently proposed a triple translation loss, similar to our transitive loss, but they focus on face generation based on age in a single domain which does not require translation between multiple domains.

\section{Data}
\begin{figure}[t!]
\begin{center}
\setlength{\tabcolsep}{10pt}
\begin{tabular}{cc}
\includegraphics[width=0.4\textwidth]{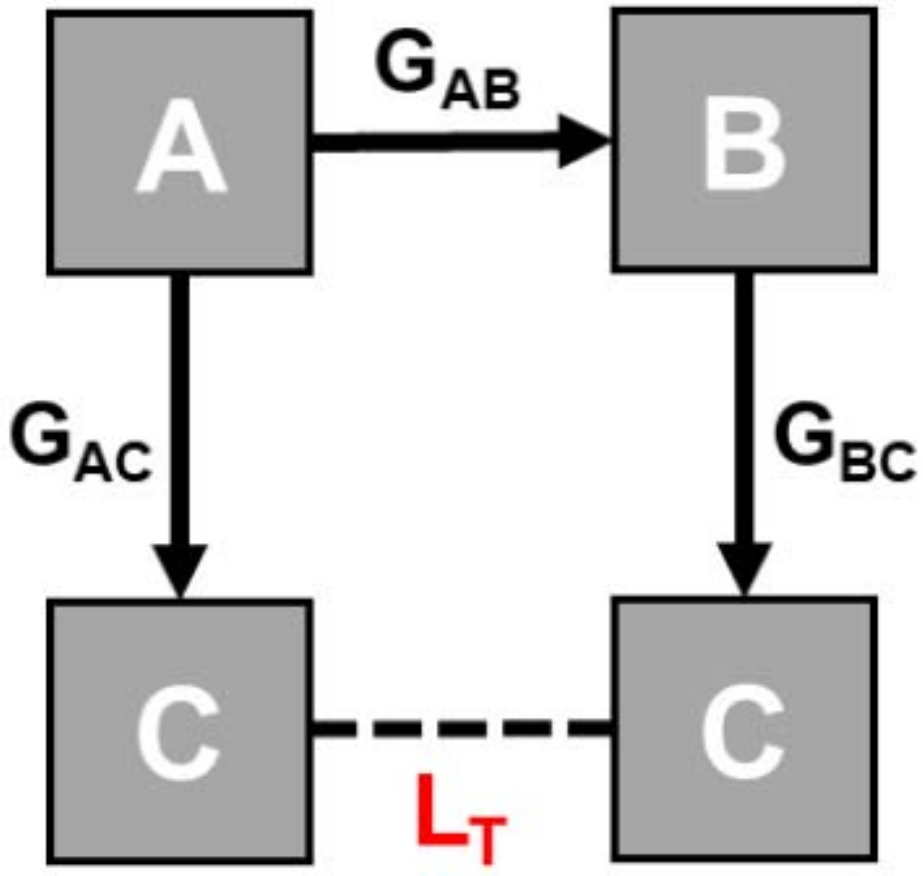}&
\includegraphics[width=0.5\textwidth]{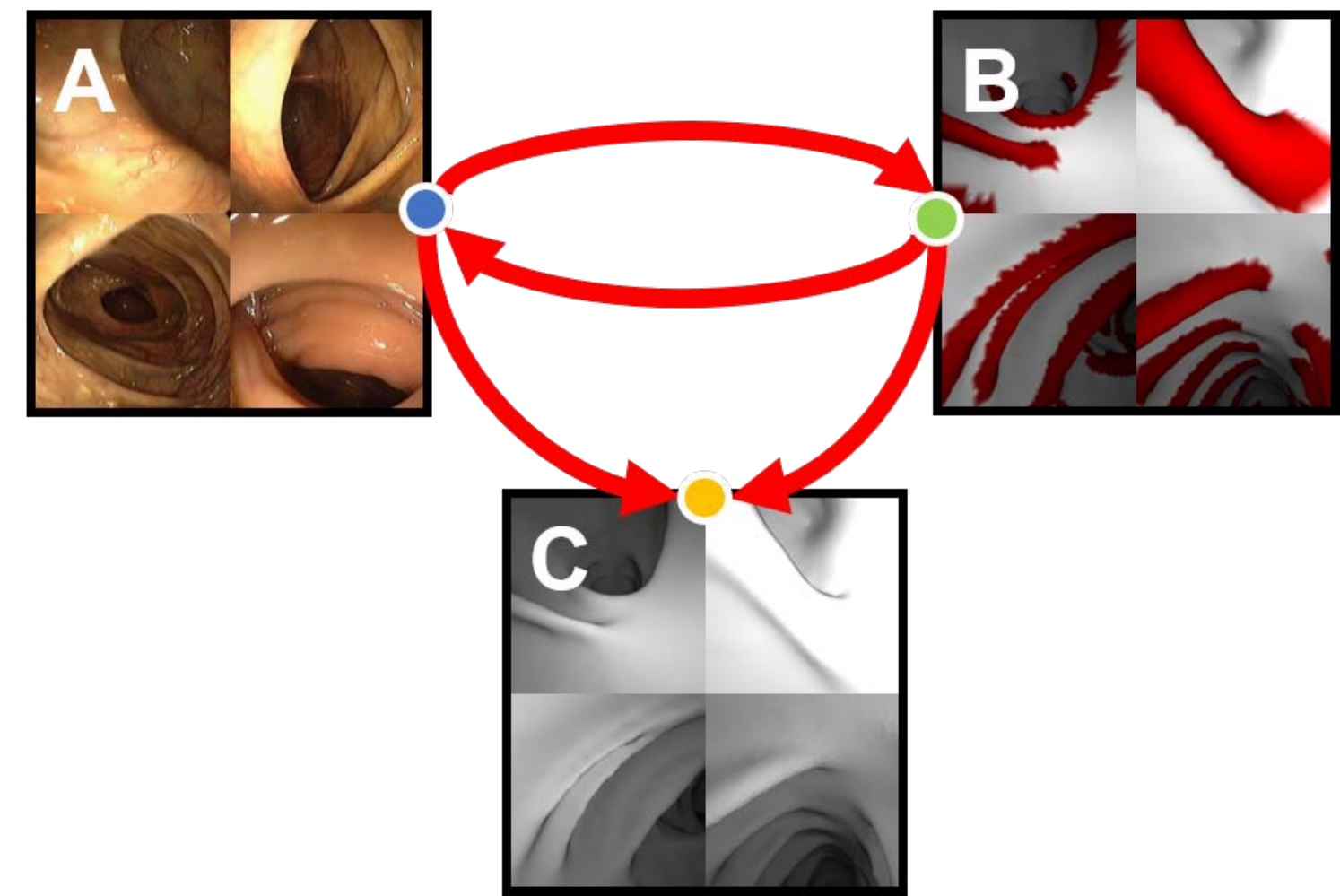}\\
(a) & (b)
\end{tabular}
\caption{(a) An image from domain $A$ is converted to domain $B$ and then $C$ through generators $G_{AB}$ and $G_{BC}$. This resulting image is compared against the image directly translated to domain $C$ via $G_{AC}$ to compute the transitive loss. (b) 
The relationship between the domains A, B, and C. We can translate between domains A and B, however, the result in domain C must remain the same. For FoldIt, A is the OC domains, B is the Haustral Fold annotation (red overlay) domains, and C is the VC domain.}
\vspace{-1mm}
\label{fig:TCS_loss}
\end{center}
\end{figure}

The OC and VC data were obtained from 10 patients who underwent a VC procedure followed by an OC procedure. Data from 7 patients were used for training and 3 for testing. The OC frames are cropped and rescaled to 256x256 to remove the borders in the frame. VC meshes are synthesized from abdominal CT scans using a pipeline described in \cite{nadeem2016computer}. These meshes do not align with the OC videos, as the shape of the colon changes between the two procedures. Flythroughs are created in the VC mesh along the centerline with random camera rotations, and 2 light sources on left and right side of the camera to replicate the OC procedure. These lights follow the inverse square fall-off property to create realistic lighting \cite{mahmood2018unsupervised}. The haustral folds on VC meshes were segmented using Fiedler vector and corresponding levelset representations, as outlined in \cite{nadeem2016corresponding}, and verified by a clinician. The segmented folds were overlaid on the VC mesh rendering and used as haustral fold annotation frames. Each video was split into 300 frames for a total of 3000 frames. Example frames from each domain are shown in Fig. \ref{fig:TCS_loss}b.

\section{Methods}

FoldIt uses 4 generators and discriminators. $G_{AB}$ translates images from domain $A$ to domain $B$, and $G_{BA}$ acts in the opposite direction. $G_{AC}$ and $G_{BC}$ translate from input domains, $A$ and $B$, to the common domain, $C$. Our approach can be broken down into four losses: adversarial losses, transitive losses, a ground truth loss, and identity losses. The complete objective function is as follows:
\begin{equation}
     \mathcal{L}_{obj} =  \lambda_{adv}\mathcal{L}_{adv} +  \lambda_{T}\mathcal{L}_{T} +  \lambda_{GT}\mathcal{L}_{GT}
     +\lambda_{idt}\mathcal{L}_{idt}
\end{equation}

Each of the 4 generators have an associated discriminator, $D$, which applies an adversarial loss to ensure the output image encompasses the features of the output domain. The adversarial loss for each of these GANs is described as:
\begin{equation}
     \mathcal{L}_{GAN}(G,D,A,B) =  \mathds{E}_{y \backsim p(B)} \big[ \textrm{log} (D(y))\big] + 
     \mathds{E}_{x \backsim p(A)} \big[\textrm{log}(1 - D(G(x))\big],
\end{equation}
where $y \backsim p(B)$ represents the output domain $B$ data distribution and $x \backsim p(A)$ represents the input domain $A$ data distribution. An adversarial loss is applied to each generator to create the adversarial component of the objective loss:
\begin{equation}
\begin{split}
     \mathcal{L}_{adv} = \ & \mathcal{L}_{GAN}(G_{AB},D_{AB},A,B) + \mathcal{L}_{GAN}(G_{BA},D_{BA},B,A)  \\                   & + \mathcal{L}_{GAN}(G_{AC},D_{AC},A,C) + \mathcal{L}_{GAN}(G_{BC},D_{BC},B,C)
\end{split}
\end{equation}

Cycle consistency losses have shown great results for the task of image-to-image translation. Our approach uses a modified cycle consistency loss, which resembles the transitive property, to learn the domain translation via a common domain. The main insight here is that the translation to the common domain, $C$, should be consistent between our domains $A$ and $B$. 
Here, $C$ is the VC domain. When we do the translation, $A \rightarrow B \rightarrow C$, the result should be the same as $A \rightarrow C$ (see Fig. \ref{fig:TCS_loss}a).
We can express this transitive loss as ($\|\cdot\|_1$ is $\ell 1$ norm):
\begin{equation}
    \mathcal{L}_{T}(G_{AB},G_{BC},G_{AC},A) =  \mathds{E}_{x \backsim p(A)} \| G_{BC}(G_{AB}(x) - G_{AC}(x)) \|_1
\end{equation}
This loss is applied on both directions in the total objective loss.
 
\begin{equation}
    \mathcal{L}_{T} = \mathcal{L}_{T}(G_{AB},G_{BC},G_{AC},A) + \mathcal{L}_{T}(G_{BA},G_{AC},G_{BC},B)
\end{equation}
$\mathcal{L}_{GT}$ is our ground truth loss,
meant to utilize the ground truth pairing between the common domain and one of the domains $A$ and $B$. It simply applies the $\ell1$ norm between the paired domains. Here, we have ground truth correspondence between haustral fold annotations and VC. The ground truth loss is:
\begin{equation}
  \mathcal{L}_{GT}(G_{BC},B,C) = \mathds{E}_{x,z \backsim p(B,C)} \|(G_{BC}(x) - z \|_1
\end{equation}
where $x,z \backsim p(A,C)$ represents the paired data distribution $(A,C)$.
 
Lastly, we have the identity loss, which is meant for additional stability during training as described in \cite{mathew2020augmenting,zhu2017unpaired}. An image from the output domain is expected to be unchanged when passed as input through the network under an identity loss. This is applied only to $G_{AC}$ and $G_{BC}$ as alterations in the output for $G_{AB}$ and $G_{BA}$ may still be valid. The identity loss is as follows:
\begin{equation}
  \mathcal{L}_{idt}(G_{AC},C) = \mathds{E}_{z \backsim p(C)} \|(G_{AC}(z) - z \|_1
\end{equation}
The complete identity loss is:
\begin{equation}
 \mathcal{L}_{idt} = \mathcal{L}_{idt}(G_{AC},C) + \mathcal{L}_{idt}(G_{BC},C)
\end{equation}

Each generator used up 18 MB for its ResNet architecture \cite{he2016deep} with 9 blocks. We use a PatchGAN discriminator \cite{isola2017image} similar to \cite{zhu2017unpaired}, which used 3MB each. The network was trained for 100 epochs and used the following weights: $\lambda_{adv} = 1, \lambda_{T} = 10, \lambda_{GT} = 1, \lambda_{idt} = 1$. The inference time for a single image is 0.04 seconds on an Nvidia RTX 6000 GPU.

\section{Results}

\begin{figure*}[t!]
\begin{center}
\scriptsize
\begin{tabular}{cccccccc}
\rotatebox{90}{\rlap{~Texture1-VC}}&
\includegraphics[width=0.14\textwidth]{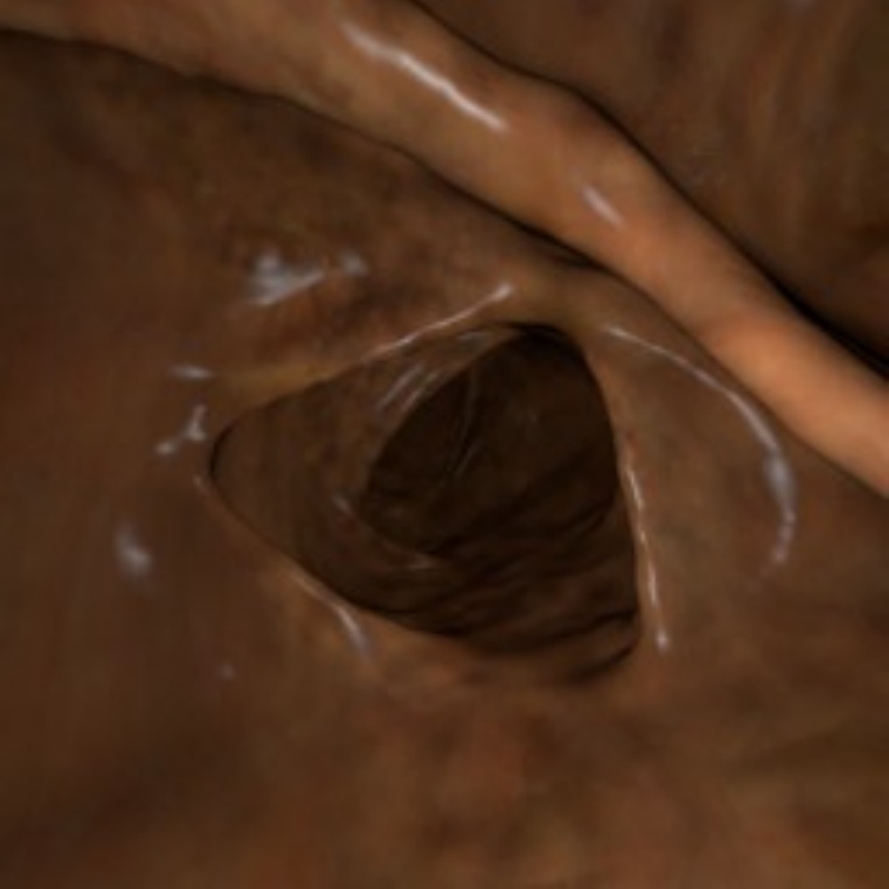}&
\includegraphics[width=0.14\textwidth]{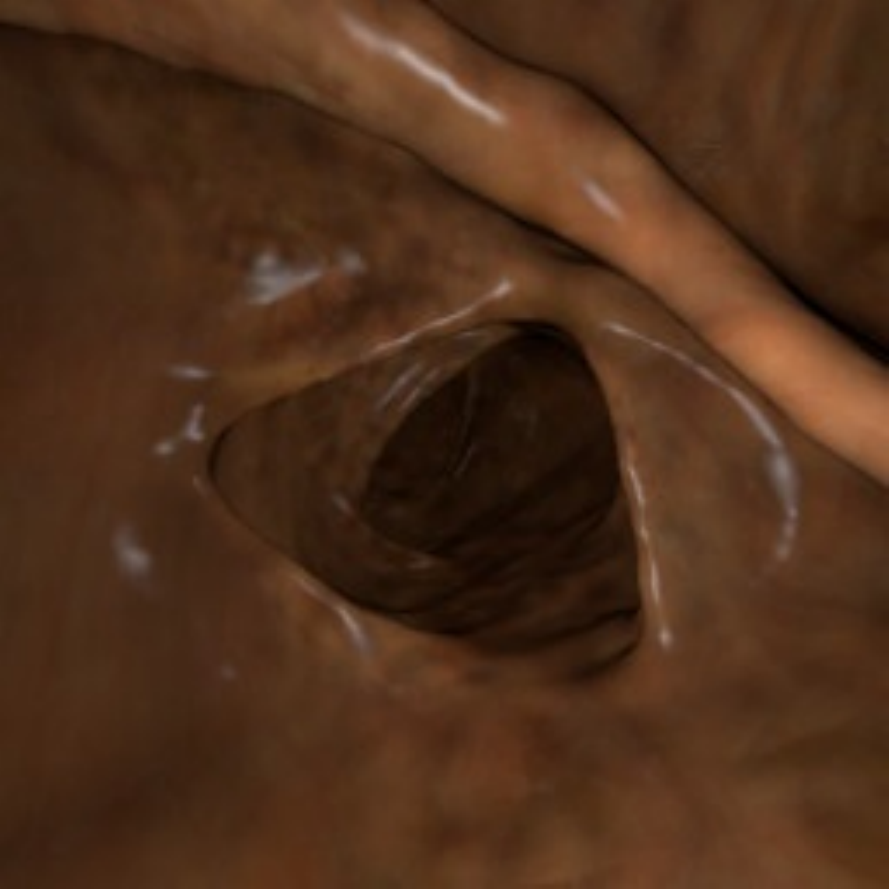}&
\includegraphics[width=0.14\textwidth]{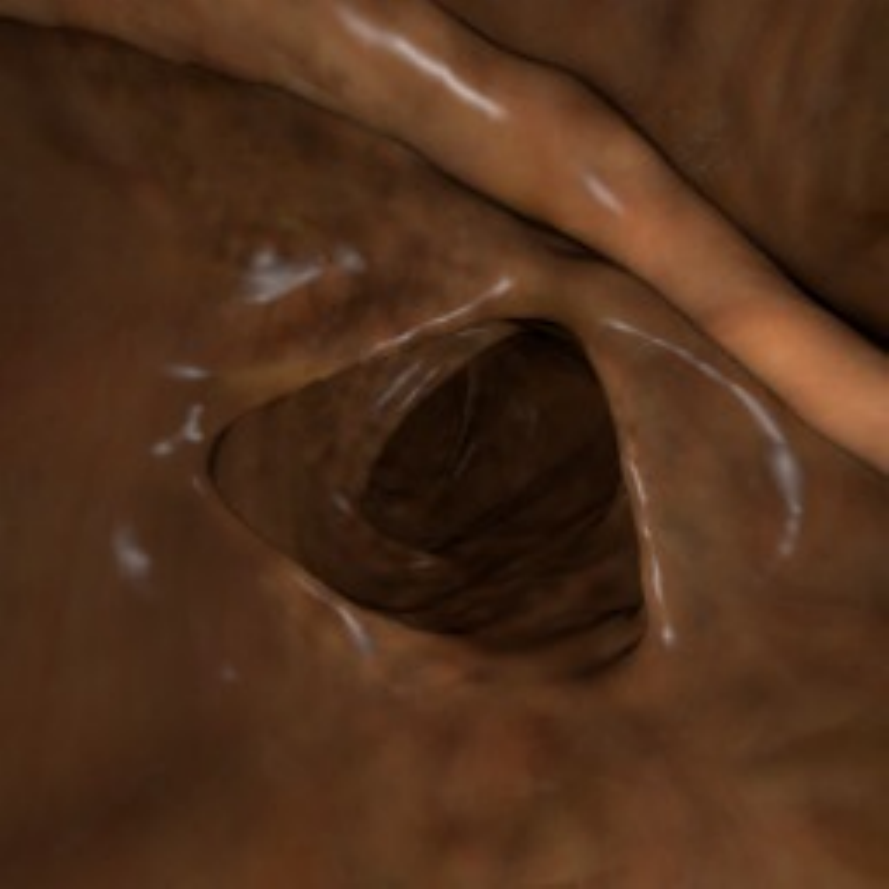}&
\ldots&
\includegraphics[width=0.14\textwidth]{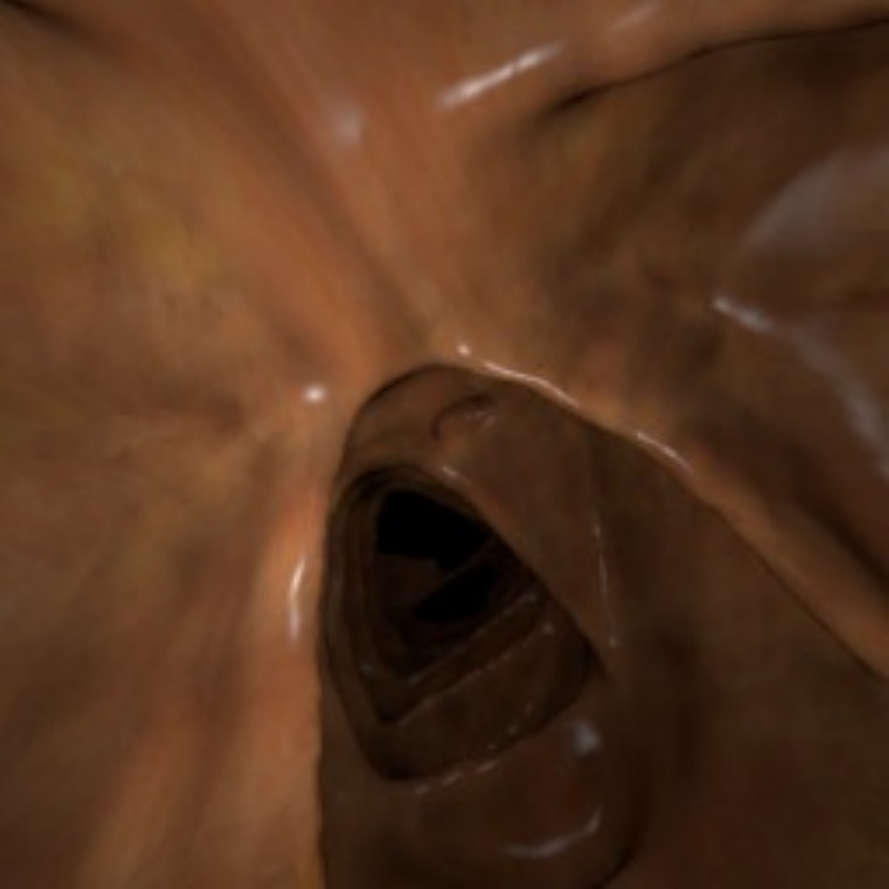}&
\includegraphics[width=0.14\textwidth]{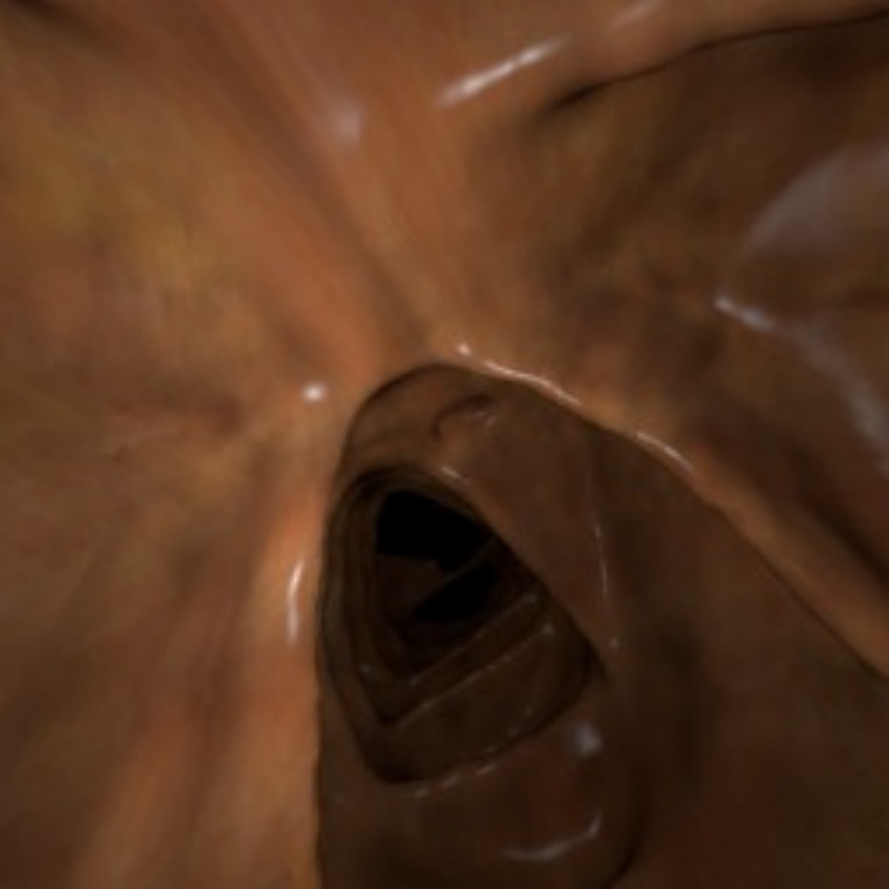}&
\includegraphics[width=0.14\textwidth]{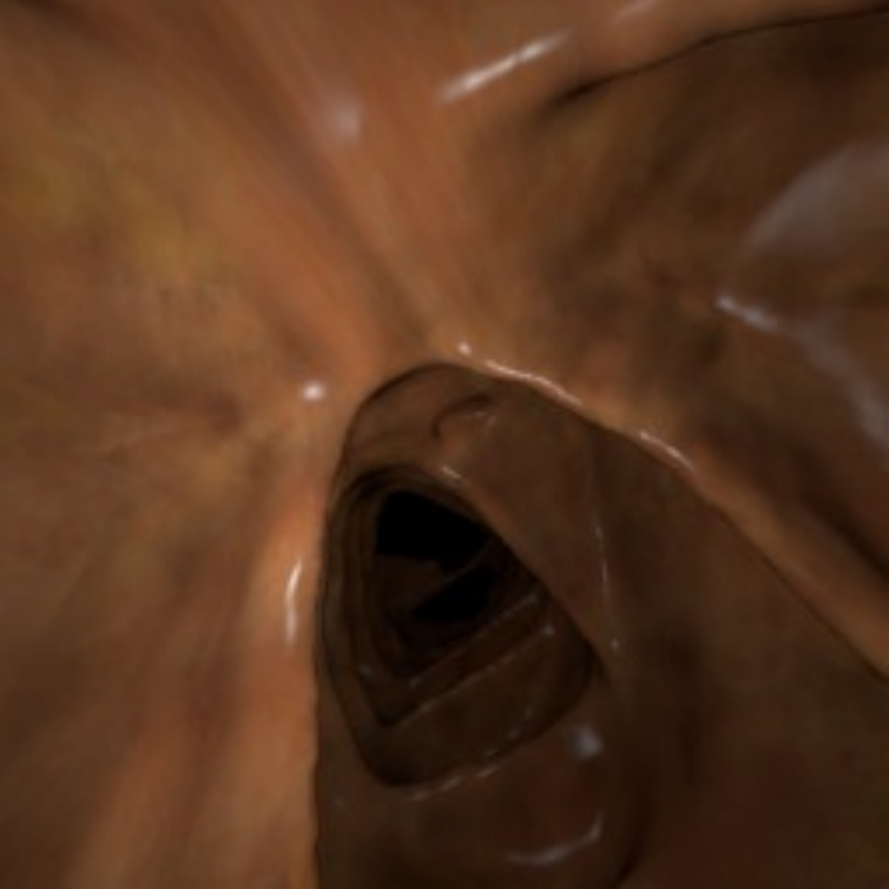}\\

\rotatebox{90}{\rlap{~Texture2-VC}}&
\includegraphics[width=0.14\textwidth]{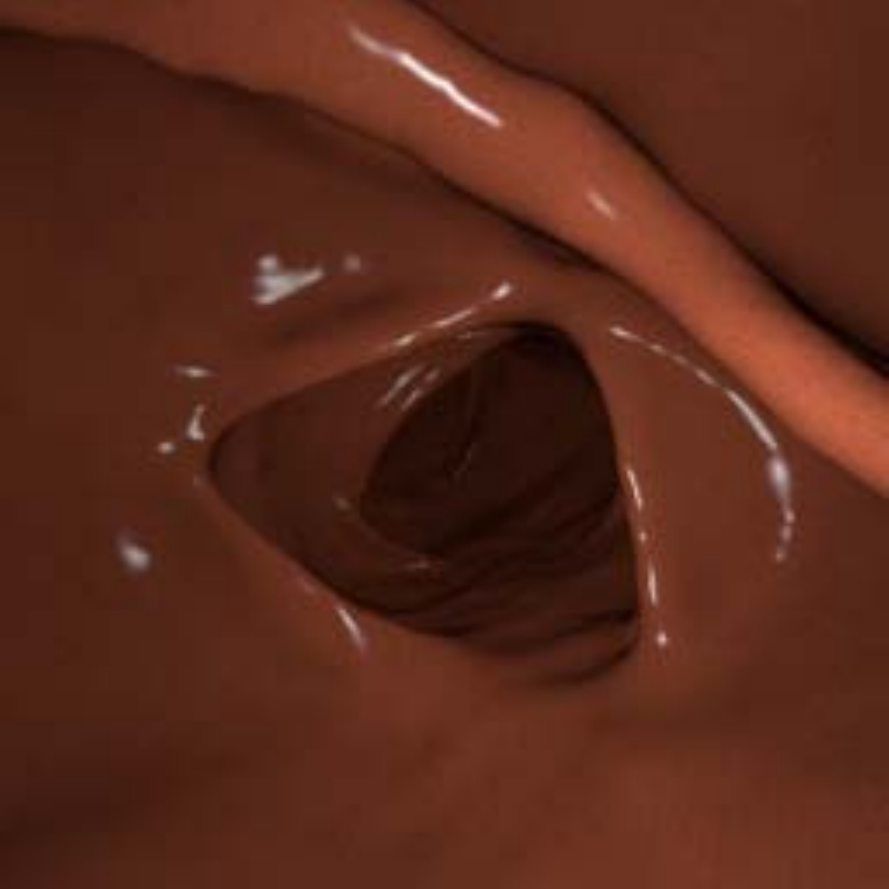}&
\includegraphics[width=0.14\textwidth]{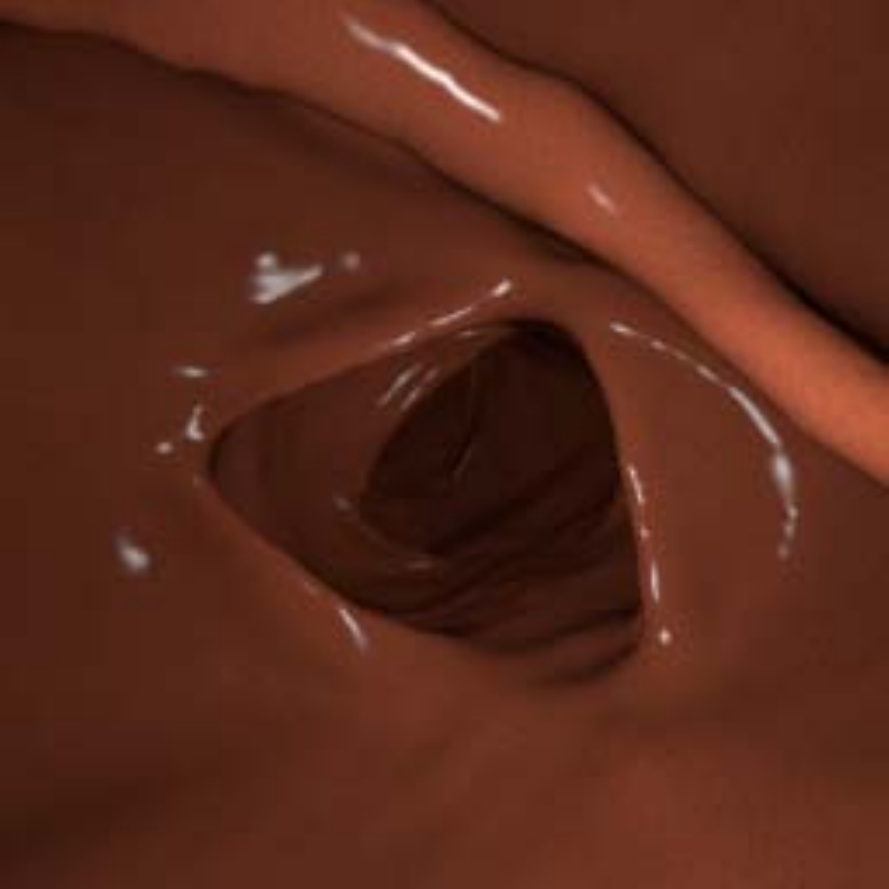}&
\includegraphics[width=0.14\textwidth]{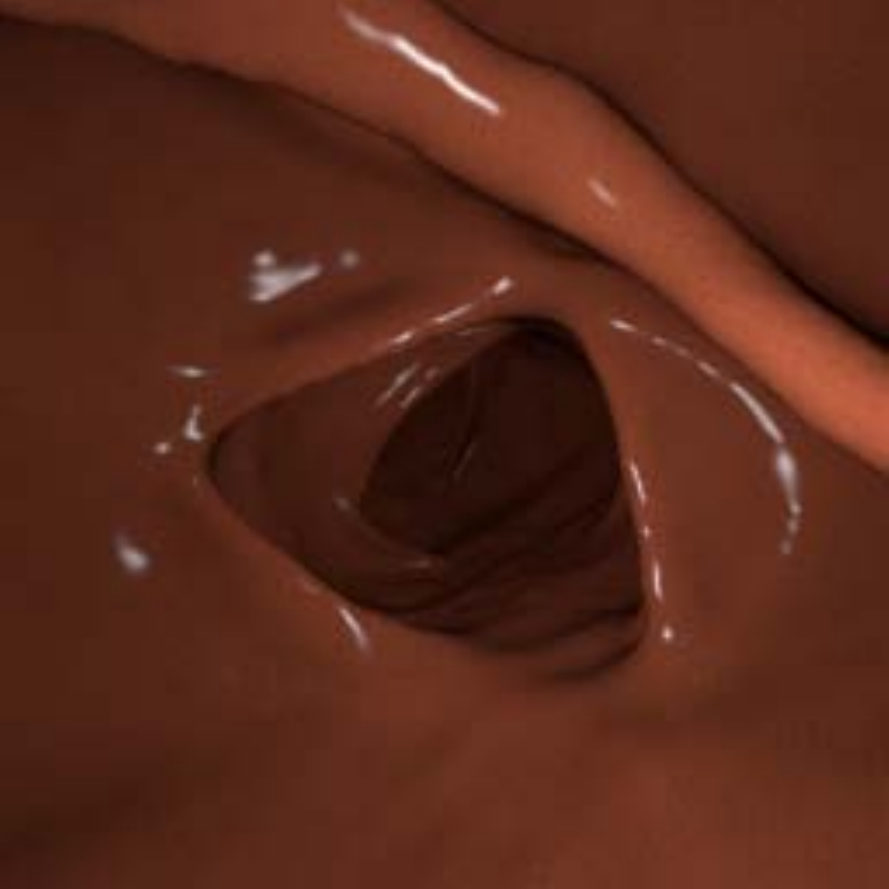}&
\ldots&
\includegraphics[width=0.14\textwidth]{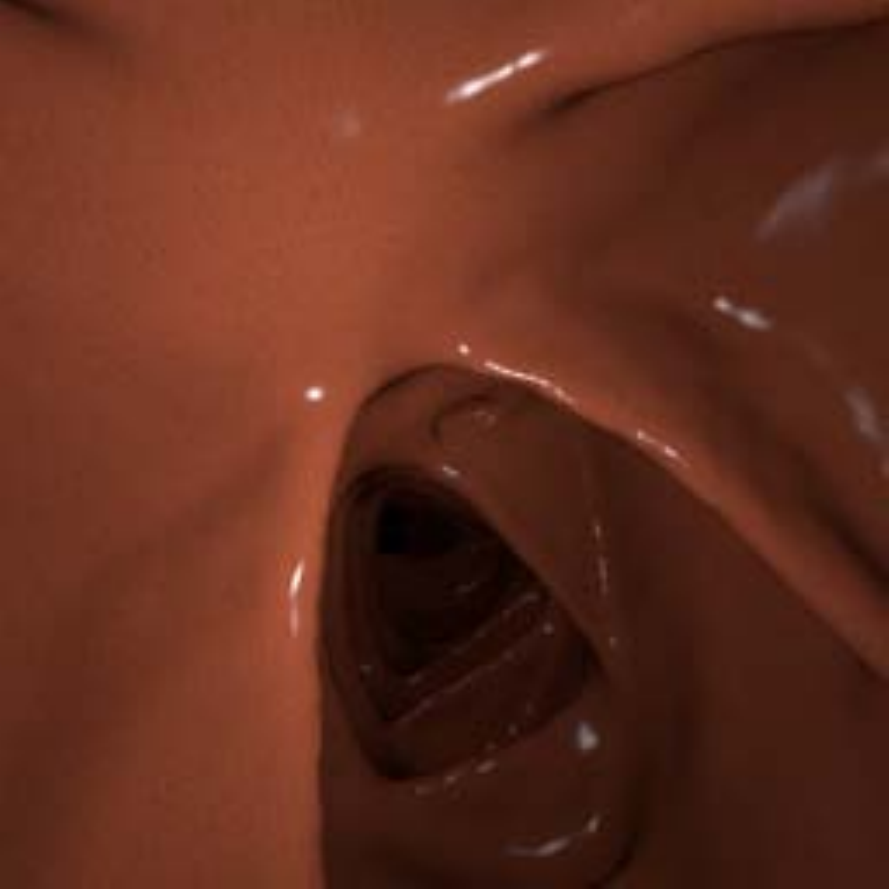}&
\includegraphics[width=0.14\textwidth]{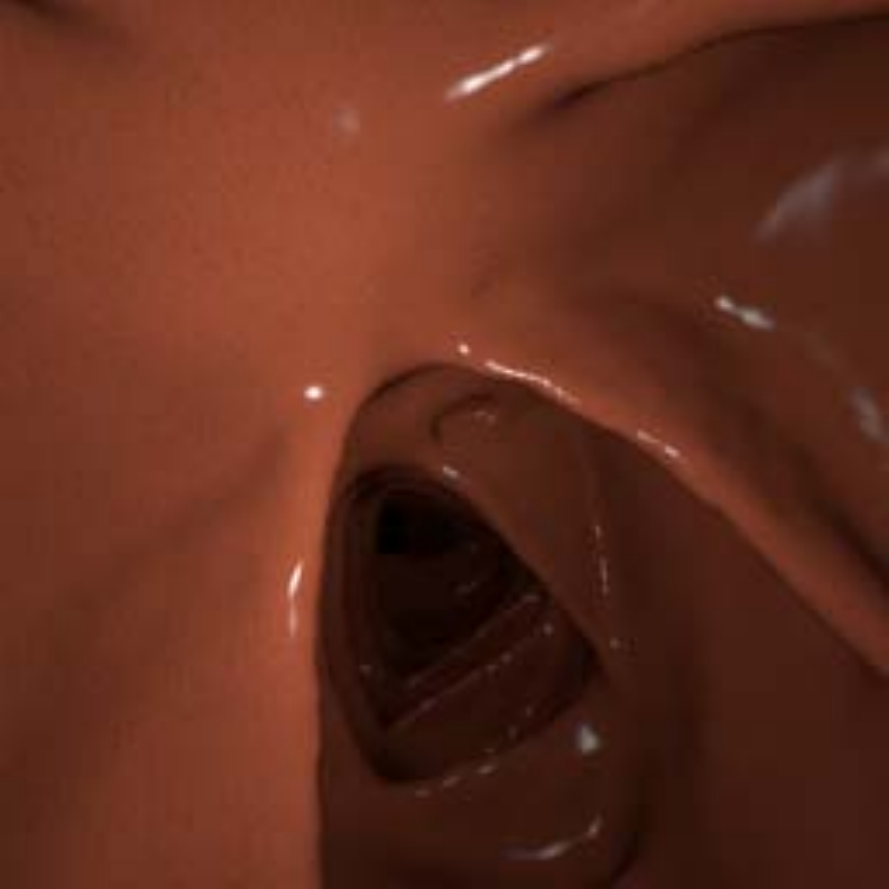}&
\includegraphics[width=0.14\textwidth]{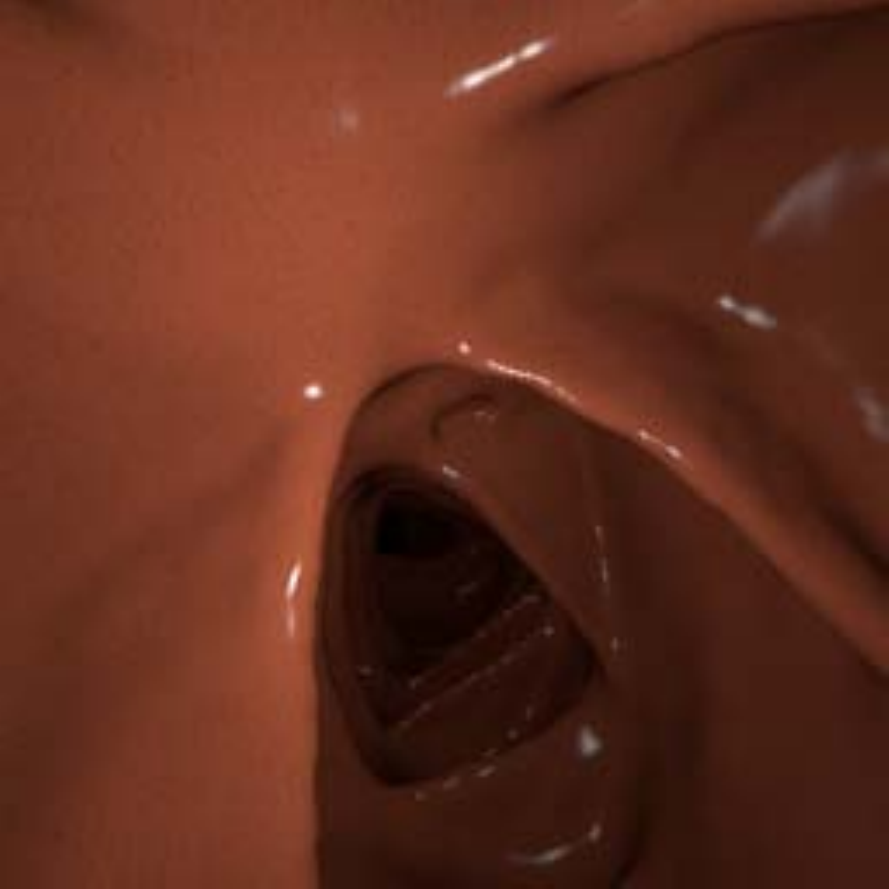}\\

\rotatebox{90}{\rlap{Ground Truth}}&
\includegraphics[width=0.14\textwidth]{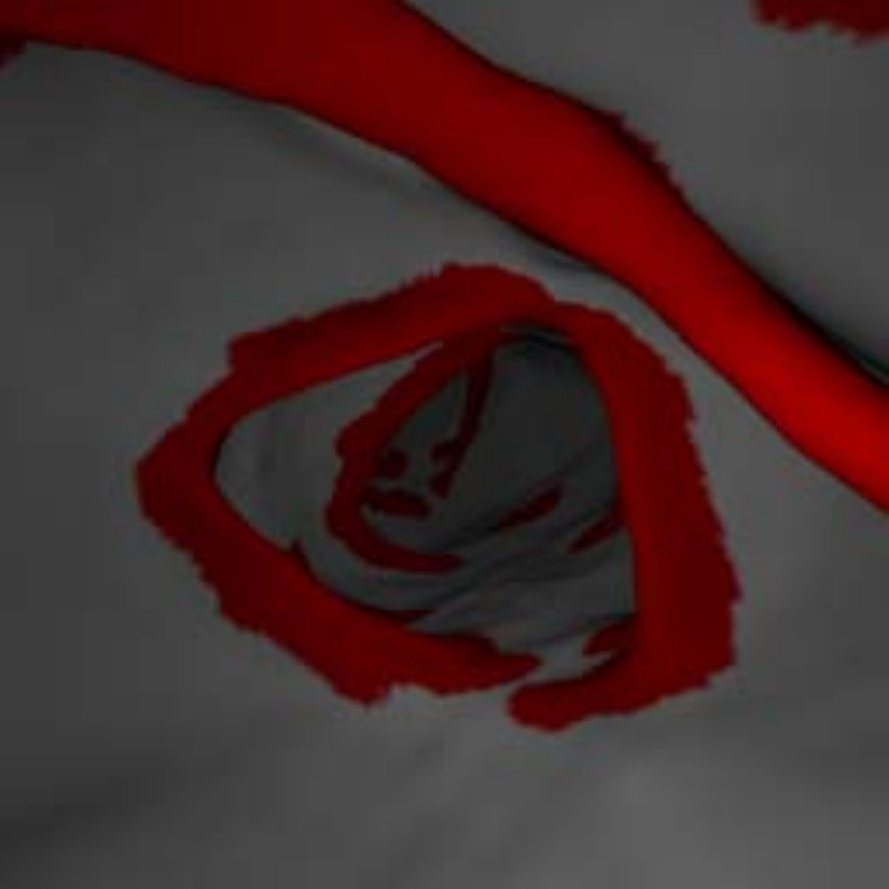}&
\includegraphics[width=0.14\textwidth]{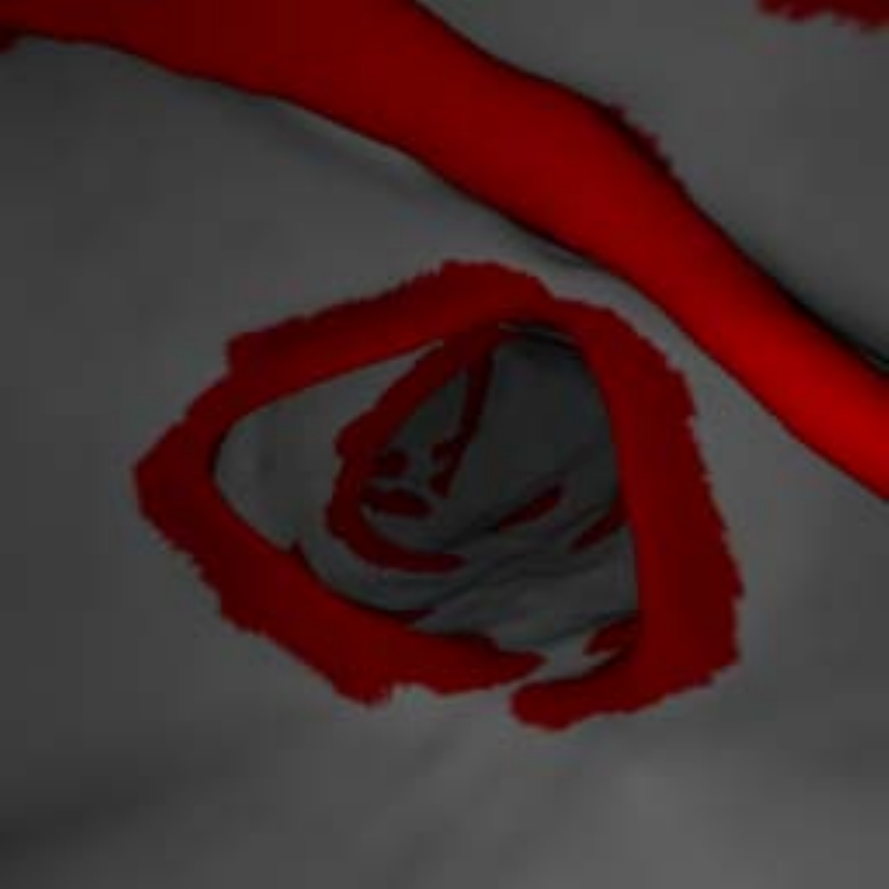}&
\includegraphics[width=0.14\textwidth]{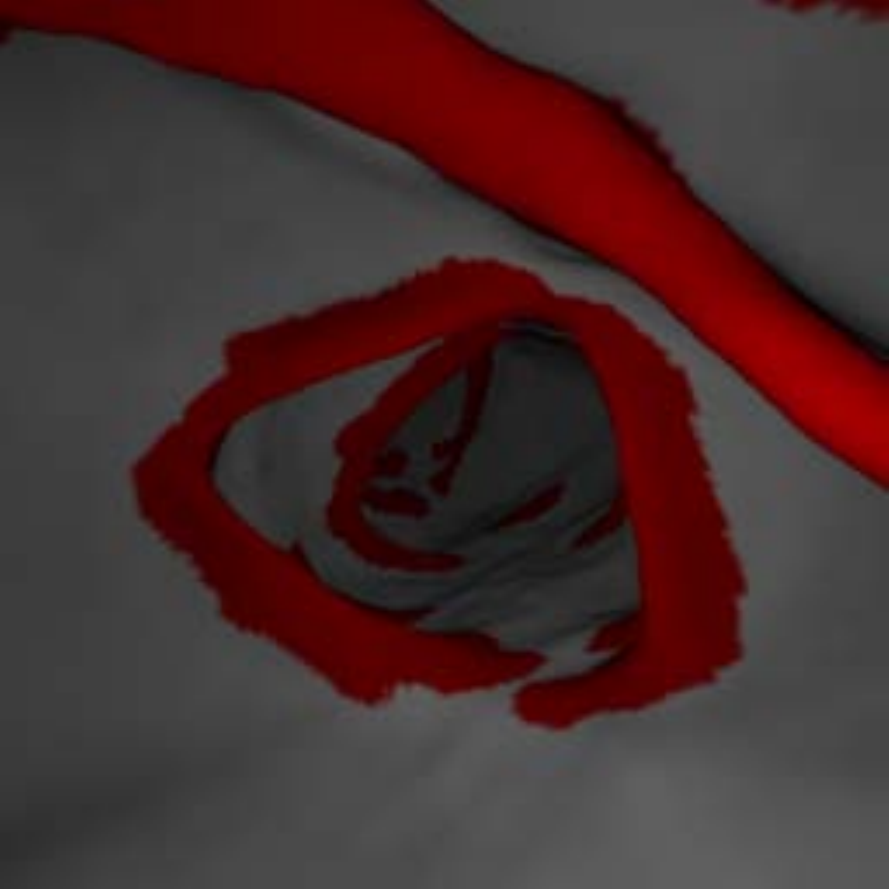}&
\ldots&
\includegraphics[width=0.14\textwidth]{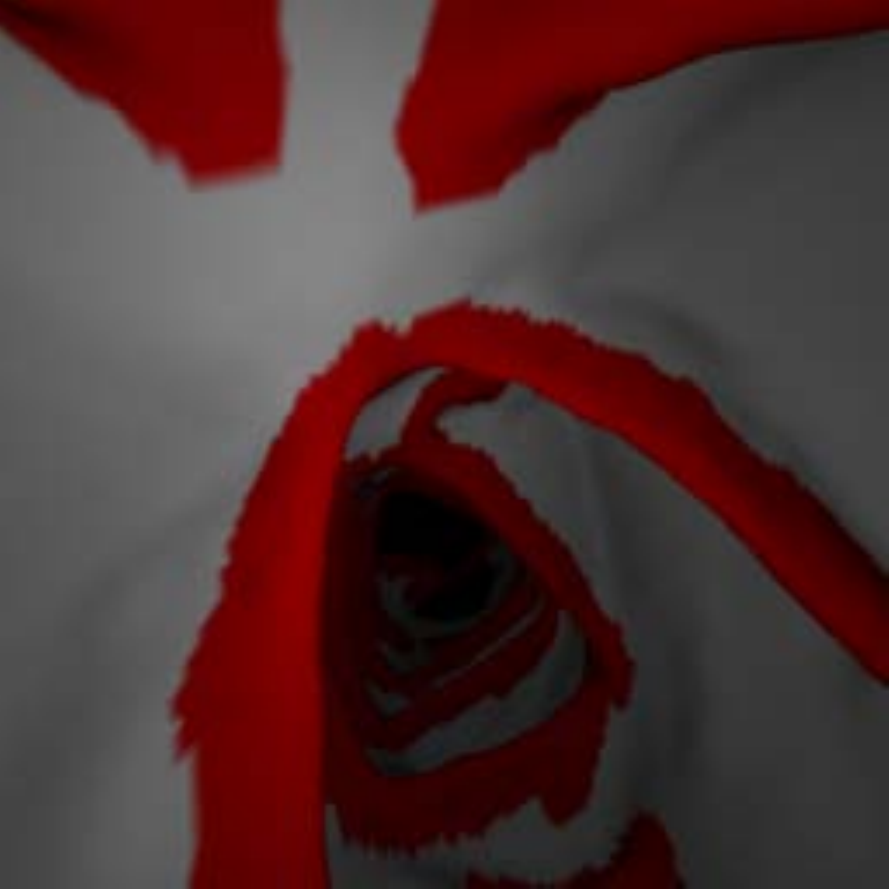}&
\includegraphics[width=0.14\textwidth]{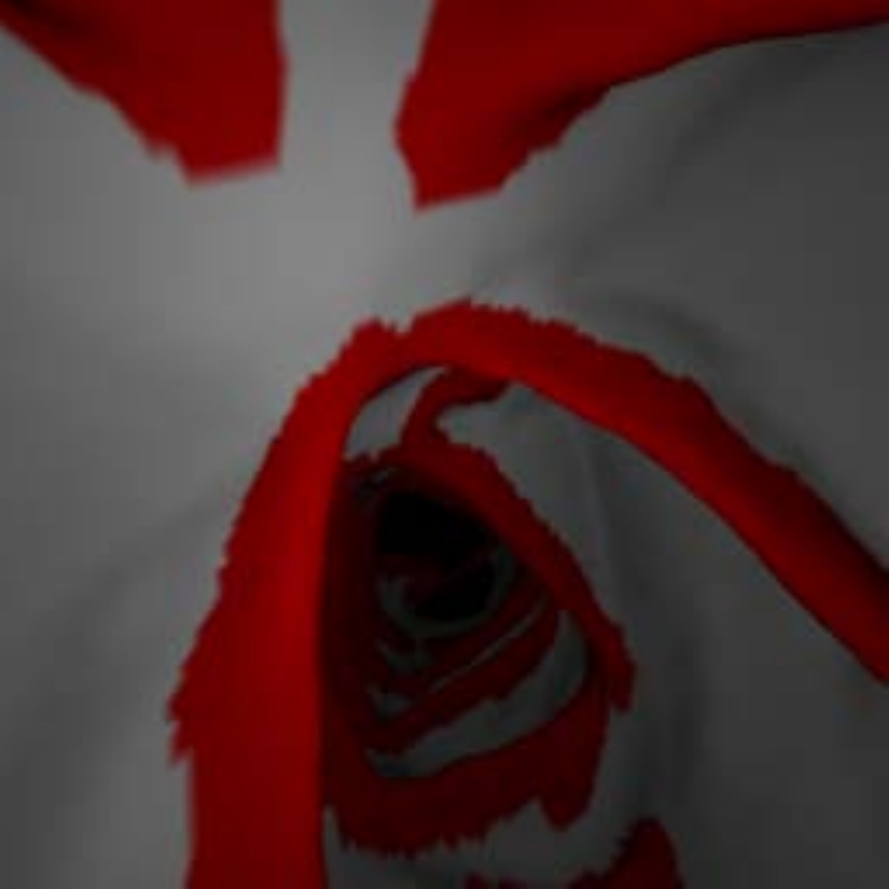}&
\includegraphics[width=0.14\textwidth]{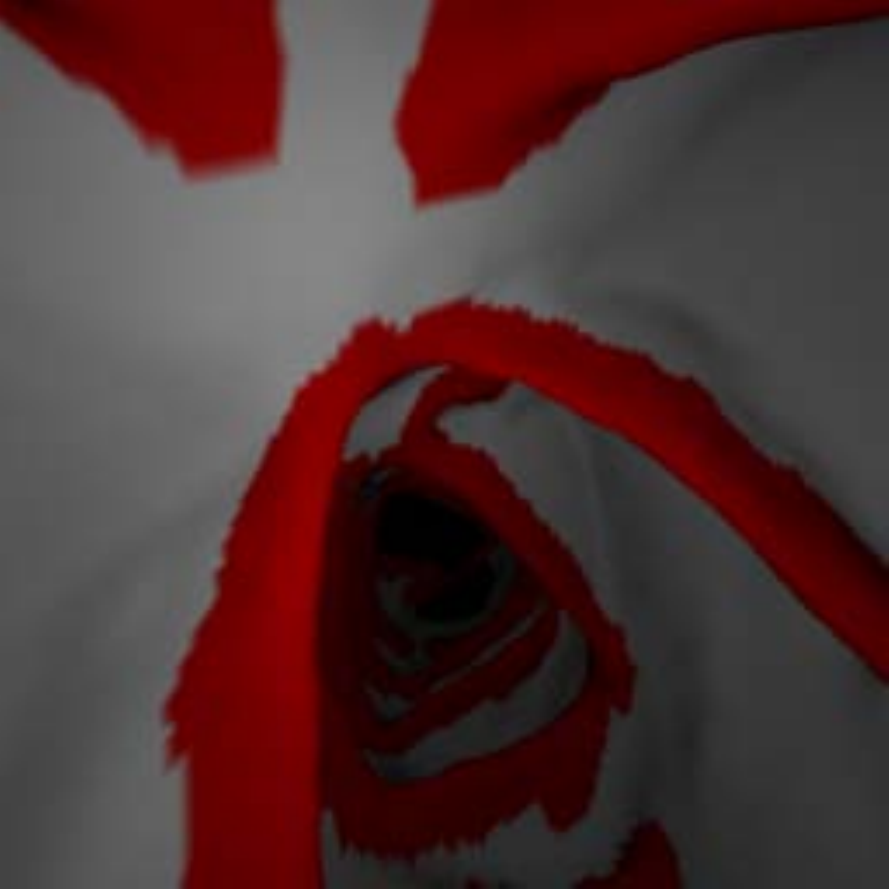}\\

\hline
\hline 

\rotatebox{90}{\rlap{\textbf{~Tex1-Ours}}}&
\includegraphics[width=0.14\textwidth]{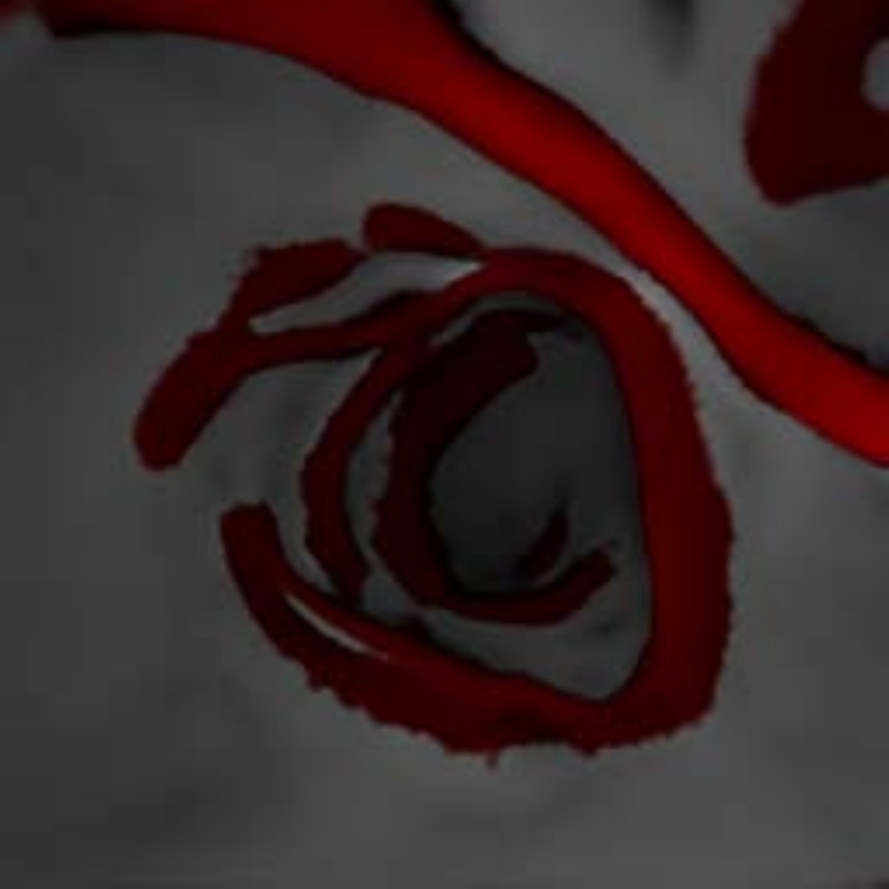}&
\includegraphics[width=0.14\textwidth]{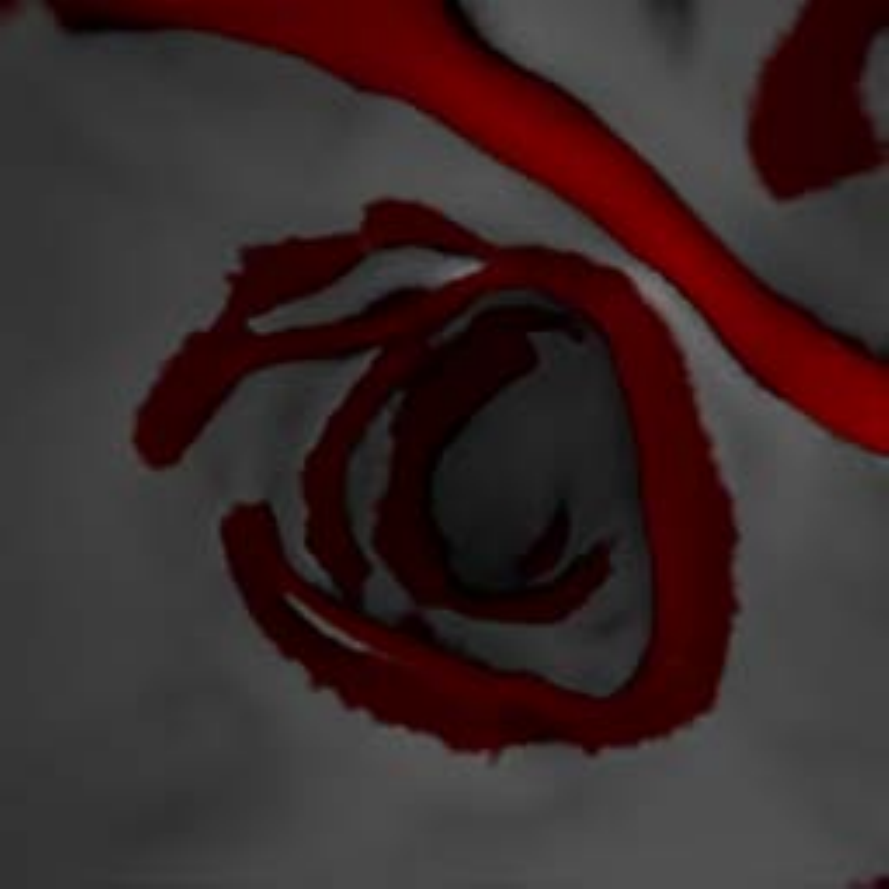}&
\includegraphics[width=0.14\textwidth]{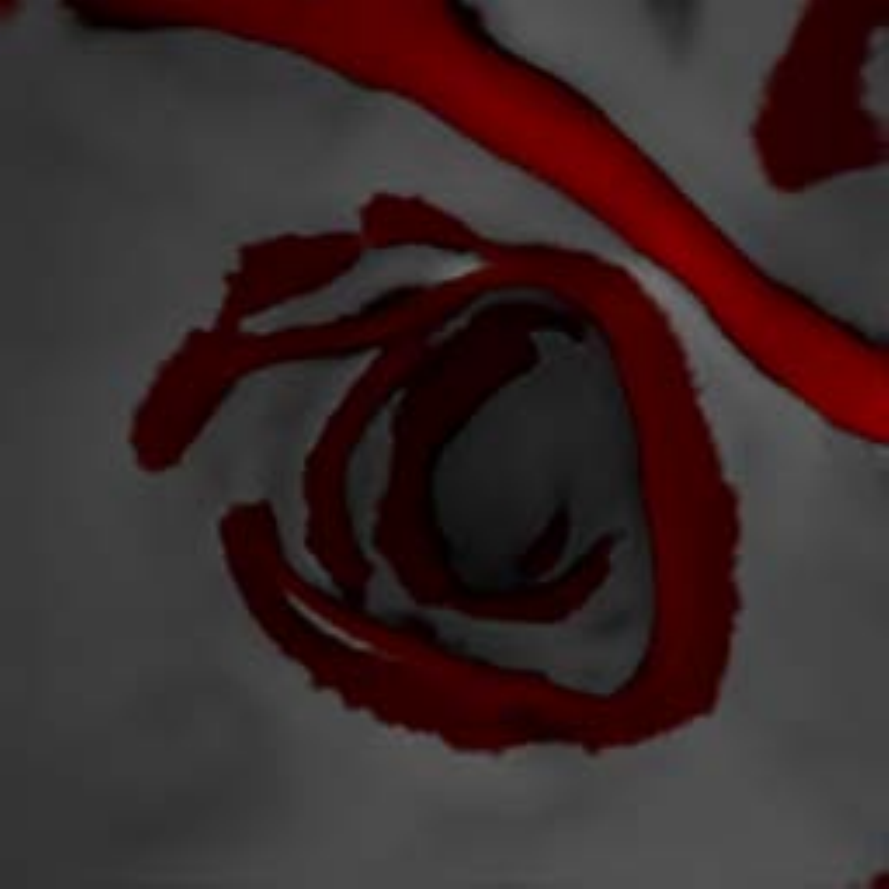}&
\ldots&
\includegraphics[width=0.14\textwidth]{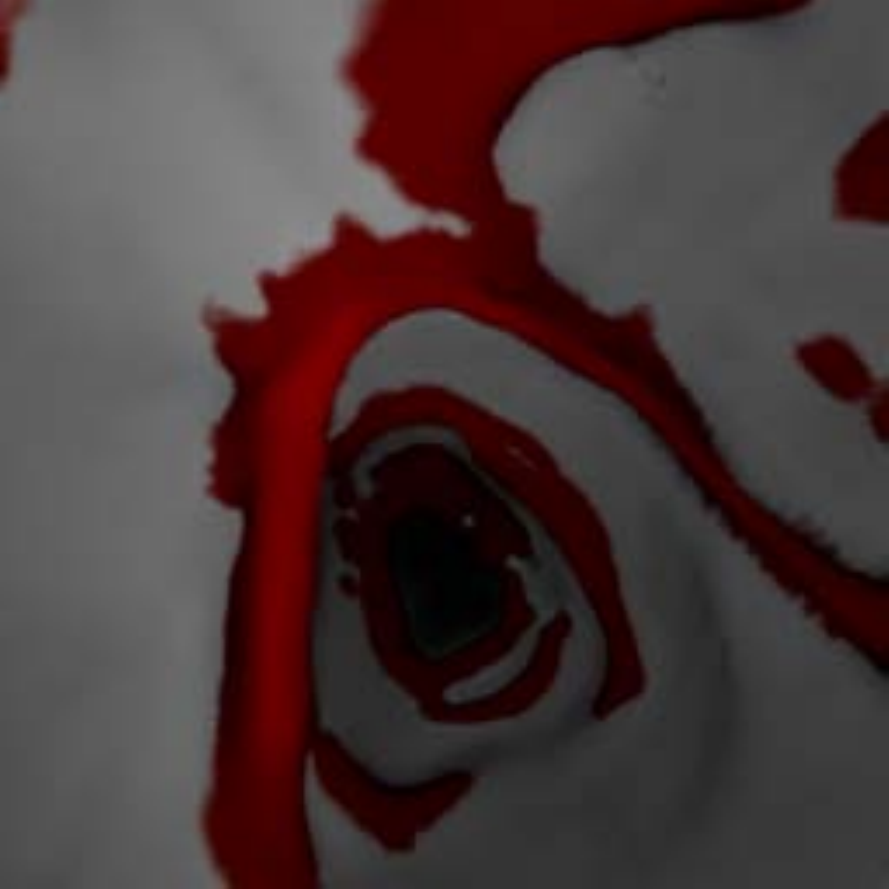}&
\includegraphics[width=0.14\textwidth]{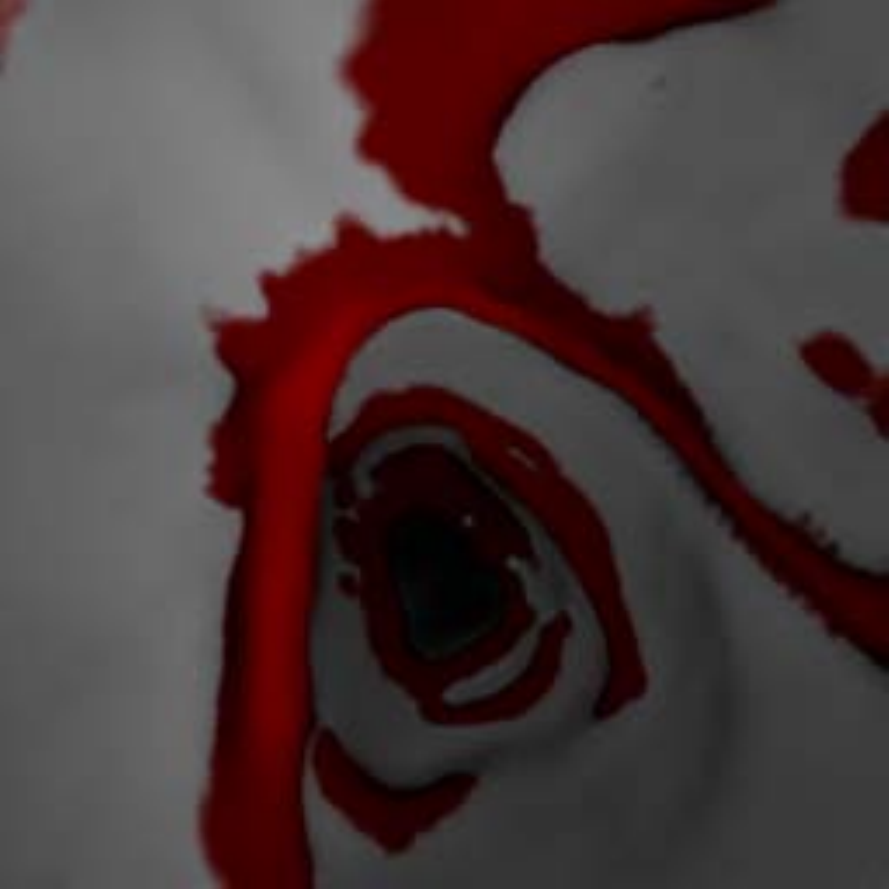}&
\includegraphics[width=0.14\textwidth]{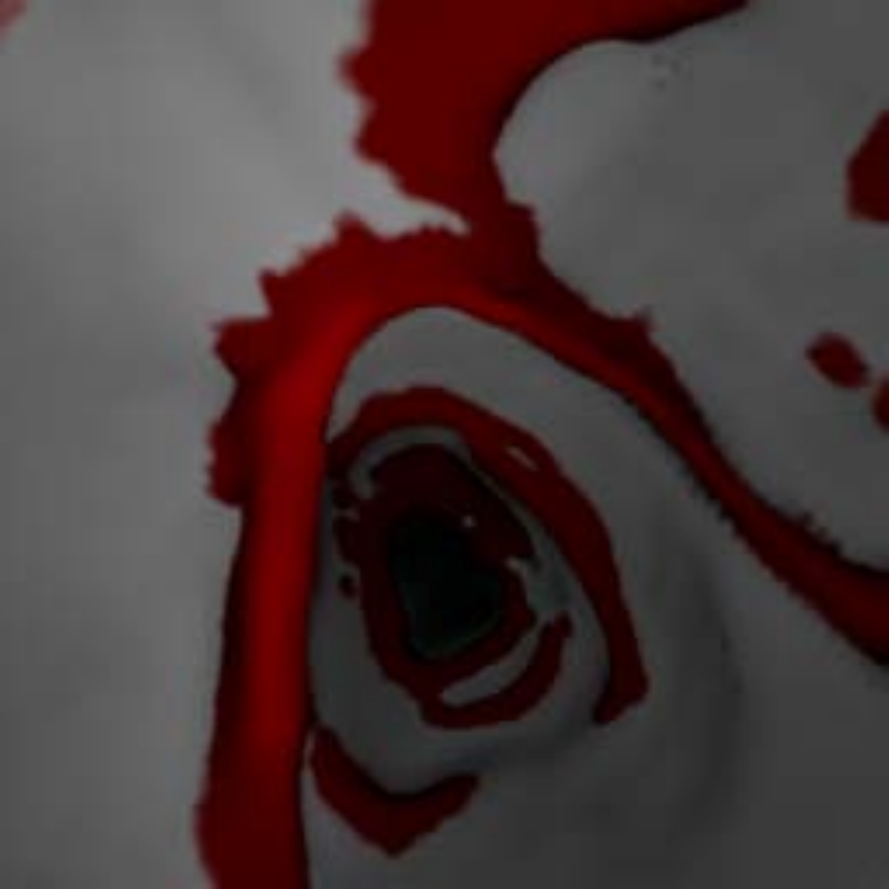}\\

\rotatebox{90}{\rlap{\textbf{~Tex2-Ours}}}&
\includegraphics[width=0.14\textwidth]{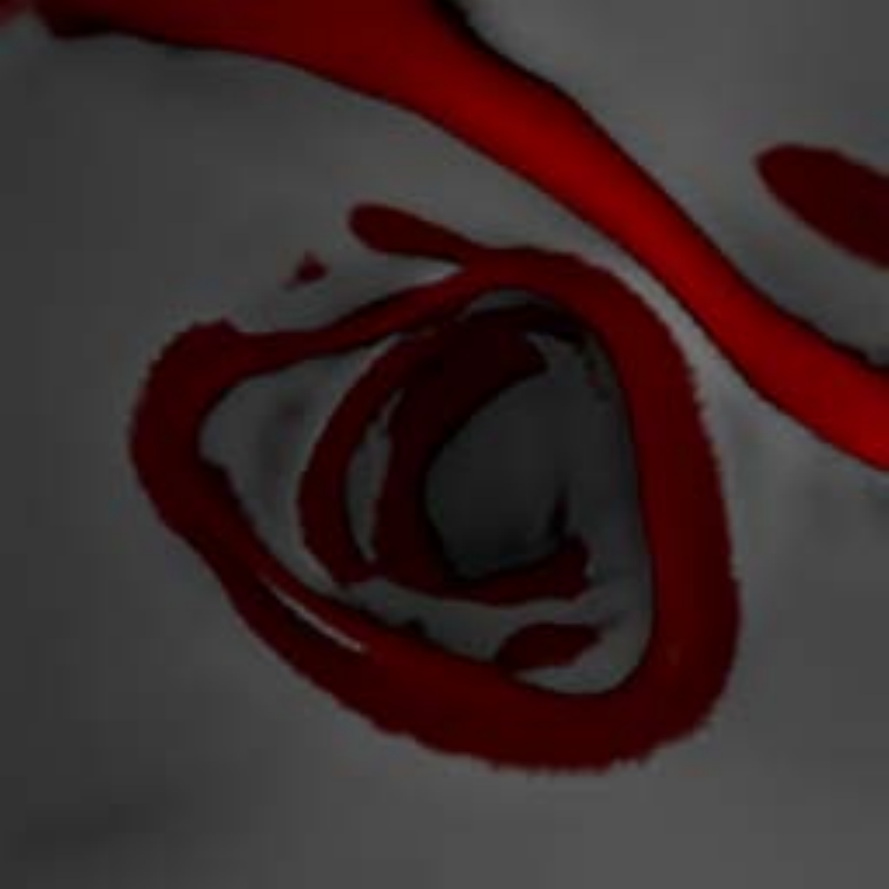}&
\includegraphics[width=0.14\textwidth]{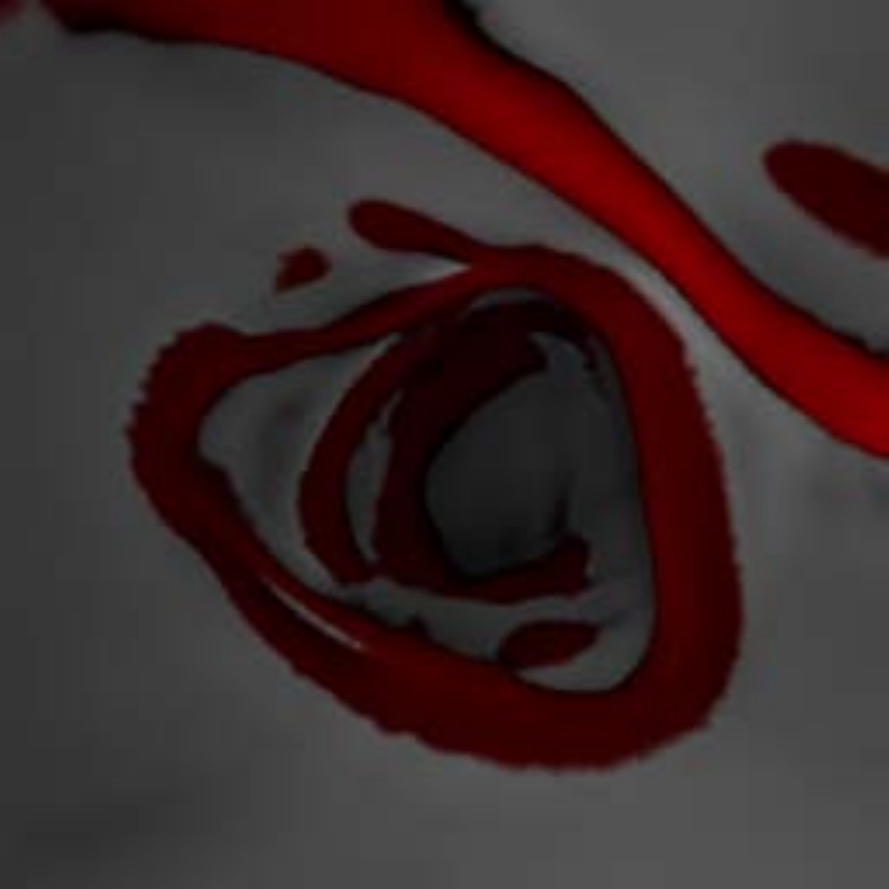}&
\includegraphics[width=0.14\textwidth]{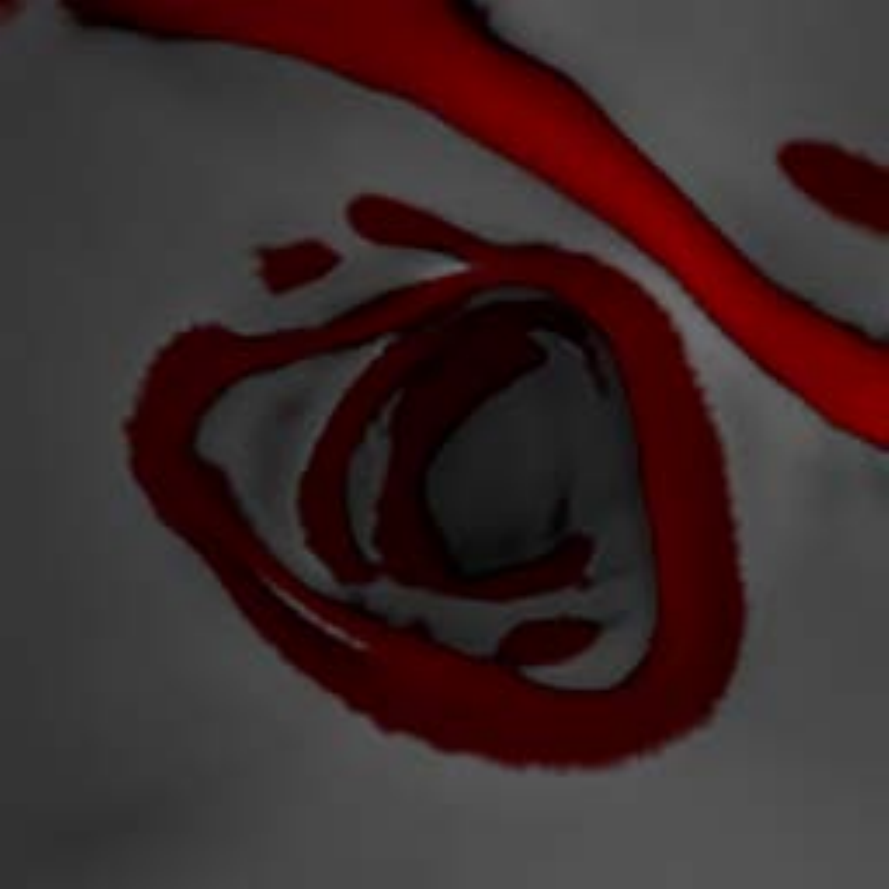}&
\ldots&
\includegraphics[width=0.14\textwidth]{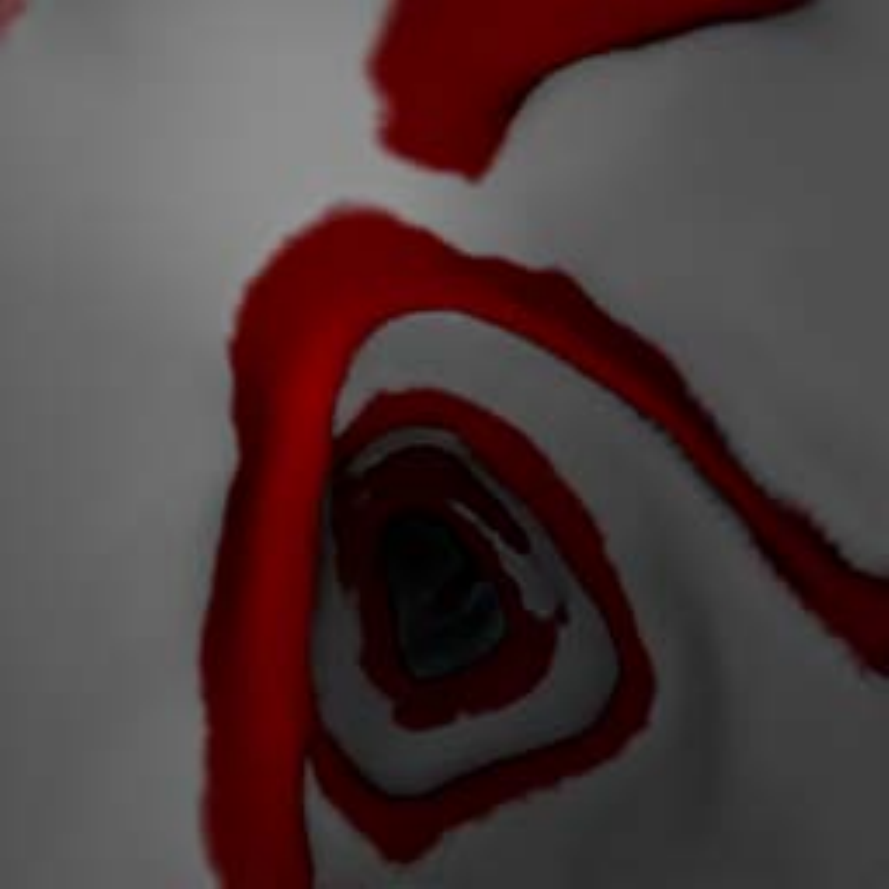}&
\includegraphics[width=0.14\textwidth]{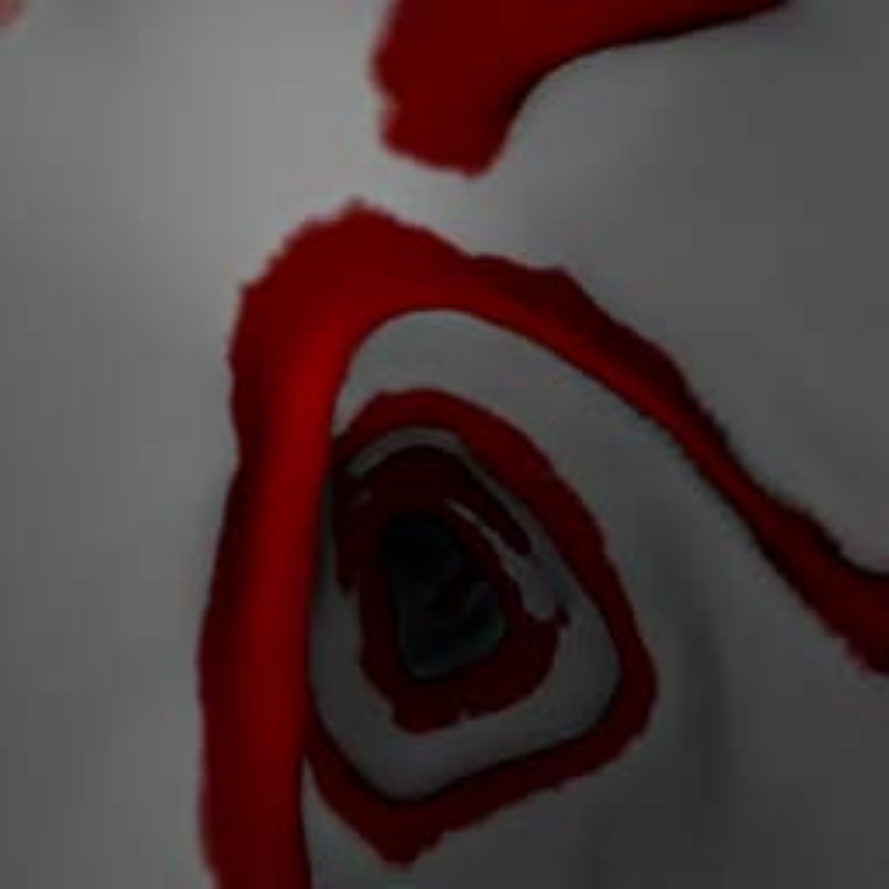}&
\includegraphics[width=0.14\textwidth]{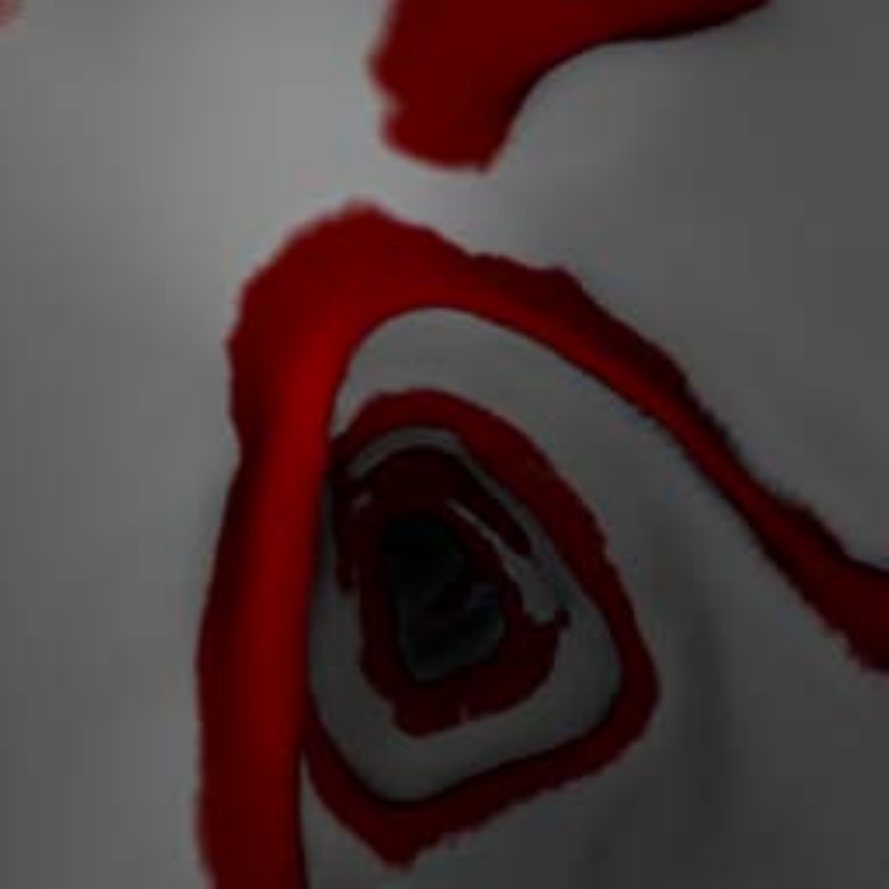}\\

\hline
\hline

\rotatebox{90}{\rlap{{Texture1 \cite{mathew2020augmenting}}}}&
\includegraphics[width=0.14\textwidth]{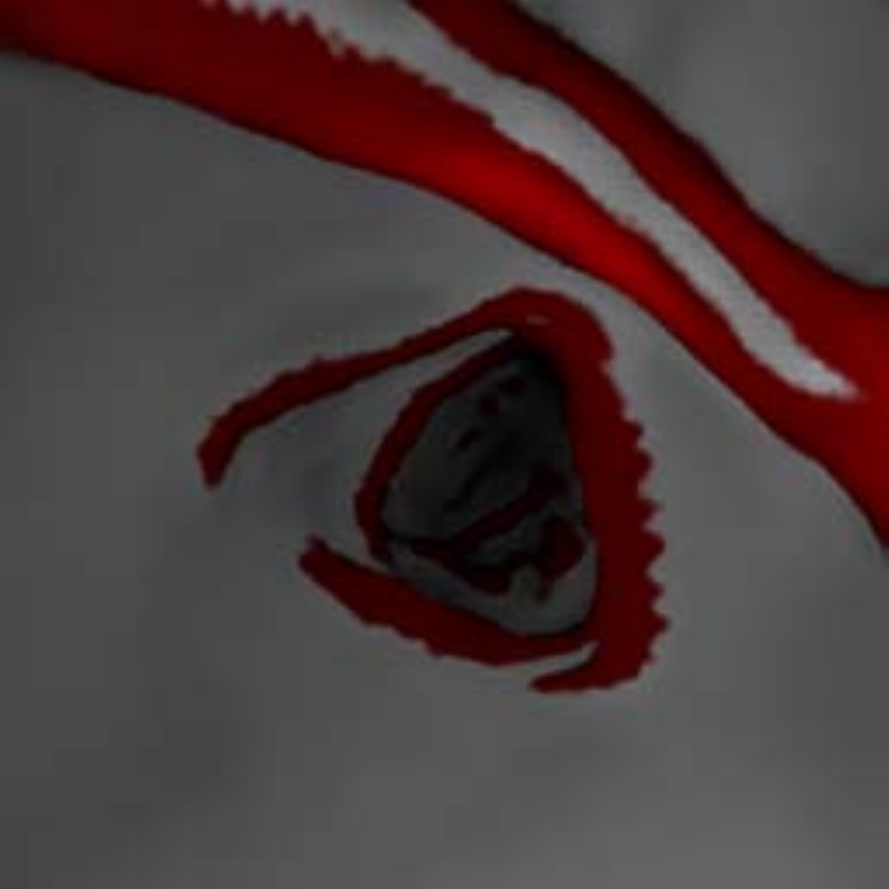}&
\includegraphics[width=0.14\textwidth]{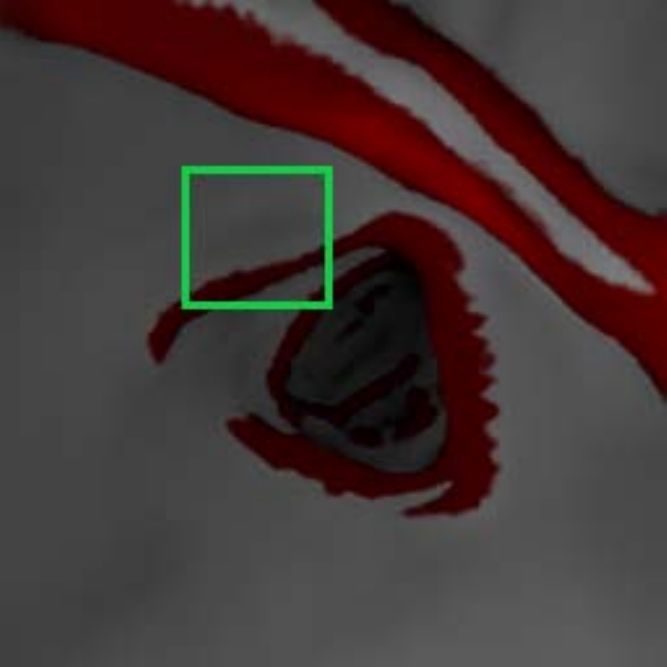}&
\includegraphics[width=0.14\textwidth]{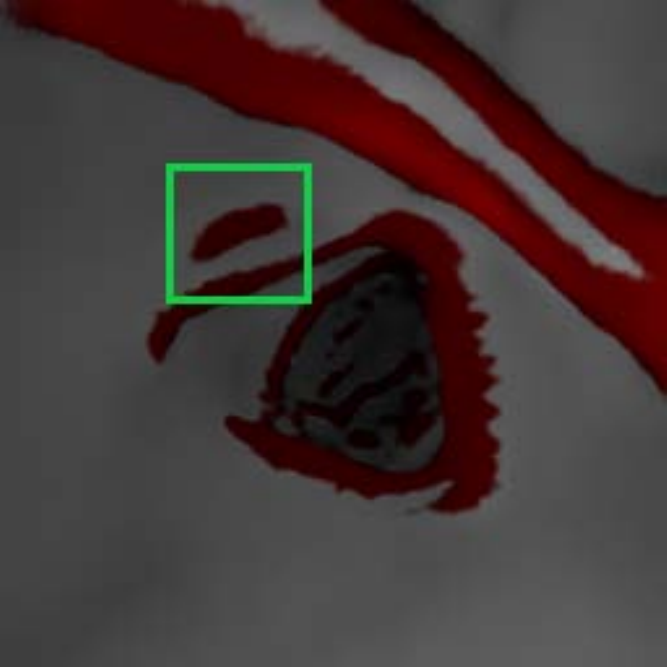}&
\ldots&
\includegraphics[width=0.14\textwidth]{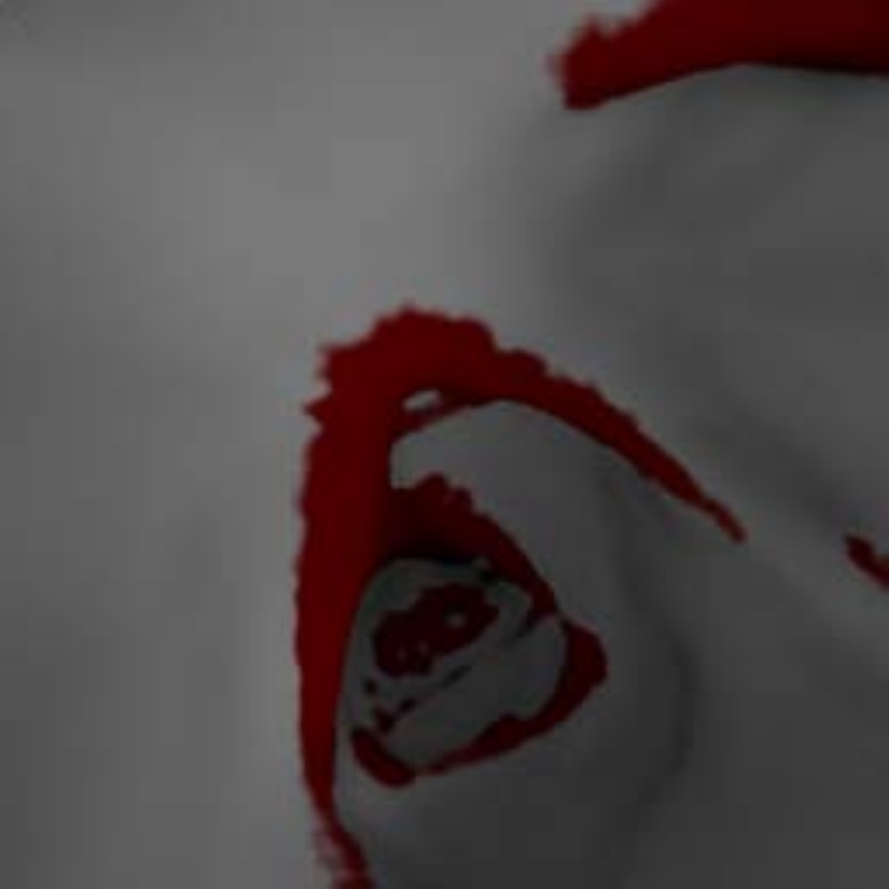}&
\includegraphics[width=0.14\textwidth]{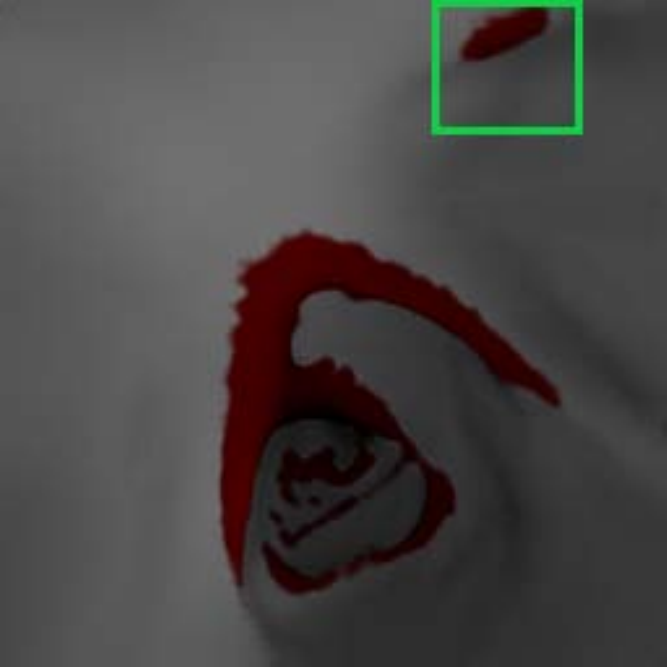}&
\includegraphics[width=0.14\textwidth]{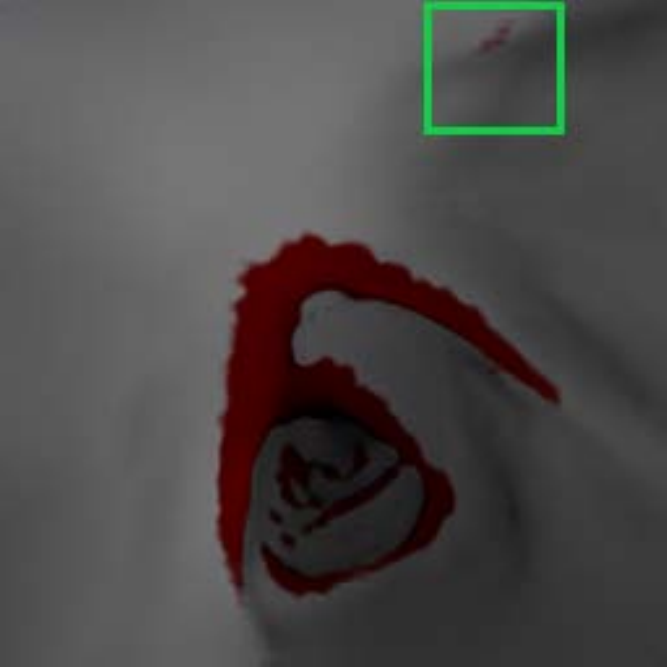}\\

\rotatebox{90}{\rlap{{Texture2 \cite{mathew2020augmenting}}}}&
\includegraphics[width=0.14\textwidth]{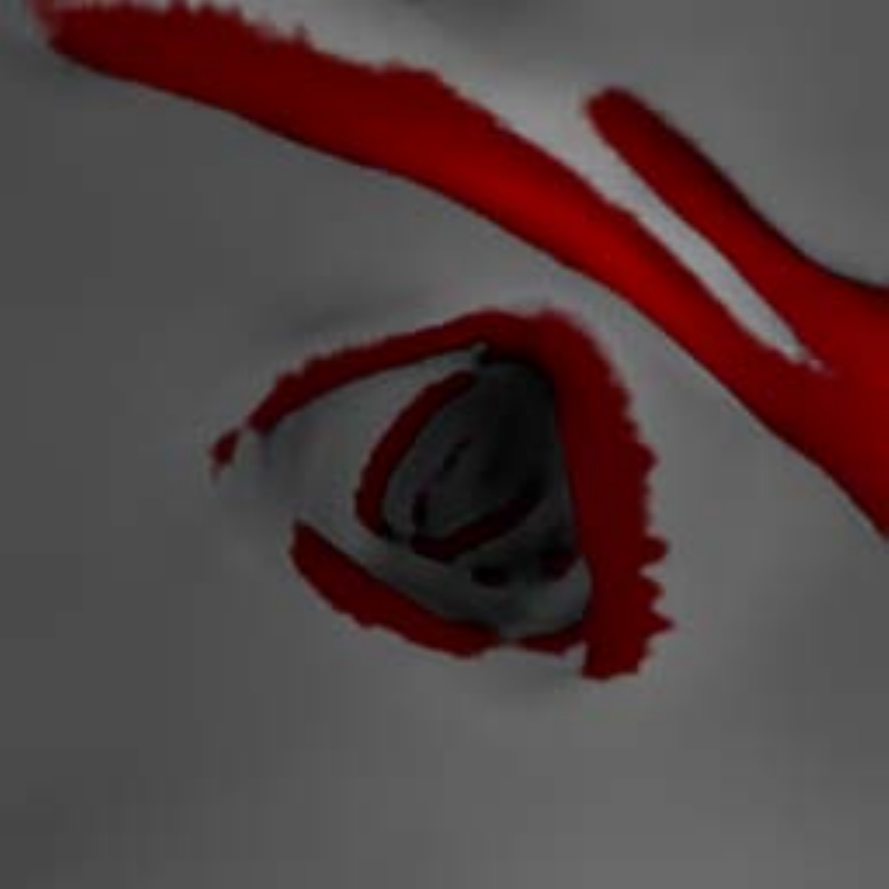}&
\includegraphics[width=0.14\textwidth]{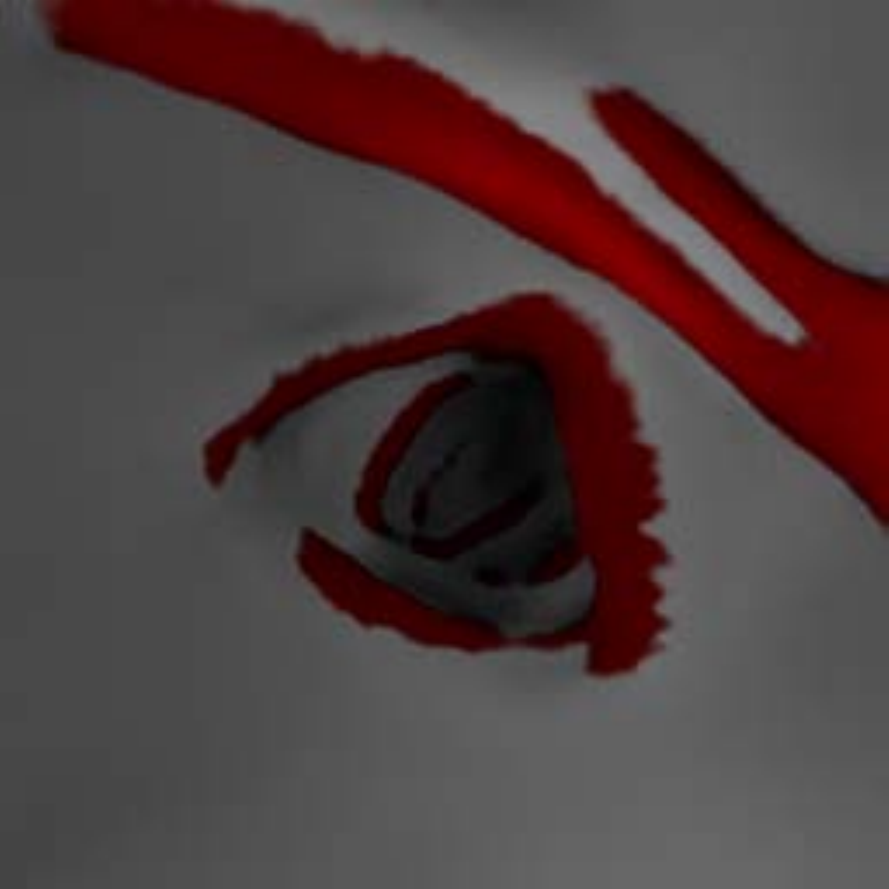}&
\includegraphics[width=0.14\textwidth]{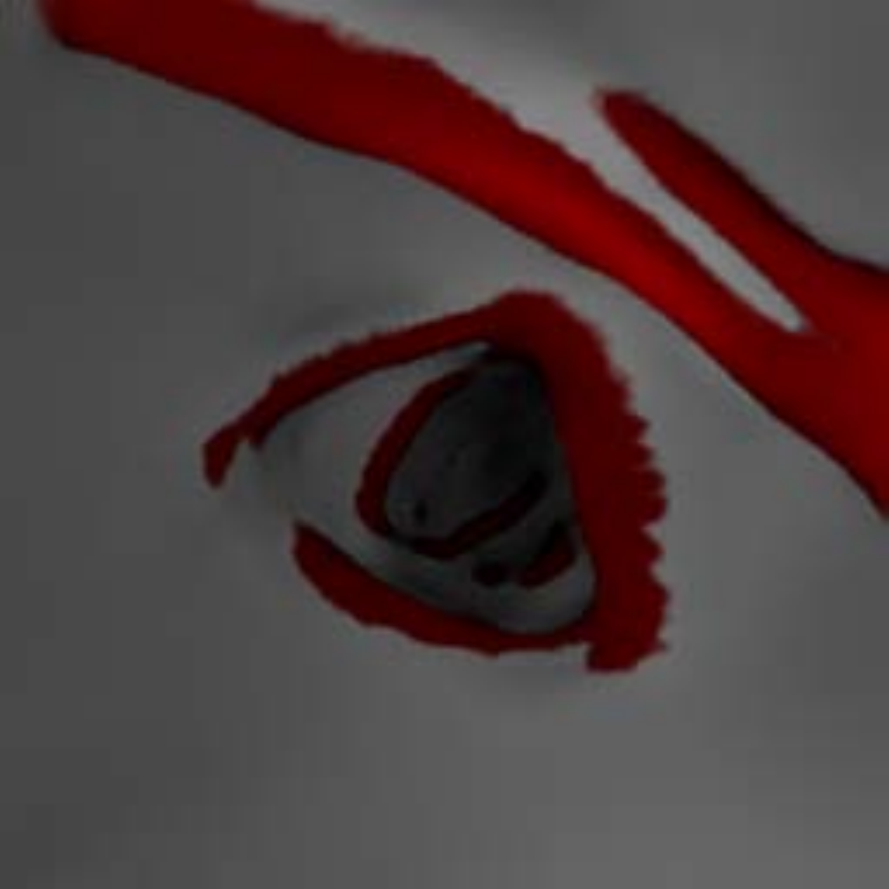}&
\ldots&
\includegraphics[width=0.14\textwidth]{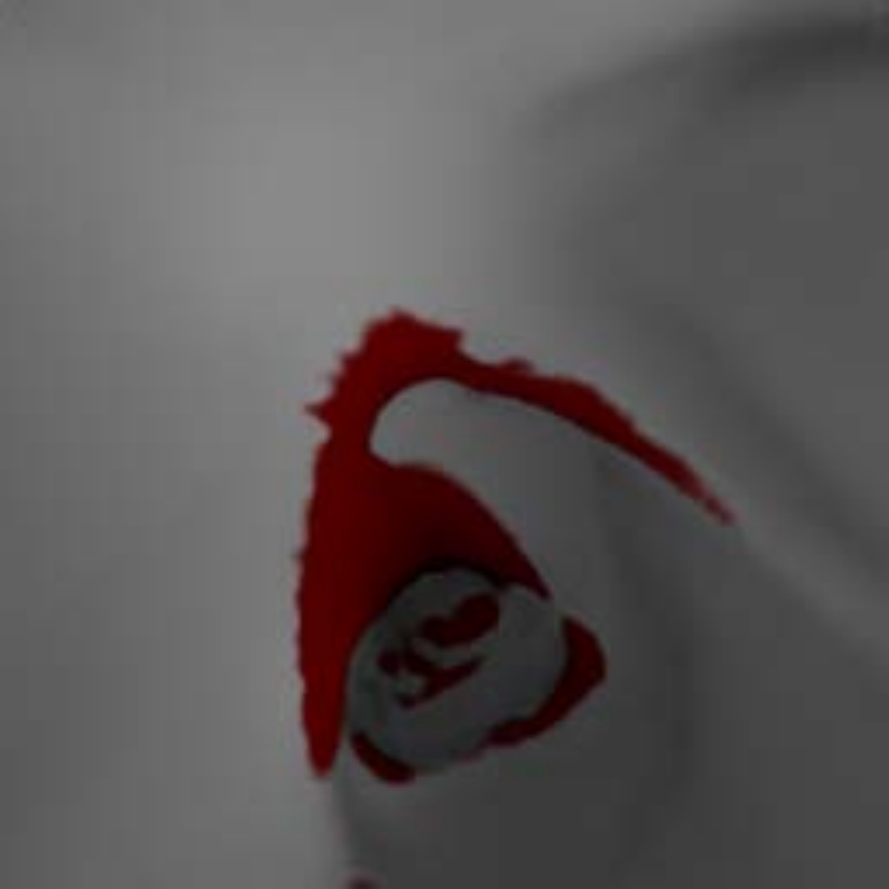}&
\includegraphics[width=0.14\textwidth]{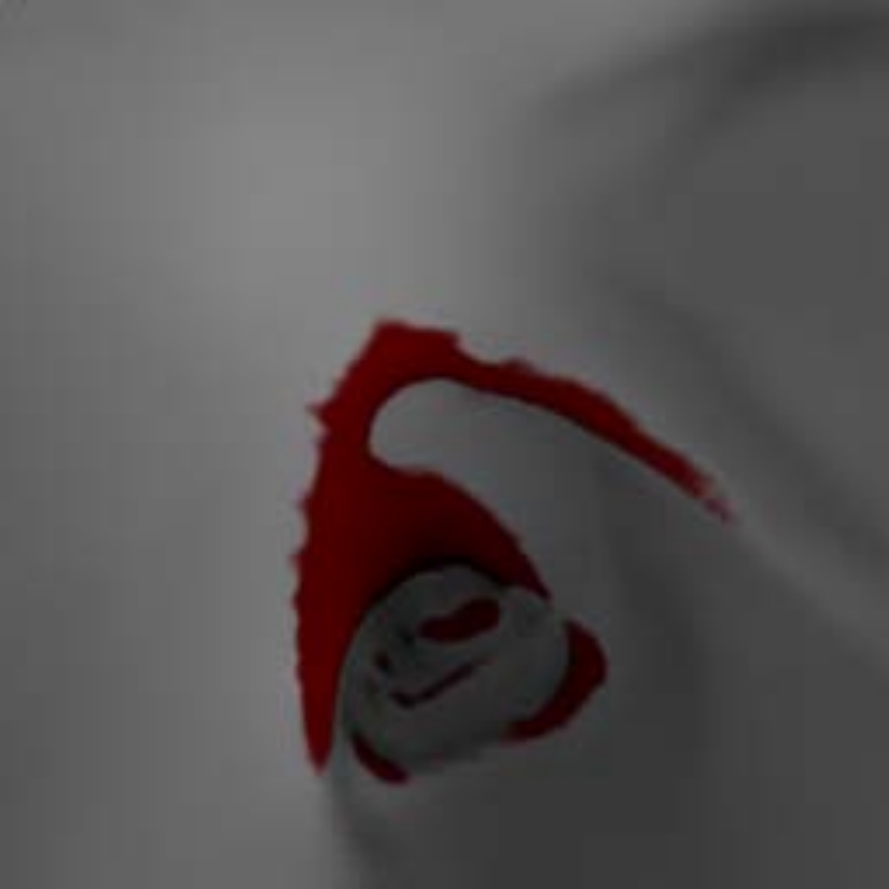}&
\includegraphics[width=0.14\textwidth]{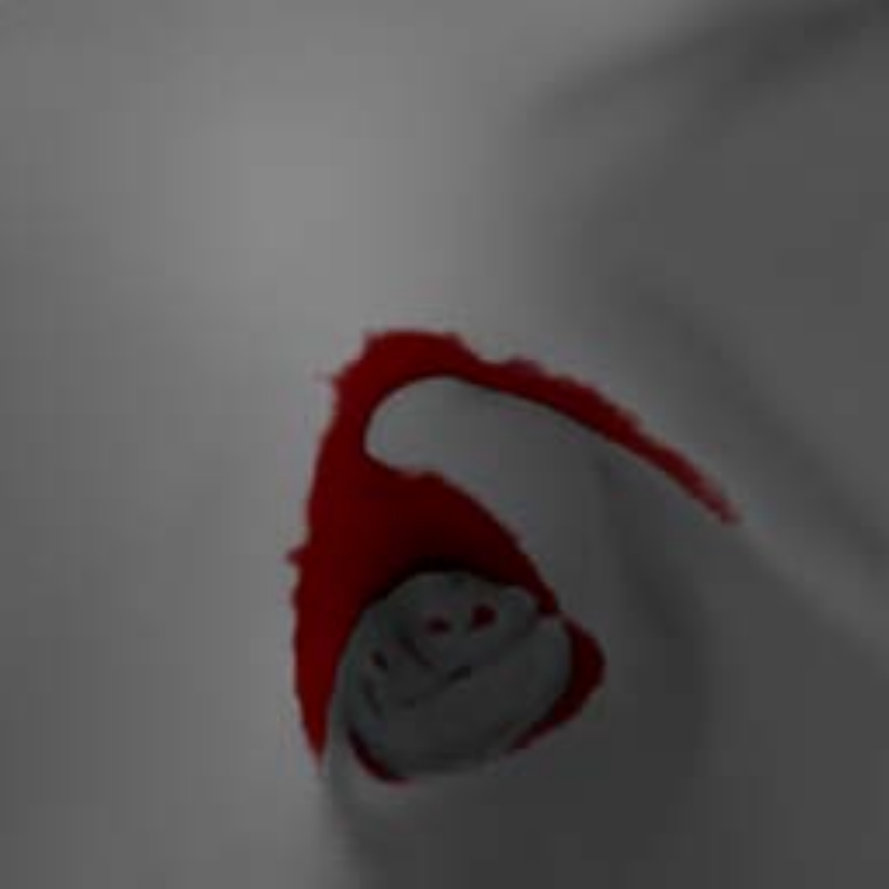}\\

\end{tabular}

\caption{Evaluation of FoldIt (ours) and XDCycleGAN model \cite{mathew2020augmenting} on VC with two different textures and ground truth/clinician-verified haustral fold (red) overlays. Green bounding boxes indicate locations where XDCycleGAN is not feature consistent and drops folds between neighboring frames. Complete video sequences are provided in the \textbf{supplementary video}.}
\vspace{-3mm}
\label{fig:textured}
\end{center}
\end{figure*}

There is no ground truth correspondence between OC and VC due to the colon shape changes between the two procedures making quantitative analysis difficult. We use textured VC mesh with clinician-verified haustral fold annotations to evaluate our results. We render the VC mesh with two different textures to test the sensitivity of our model to texture and lighting variations. The haustral folds were segmented and overlaid on the VC rendering. Fig. \ref{fig:textured} shows results from our FoldIt versus XDCycleGAN \cite{mathew2020augmenting} on these textured VC meshes.

FoldIt results are closer to the ground truth for both textures, while the haustral fold features remain preserved/consistent across frames. The green bounding boxes indicate locations where XDCycleGAN is not feature consistent and removes/adds folds between neighboring frames. As seen in Table \ref{tab:tab}, FoldIt  achieves higher Dice and IoU scores when compared with XDCycleGAN on both textures. Since the  textured VC colons have the same underlying geometry, the network should have similar predictions for both videos. The Dice and IoU scores are calculated between Texture 1 and Texture 2 and shown in the third row of Table \ref{tab:tab}. Again, FoldIt is more consistent in its predictions with varying textures. The complete video sequences are provided in the \textbf{supplementary video}\footnote{Supplementary Video: \url{https://youtu.be/_iWBJnDMXjo}}.

\begin{table}[t!]
    \caption{Dice and IoU scores on ground truth VC model with the two textures shown in Fig. \ref{fig:textured}. Dice and IoU scores between Texture 1 and 2 is shown on the third row.}
    \vspace{-1mm}
    \label{tab:tab}
    \centering
    \setlength{\tabcolsep}{8pt}
    \begin{tabular}{|c||c|c||c|c|}
        \hline
        & \multicolumn{2}{|c||}{Dice} & \multicolumn{2}{|c|}{IoU}\\  
        \hline \hline
        & FoldIt & XDCycleGAN & FoldIt & XDCycleGAN \\
        \hline
         Texture 1 & $\mathbf{0.47\pm0.11}$ & $0.25\pm0.11$ & $\mathbf{0.31\pm0.09}$ & $0.15\pm0.07$\\ 
         \hline
         Texture 2 & $\mathbf{0.50 \pm 0.10}$ & $0.21\pm0.10$ &
         $\mathbf{0.33 \pm 0.09}$ & $0.12\pm0.06$\\ 
         \hline
         Consistency & $\mathbf{0.77 \pm 0.10}$ & $0.64 \pm 0.16$ &
         $\mathbf{0.64 \pm 0.12}$ & $0.49 \pm 0.16$\\

         \hline
    \end{tabular}
\end{table}

\begin{figure}[t!]
\begin{center}
\setlength{\tabcolsep}{1pt}
\scriptsize
\begin{tabular}{ccccc|cccc}

& \multicolumn{4}{c|}{Ma et al. \cite{ma2019real} Video Sequence 1} & \multicolumn{4}{c}{Ma et al. \cite{ma2019real} Video Sequence 2}\\

\rotatebox{90}{~~Input}&
\includegraphics[width=0.115\textwidth]{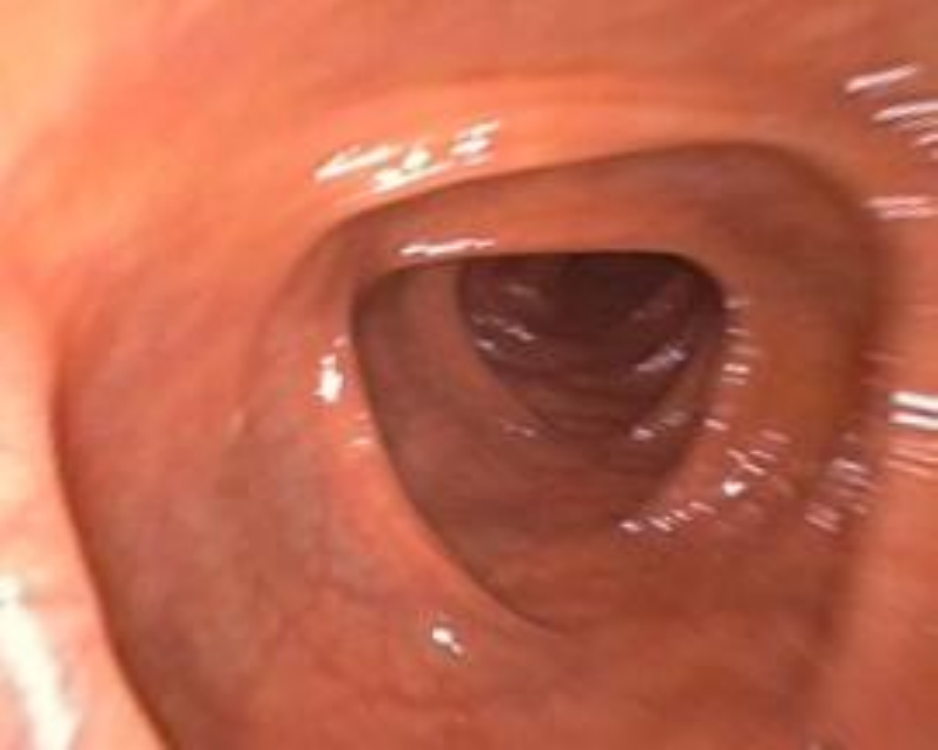}&
\includegraphics[width=0.115\textwidth]{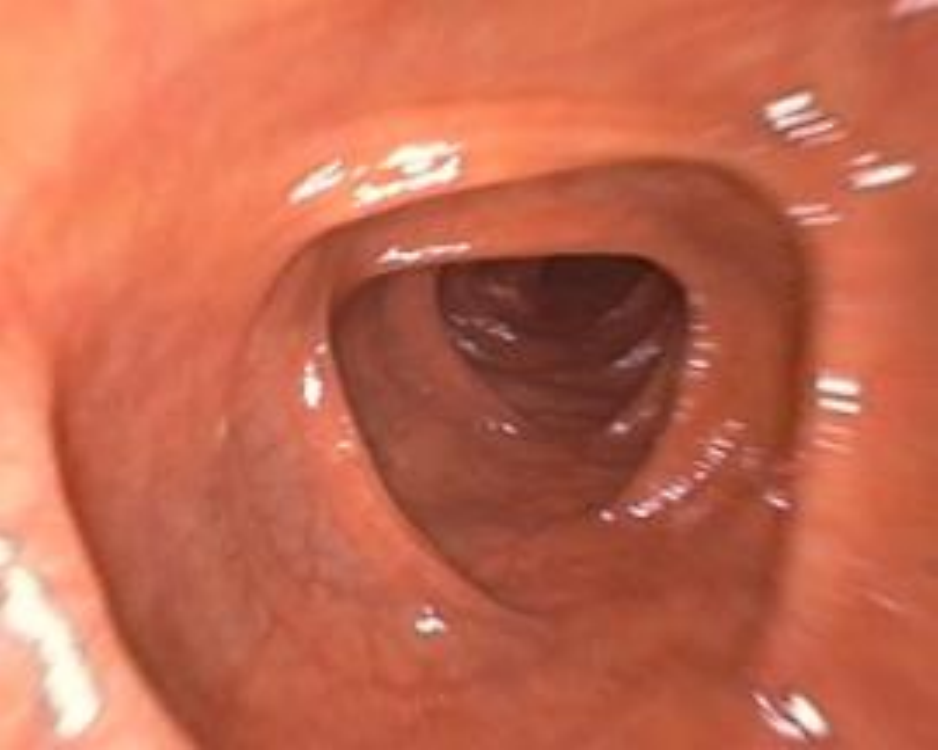}&
\includegraphics[width=0.115\textwidth]{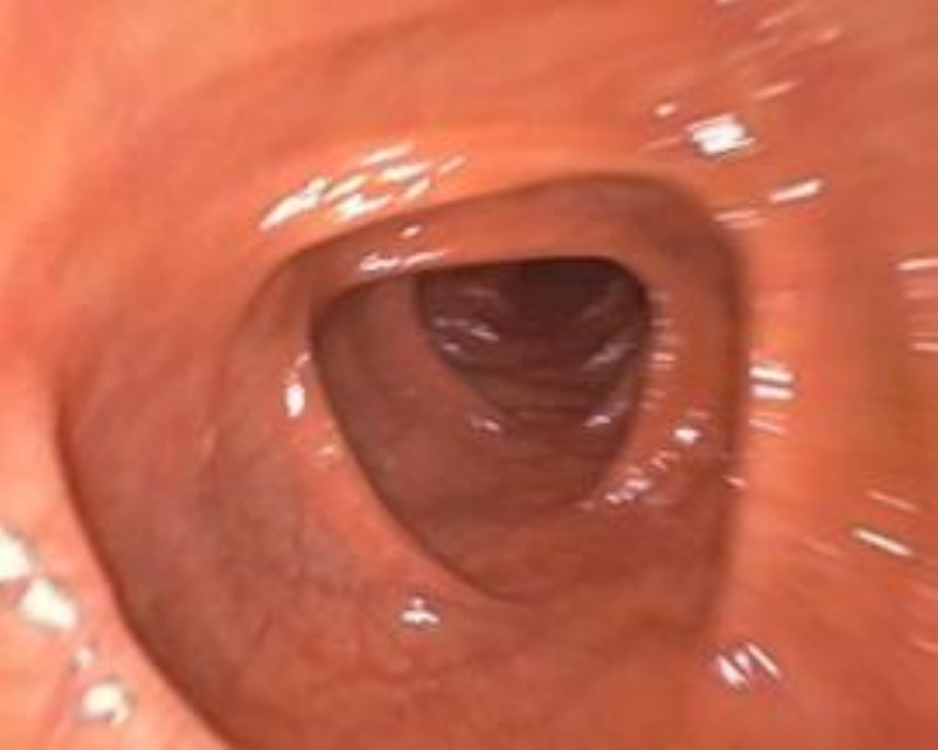}&
\includegraphics[width=0.115\textwidth]{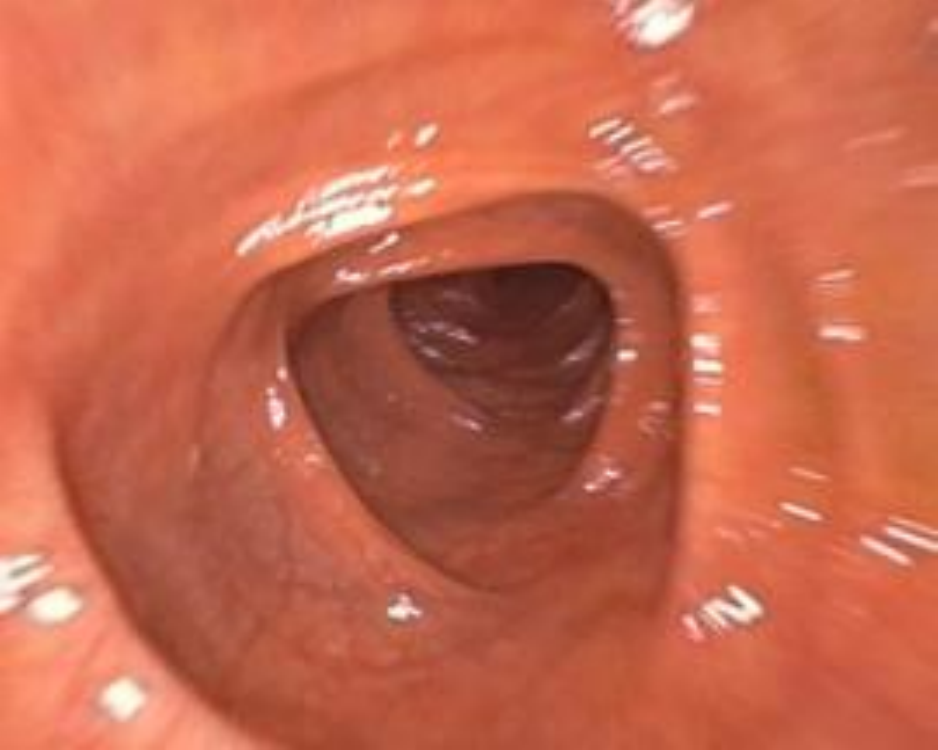}&

\includegraphics[width=0.115\textwidth]{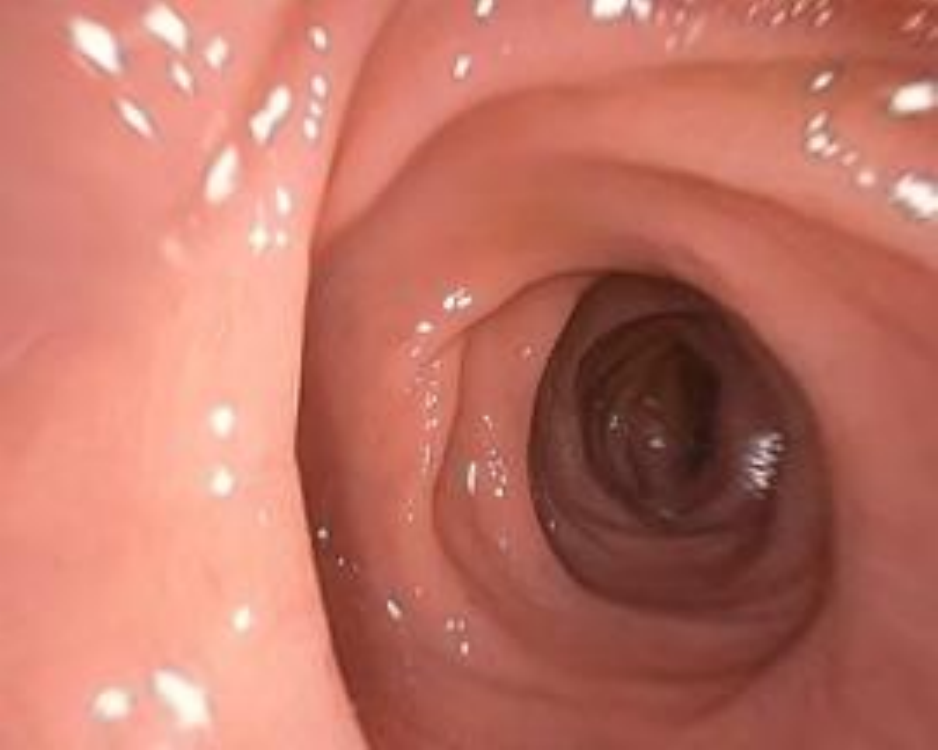}&
\includegraphics[width=0.115\textwidth]{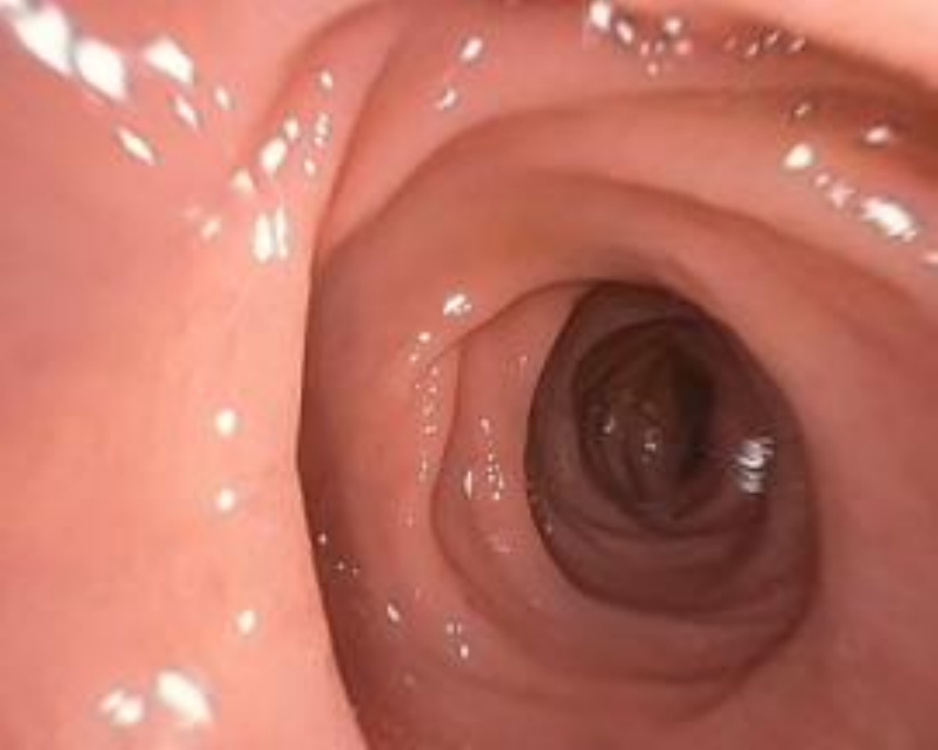}&
\includegraphics[width=0.115\textwidth]{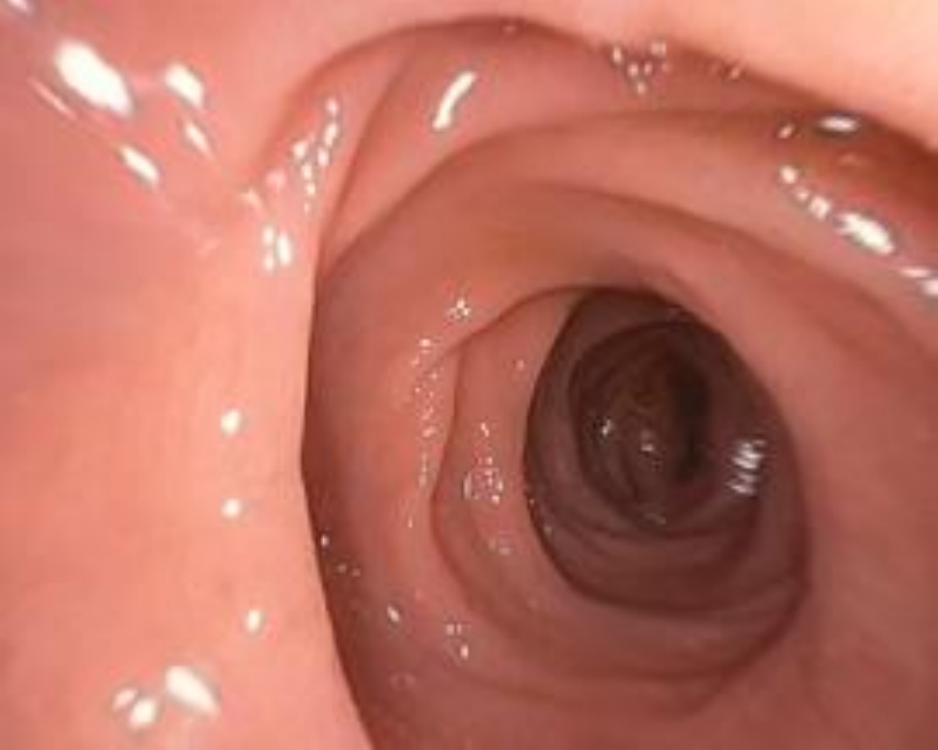}&
\includegraphics[width=0.115\textwidth]{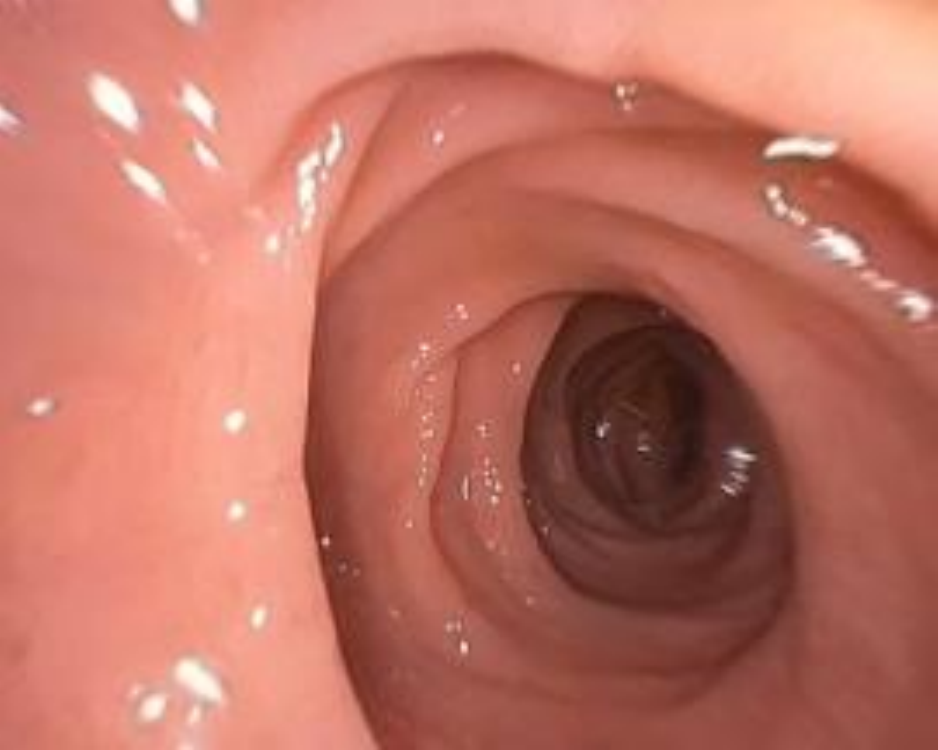}\\

\rotatebox{90}{~~~~\textbf{Ours}}&
\includegraphics[width=0.115\textwidth]{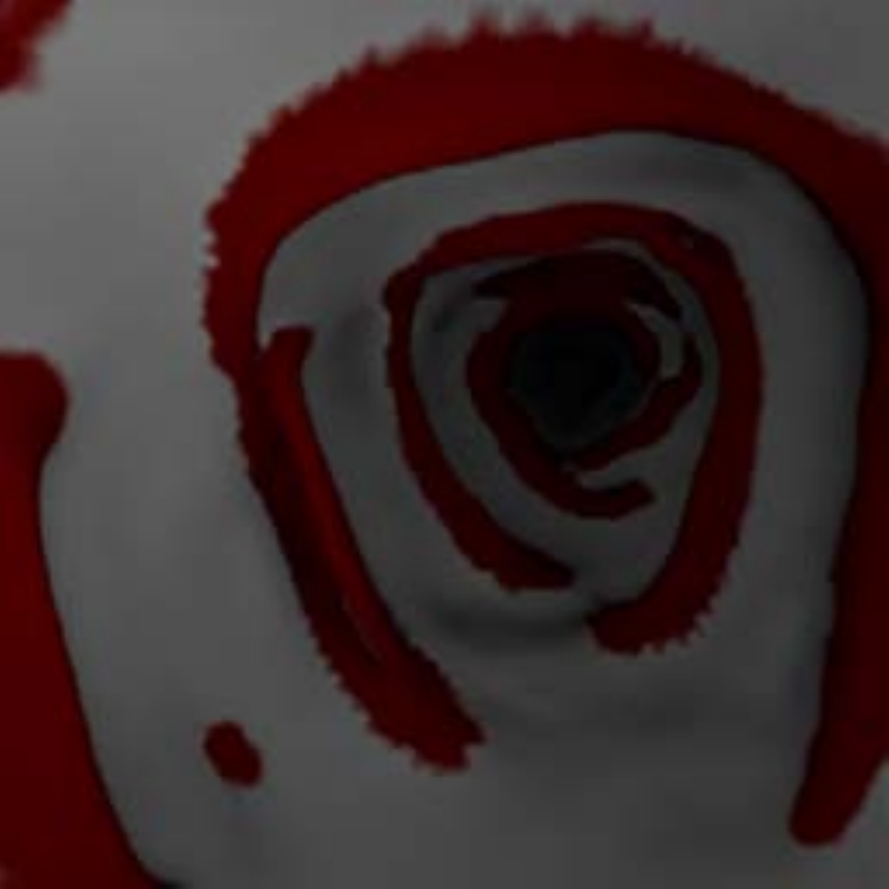}&
\includegraphics[width=0.115\textwidth]{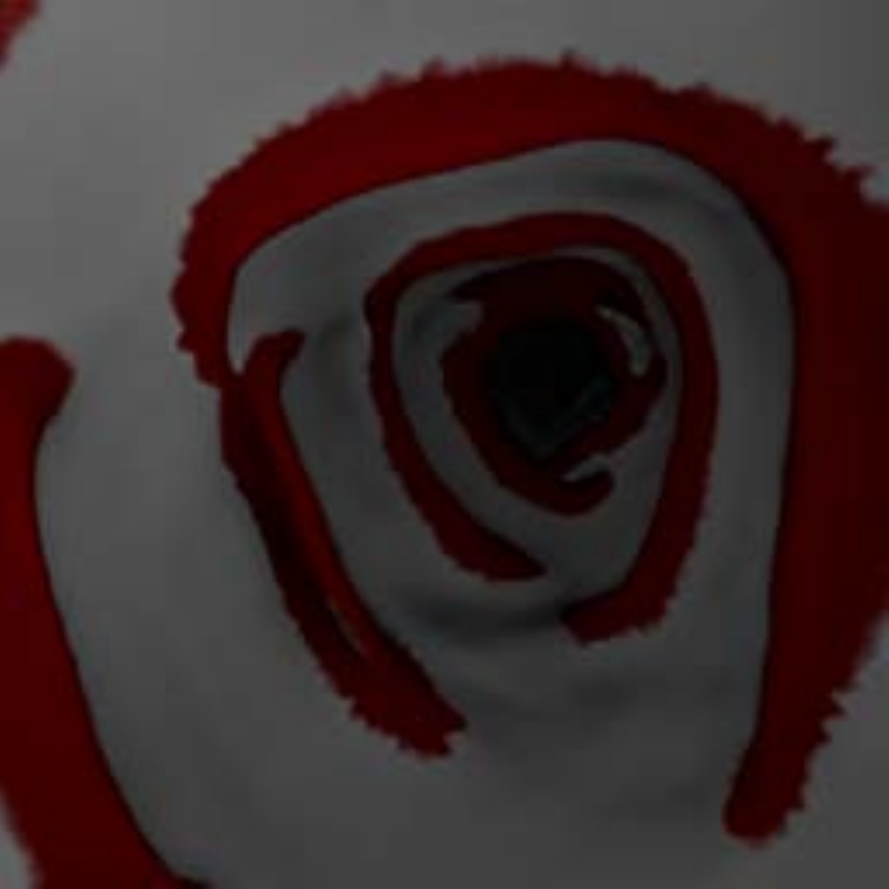}&
\includegraphics[width=0.115\textwidth]{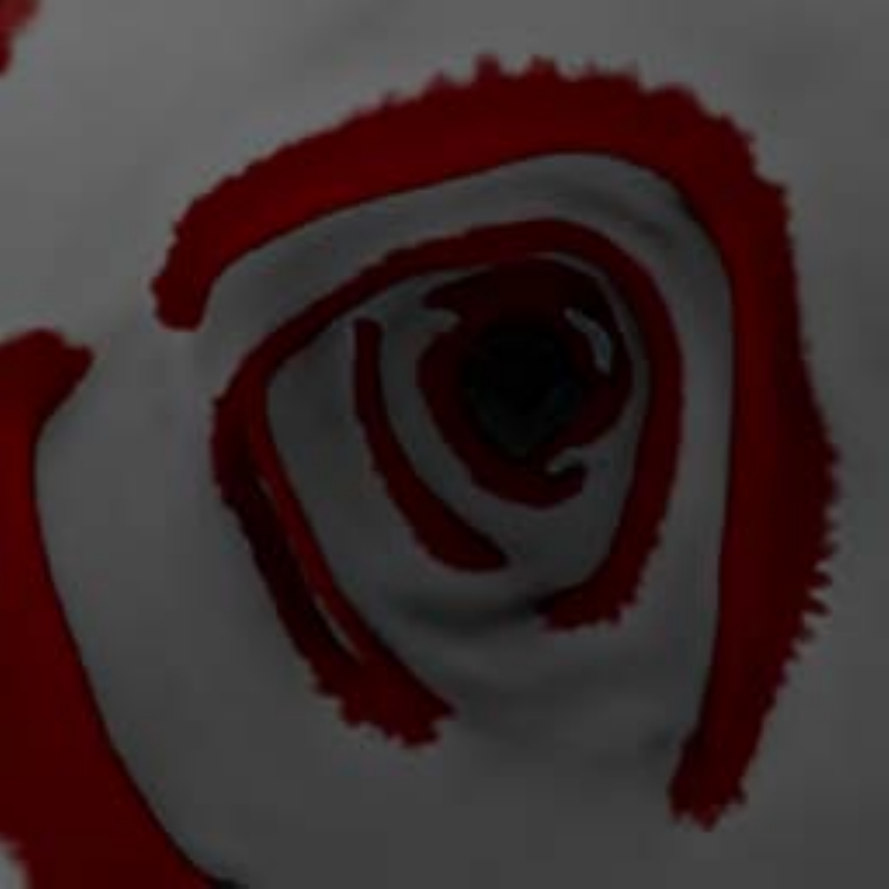}&
\includegraphics[width=0.115\textwidth]{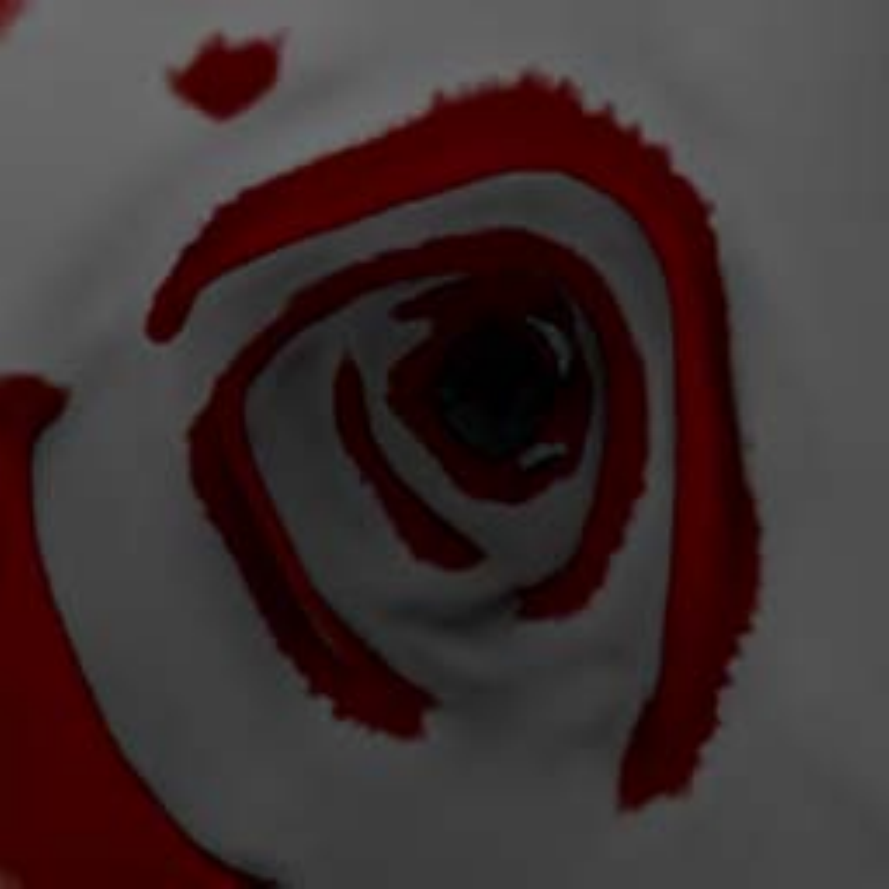}&

\includegraphics[width=0.115\textwidth]{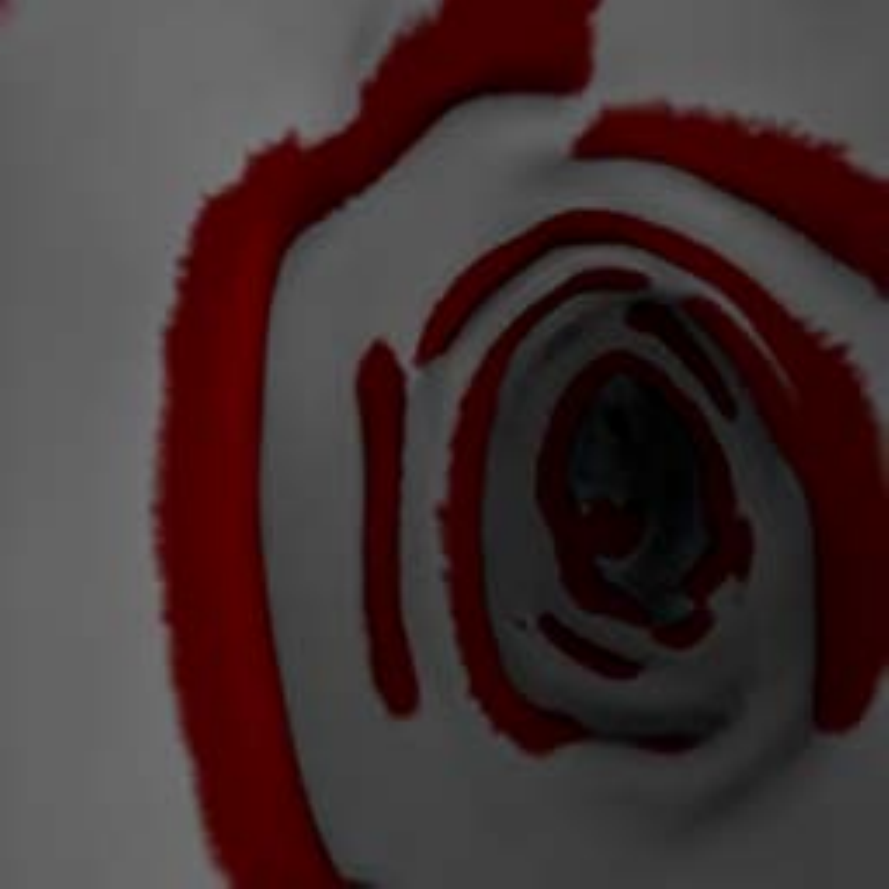}&
\includegraphics[width=0.115\textwidth]{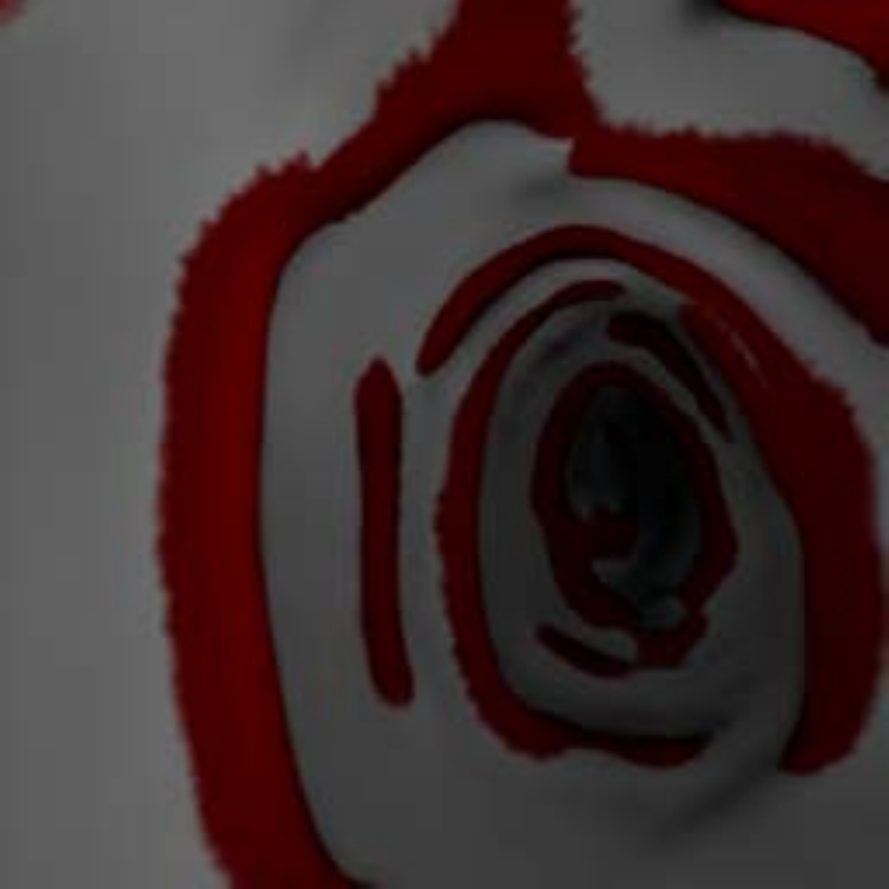}&
\includegraphics[width=0.115\textwidth]{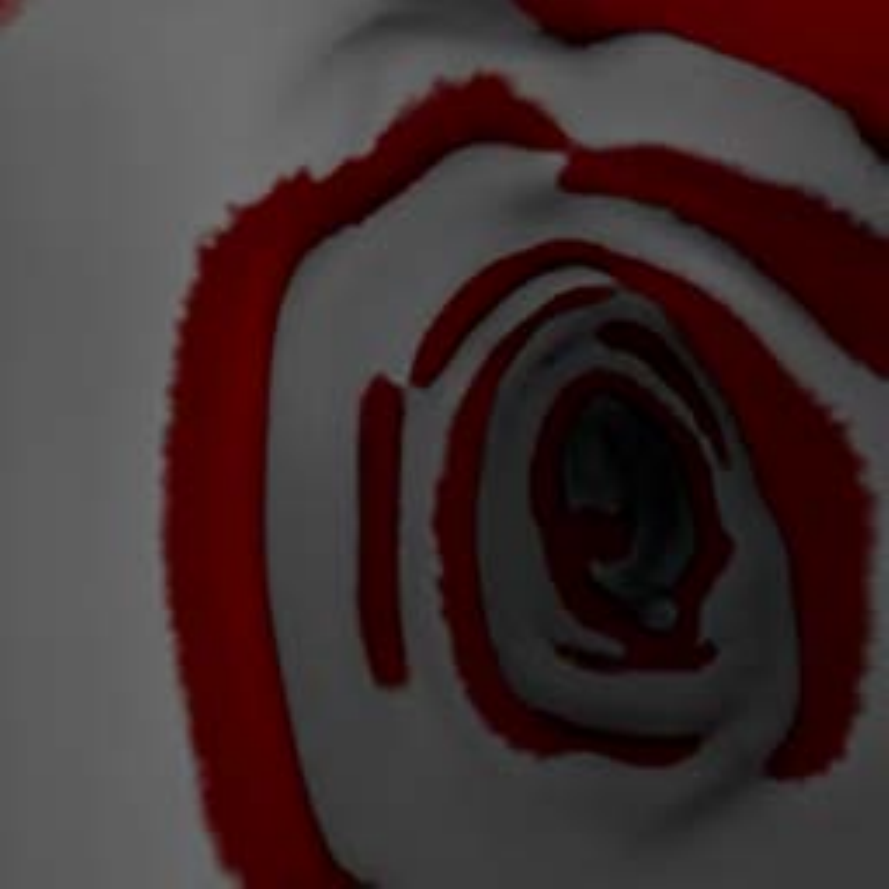}&
\includegraphics[width=0.115\textwidth]{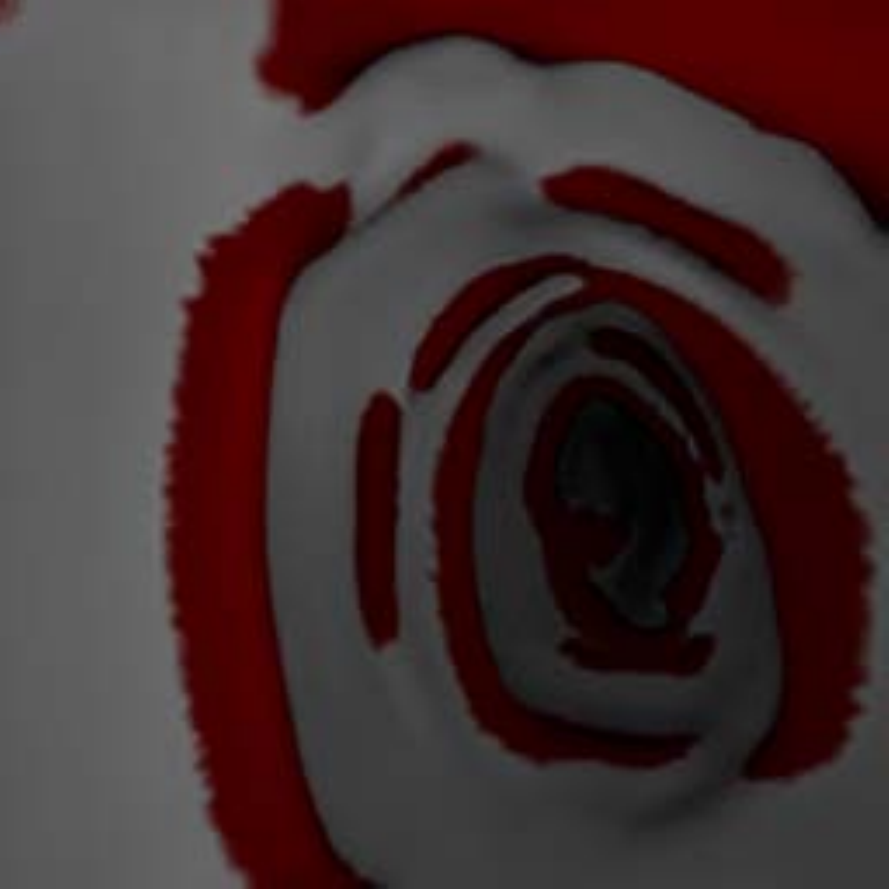}\\

\rotatebox{90}{Mathew\cite{mathew2020augmenting}}&
\includegraphics[width=0.115\textwidth]{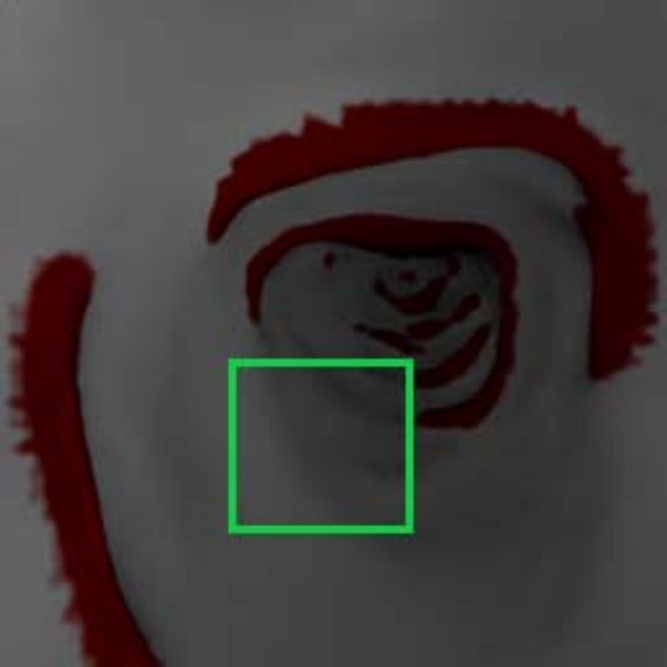}&
\includegraphics[width=0.115\textwidth]{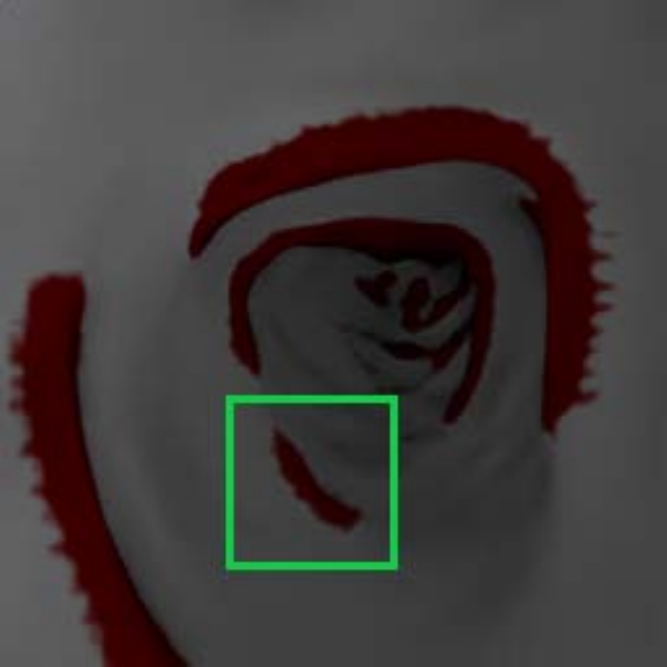}&
\includegraphics[width=0.115\textwidth]{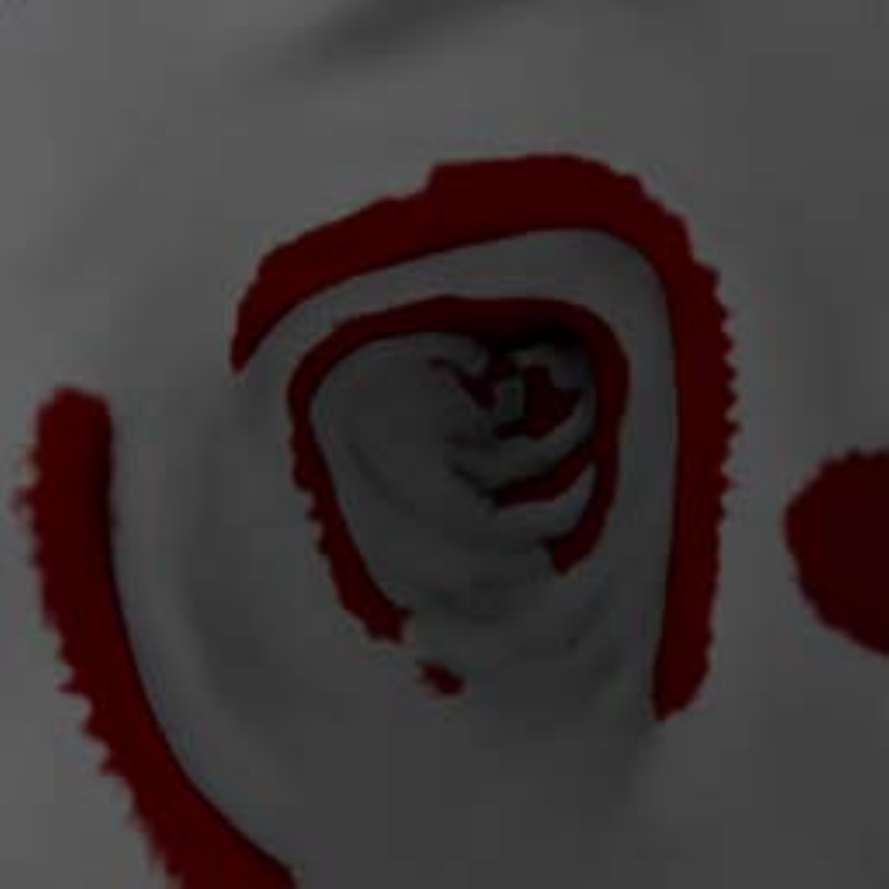}&
\includegraphics[width=0.115\textwidth]{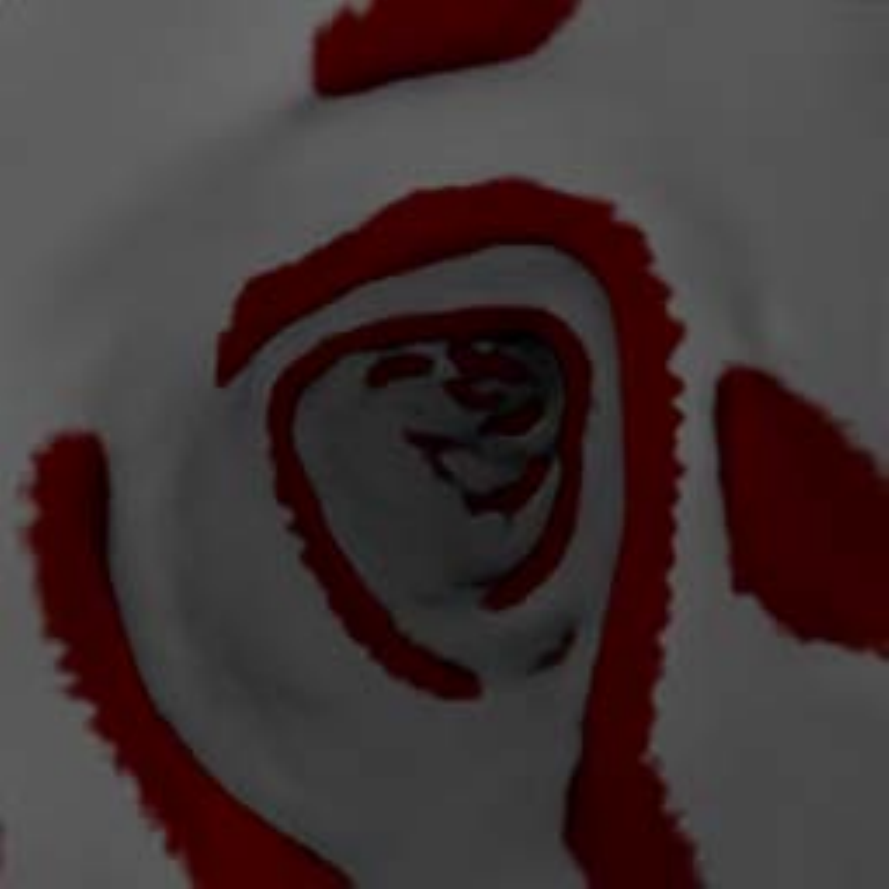}&

\includegraphics[width=0.115\textwidth]{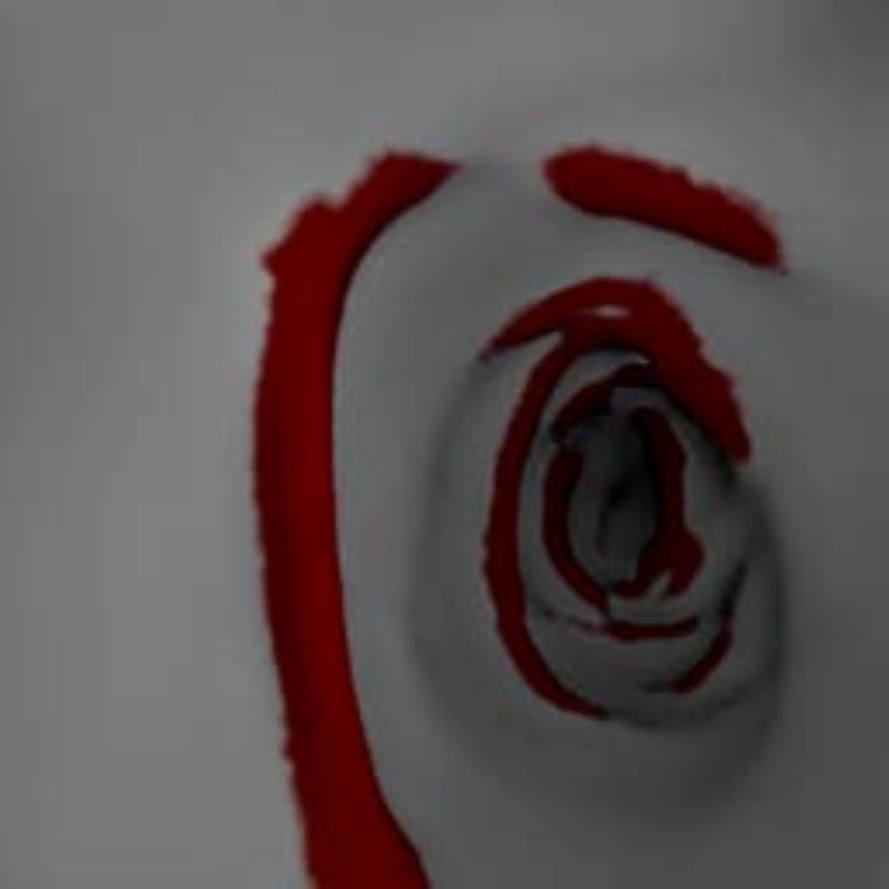}&
\includegraphics[width=0.115\textwidth]{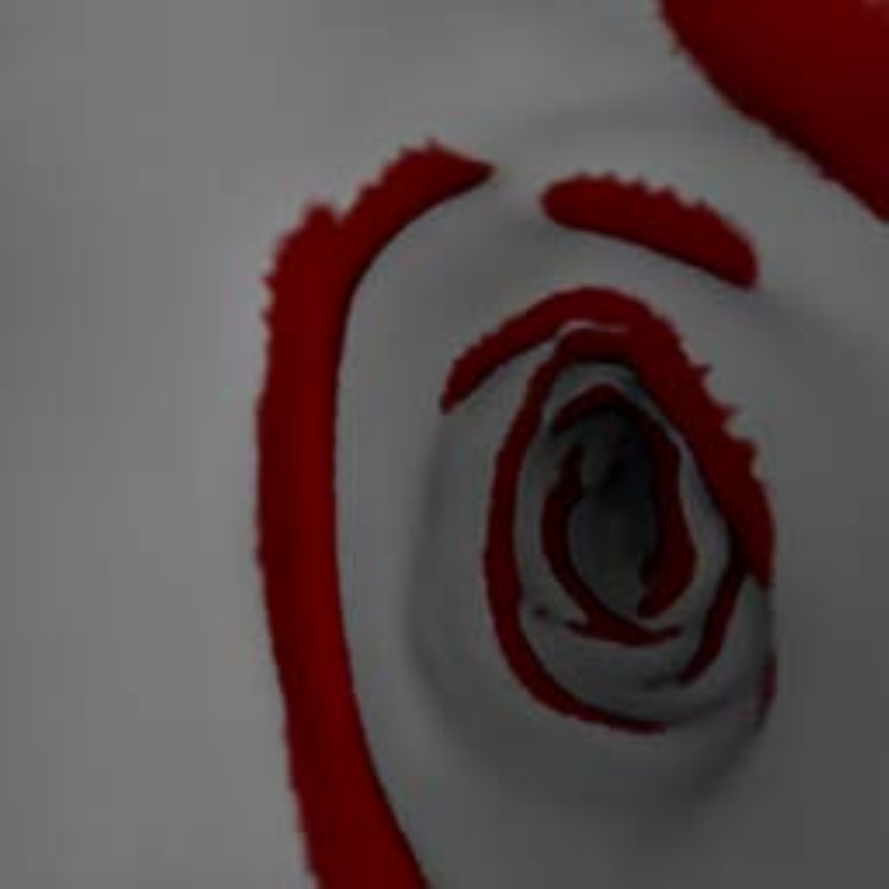}&
\includegraphics[width=0.115\textwidth]{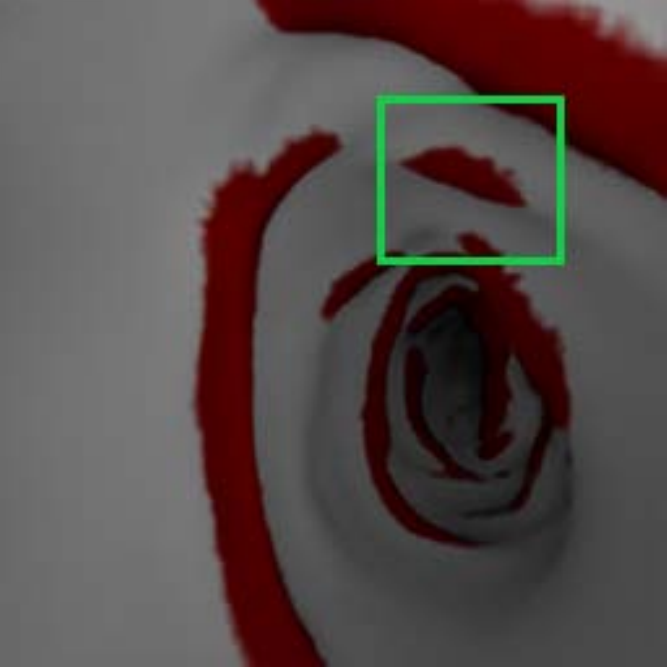}&
\includegraphics[width=0.115\textwidth]{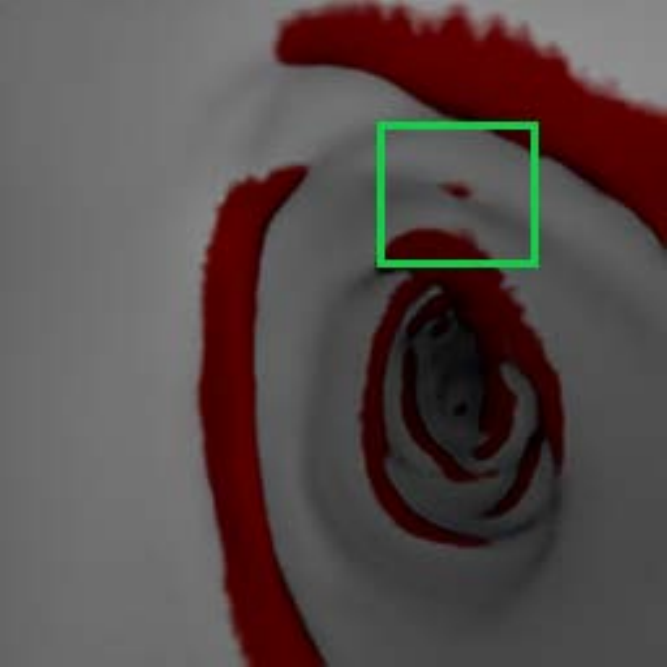}\\

\rotatebox{90}{~~~~\textbf{Ours}}&
\includegraphics[width=0.115\textwidth]{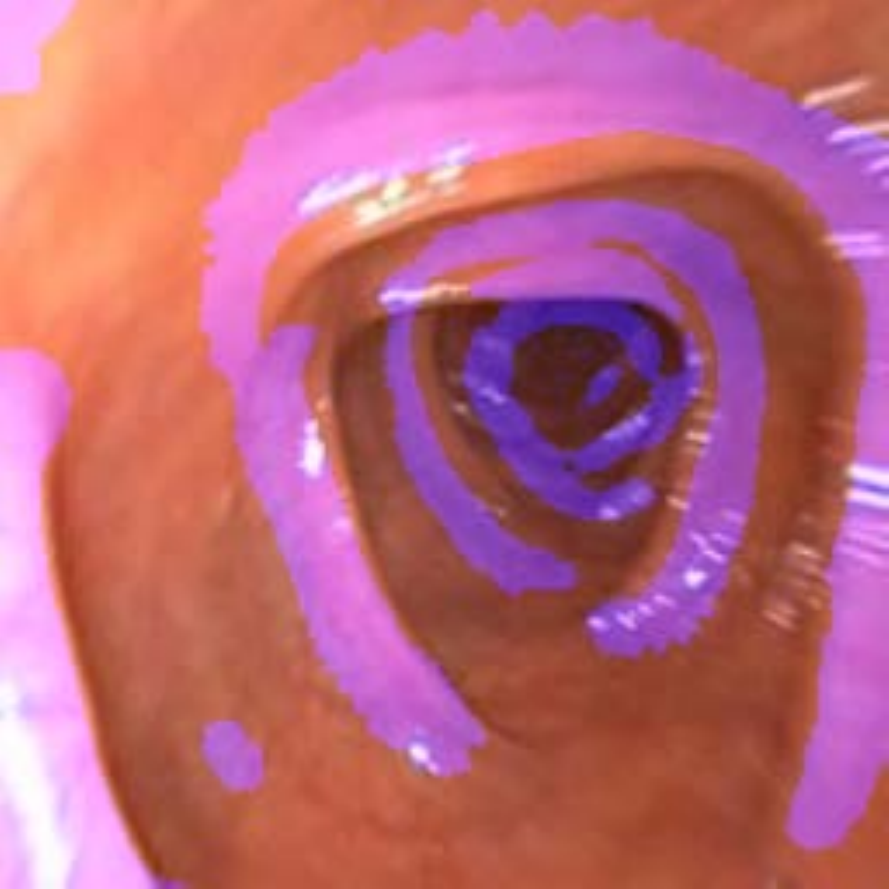}&
\includegraphics[width=0.115\textwidth]{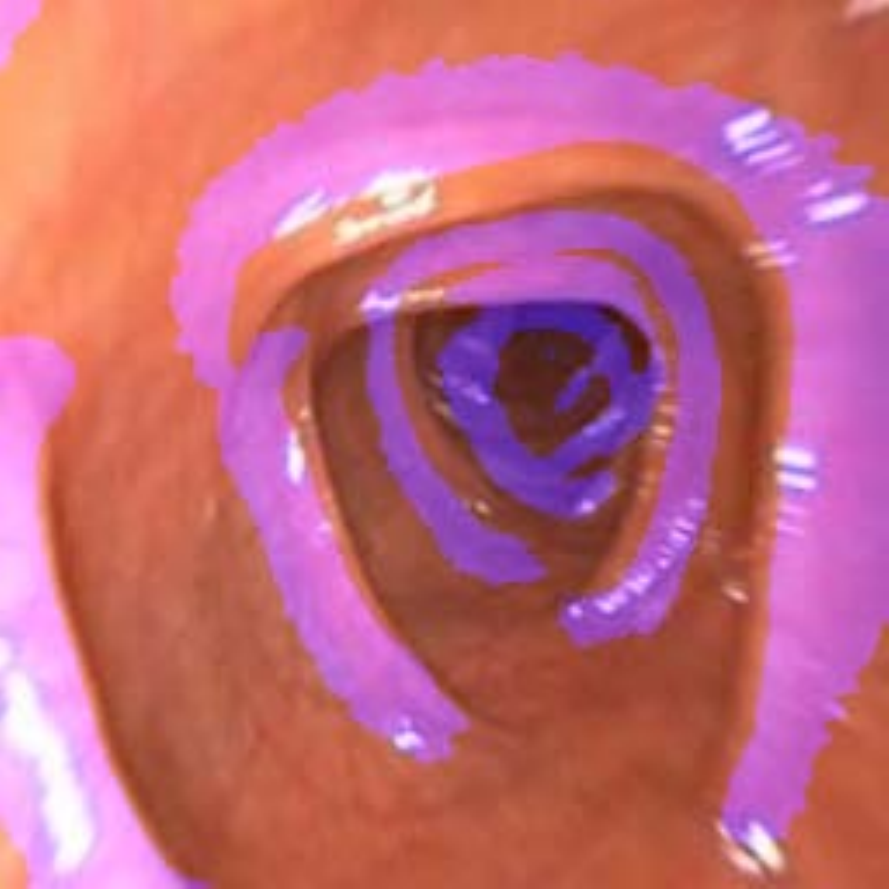}&
\includegraphics[width=0.115\textwidth]{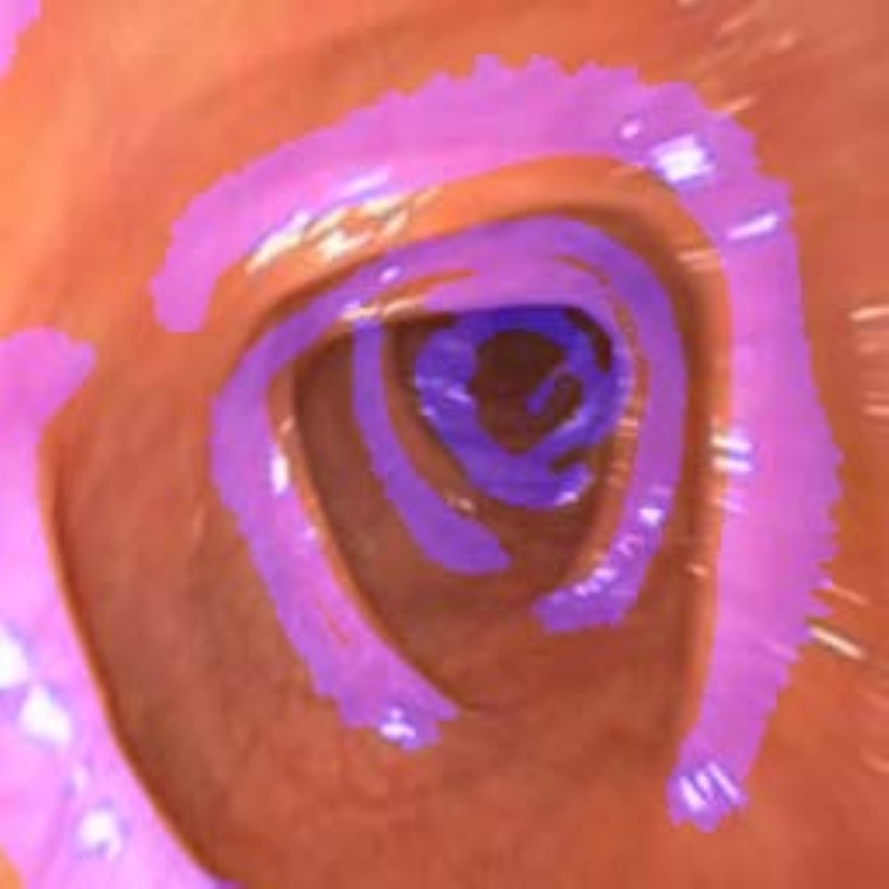}&
\includegraphics[width=0.115\textwidth]{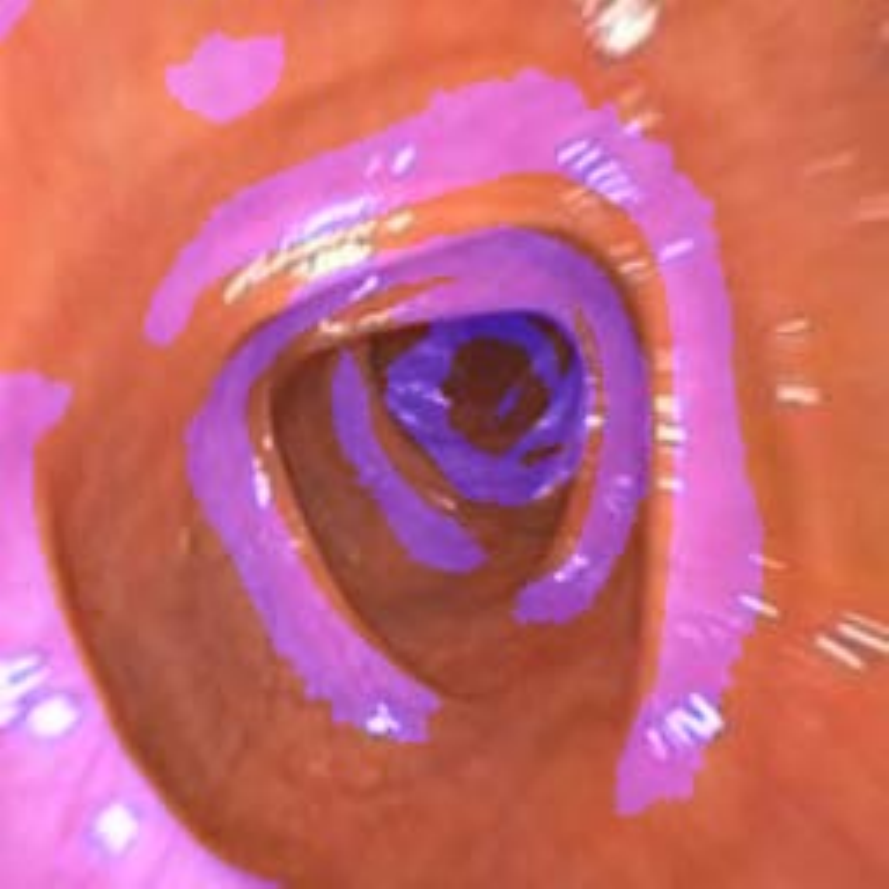}&

\includegraphics[width=0.115\textwidth]{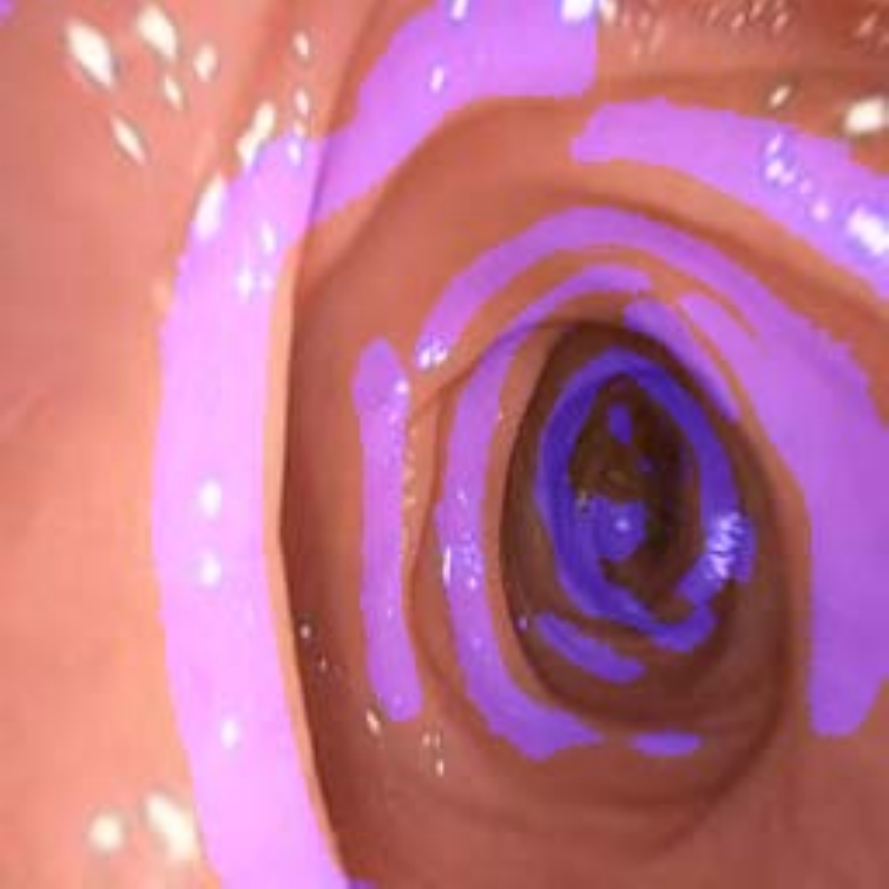}&
\includegraphics[width=0.115\textwidth]{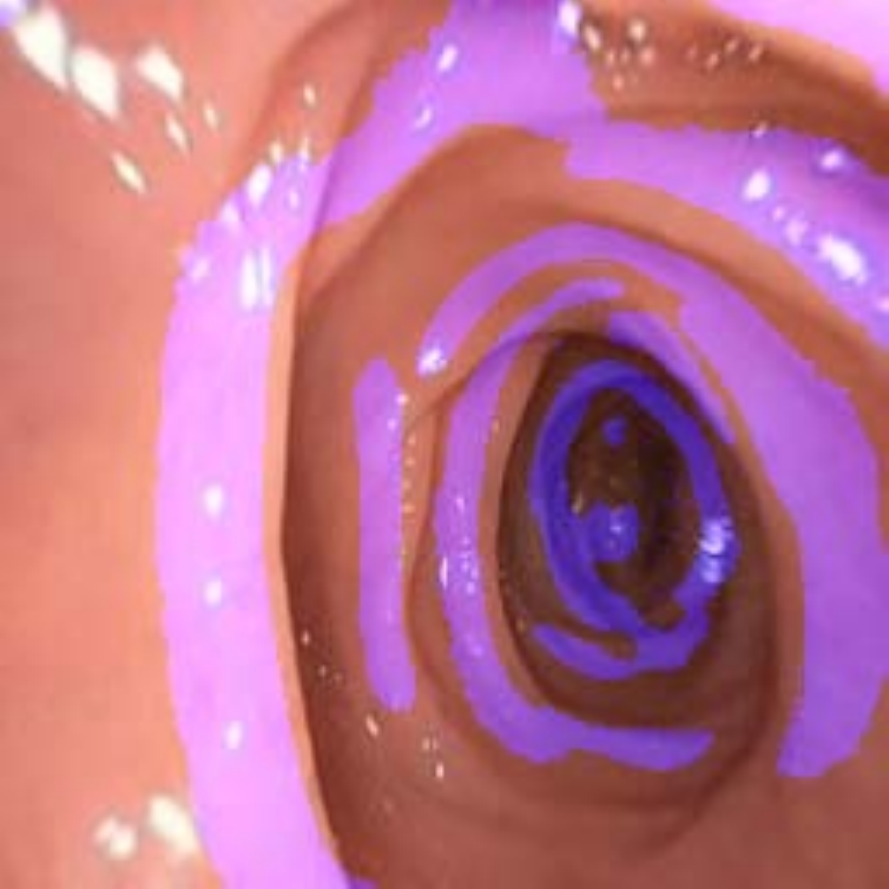}&
\includegraphics[width=0.115\textwidth]{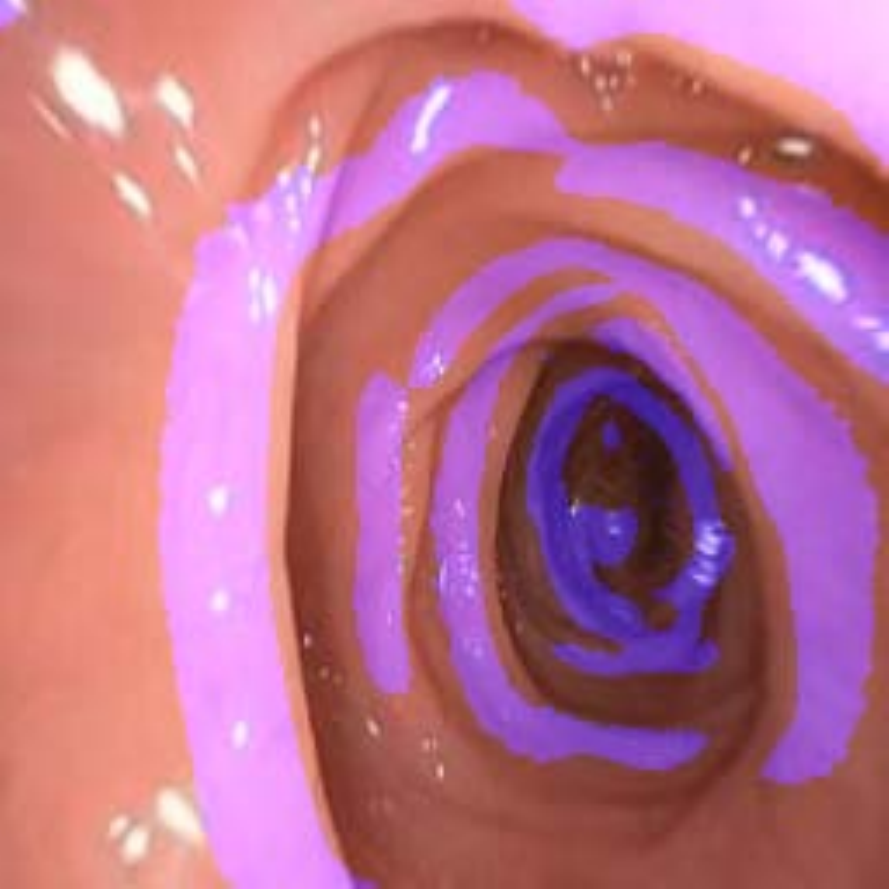}&
\includegraphics[width=0.115\textwidth]{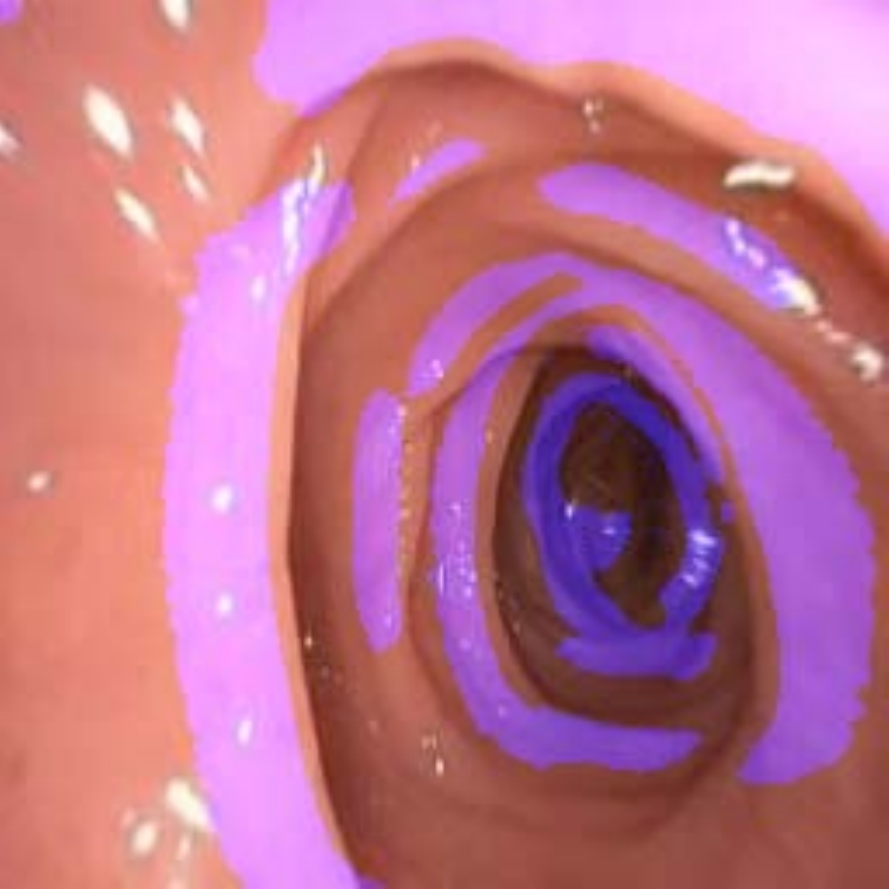}\\

\rotatebox{90}{Mathew\cite{mathew2020augmenting}}&
\includegraphics[width=0.115\textwidth]{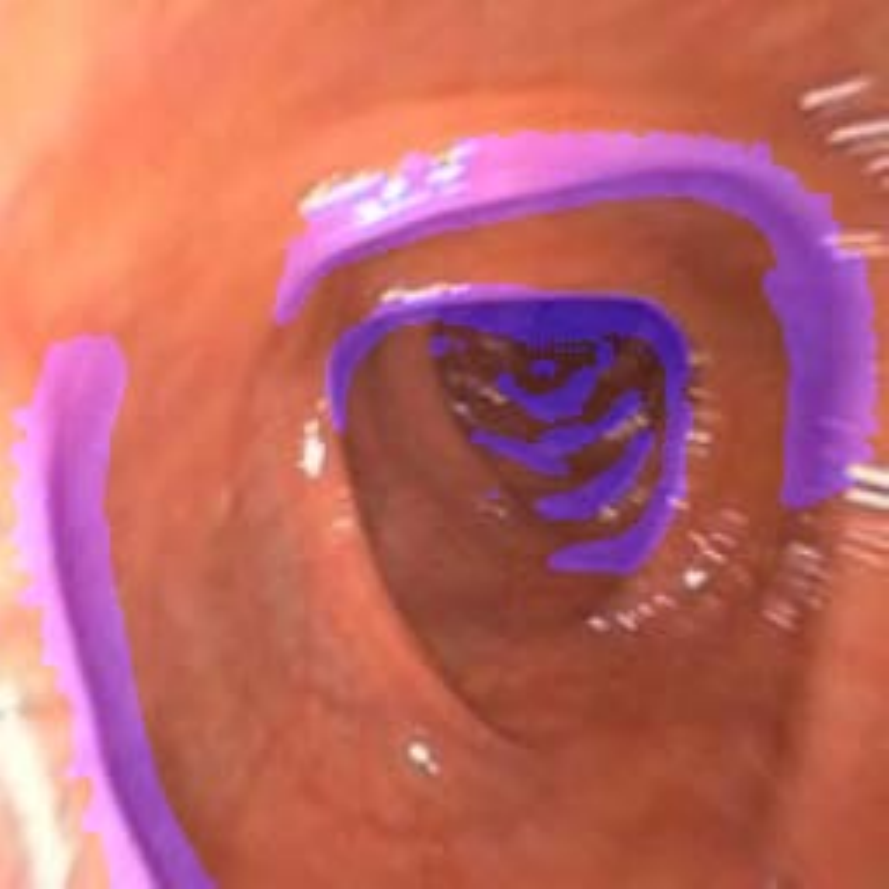}&
\includegraphics[width=0.115\textwidth]{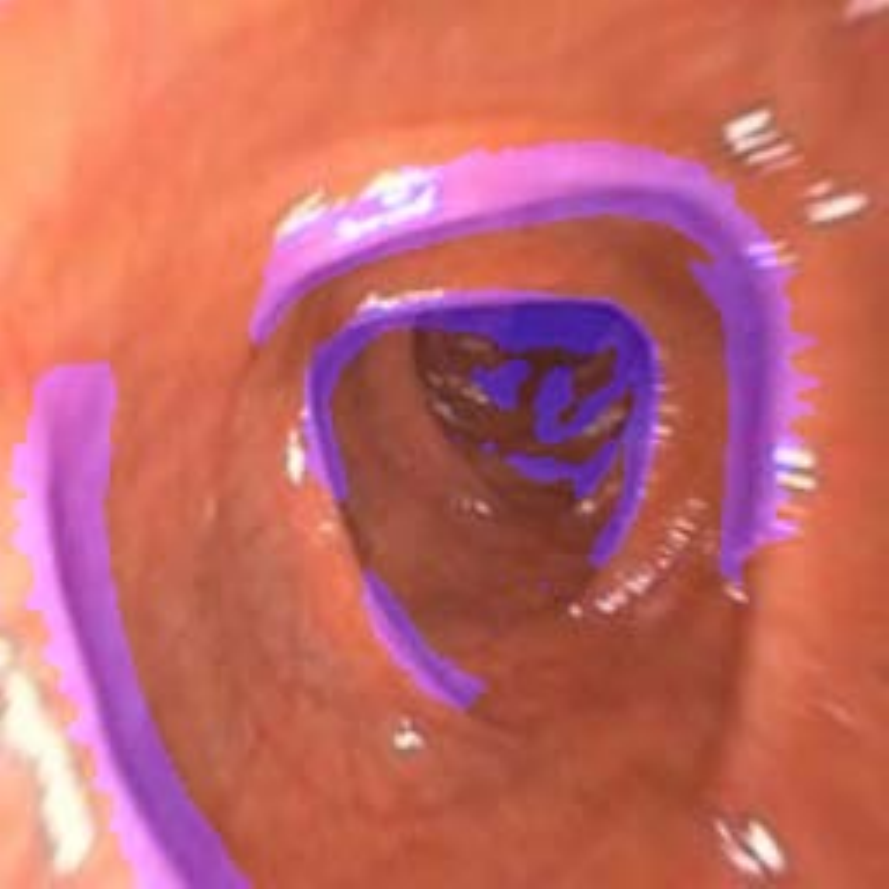}&
\includegraphics[width=0.115\textwidth]{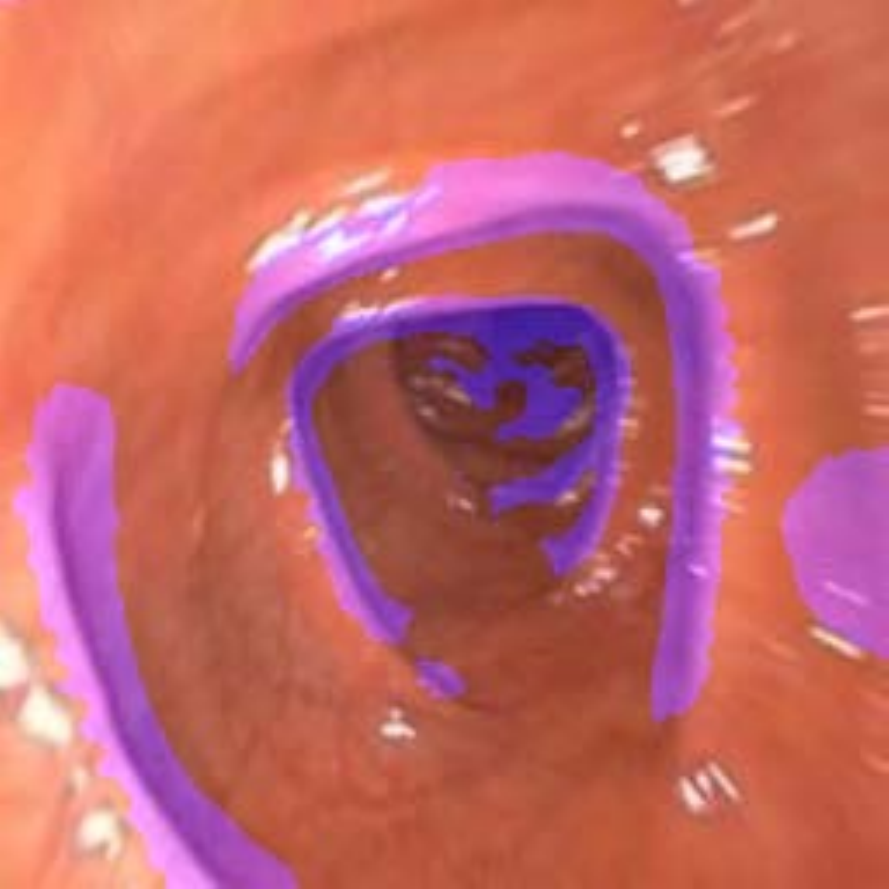}&
\includegraphics[width=0.115\textwidth]{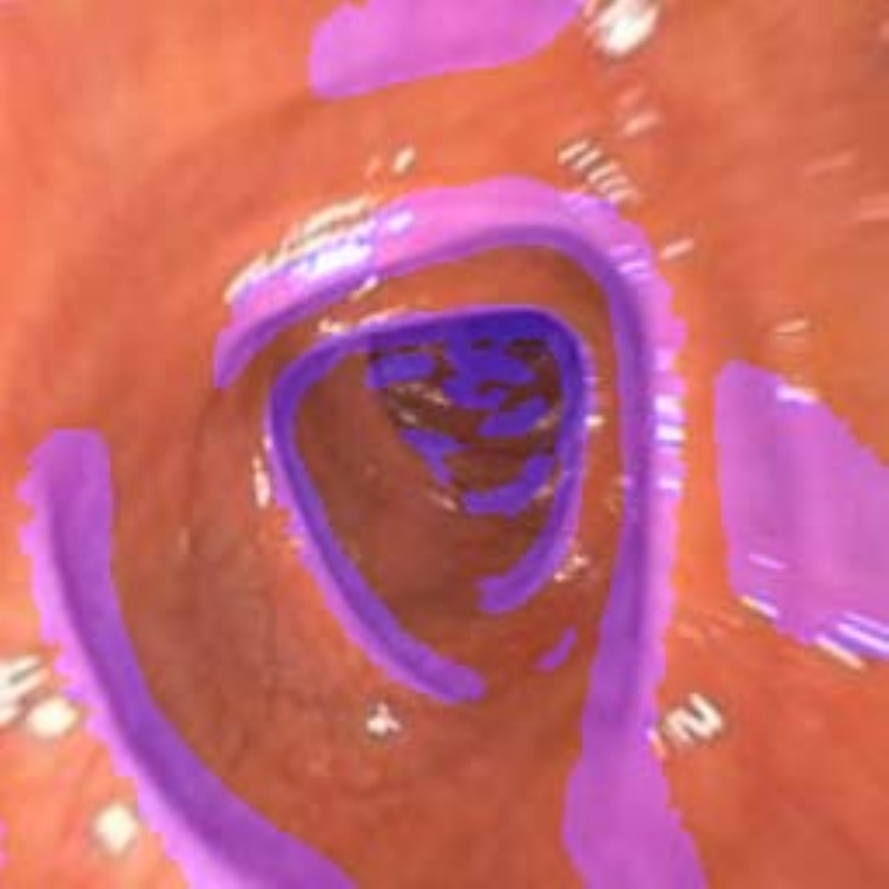}&

\includegraphics[width=0.115\textwidth]{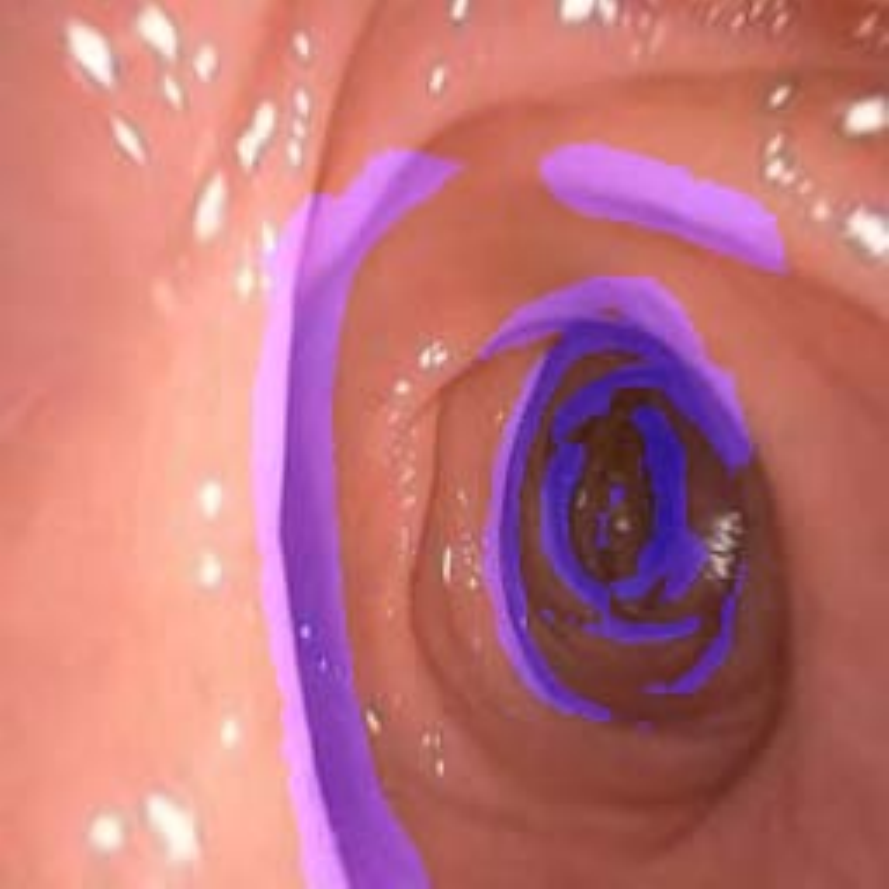}&
\includegraphics[width=0.115\textwidth]{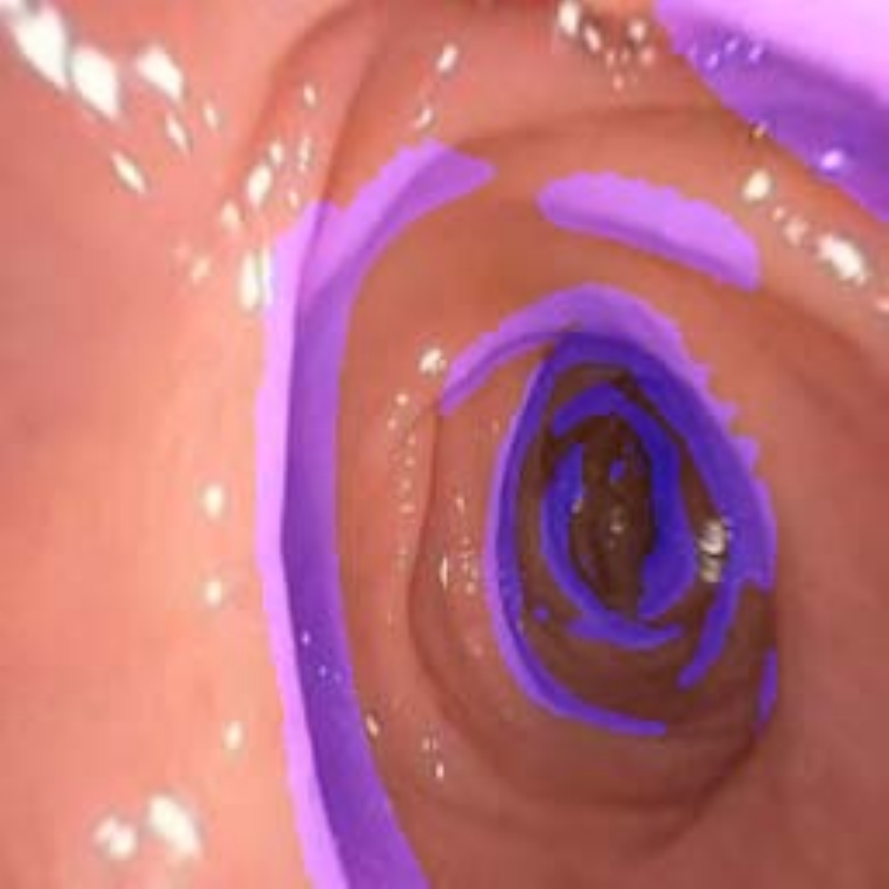}&
\includegraphics[width=0.115\textwidth]{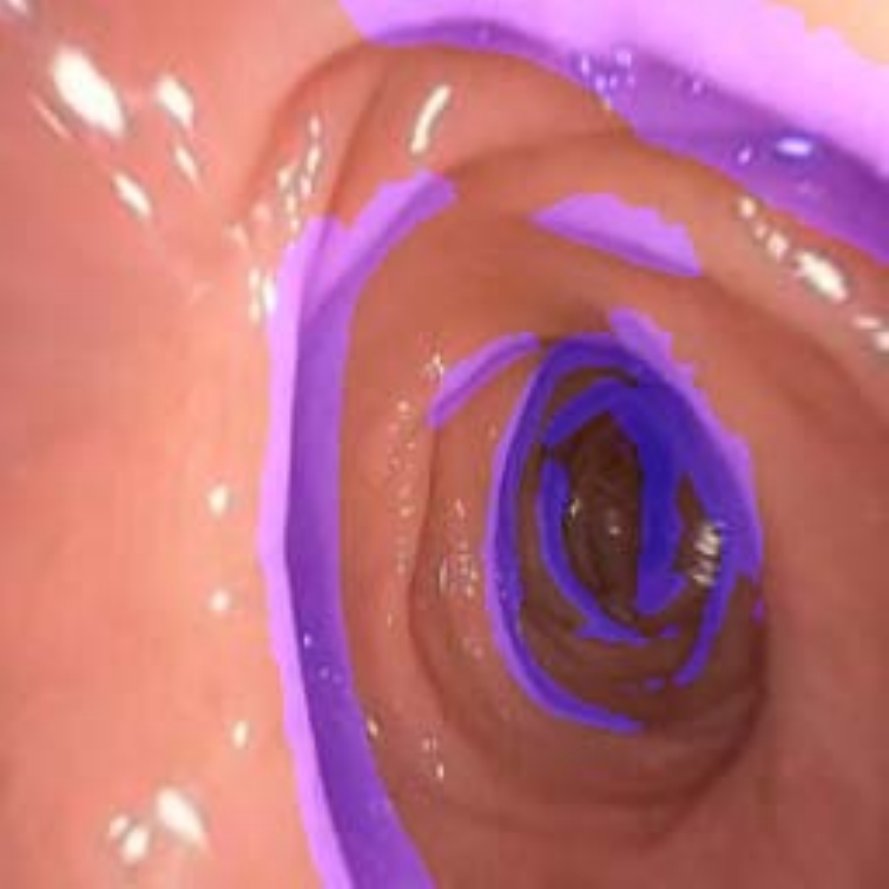}&
\includegraphics[width=0.115\textwidth]{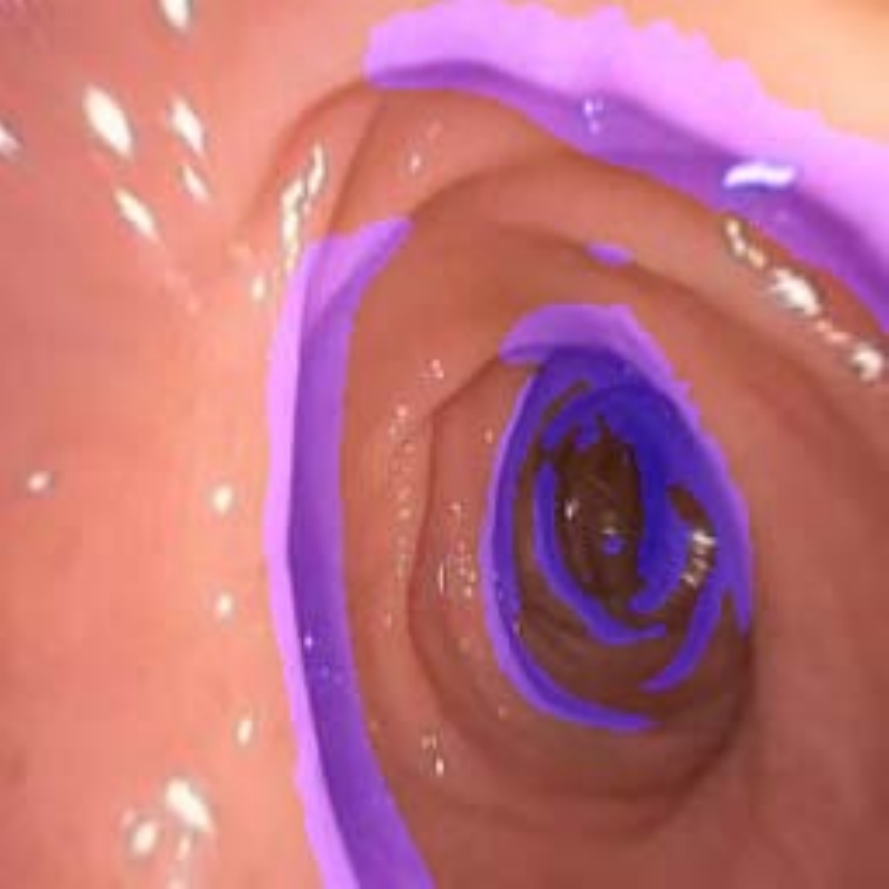}\\

\hline
& \multicolumn{4}{c|}{VR-CAPS \cite{incetan2021vr} Video Sequence 1} & \multicolumn{4}{c}{VR-CAPS \cite{incetan2021vr} Video Sequence 2}\\

\rotatebox{90}{~~~Input}&
\includegraphics[width=0.115\textwidth]{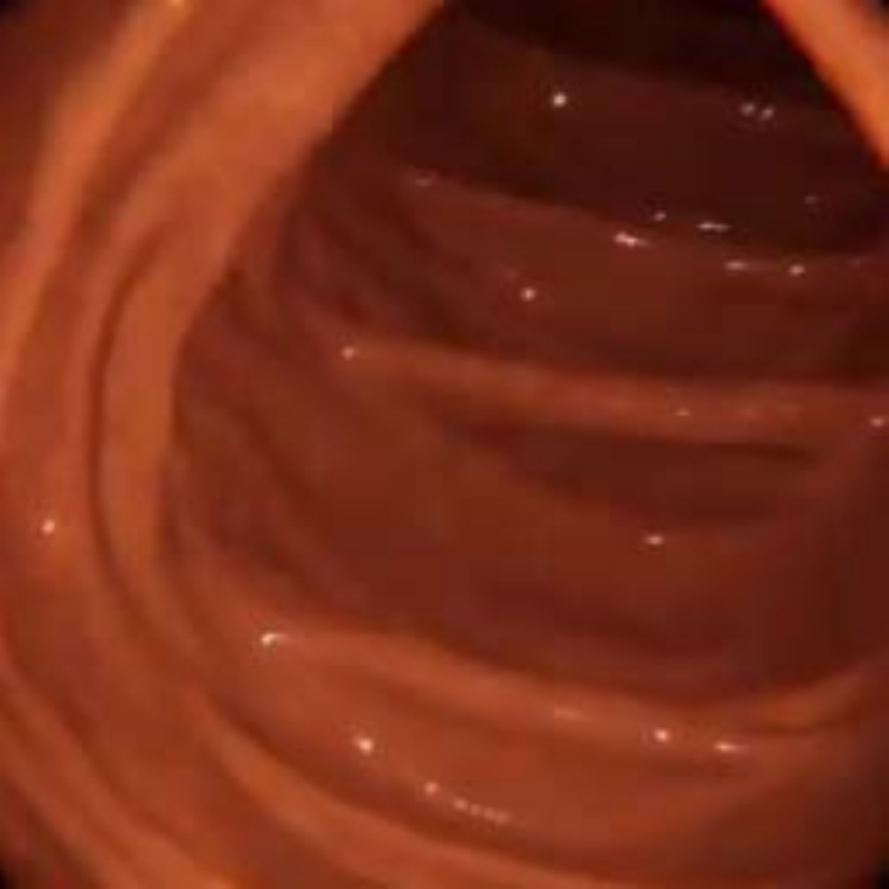}&
\includegraphics[width=0.115\textwidth]{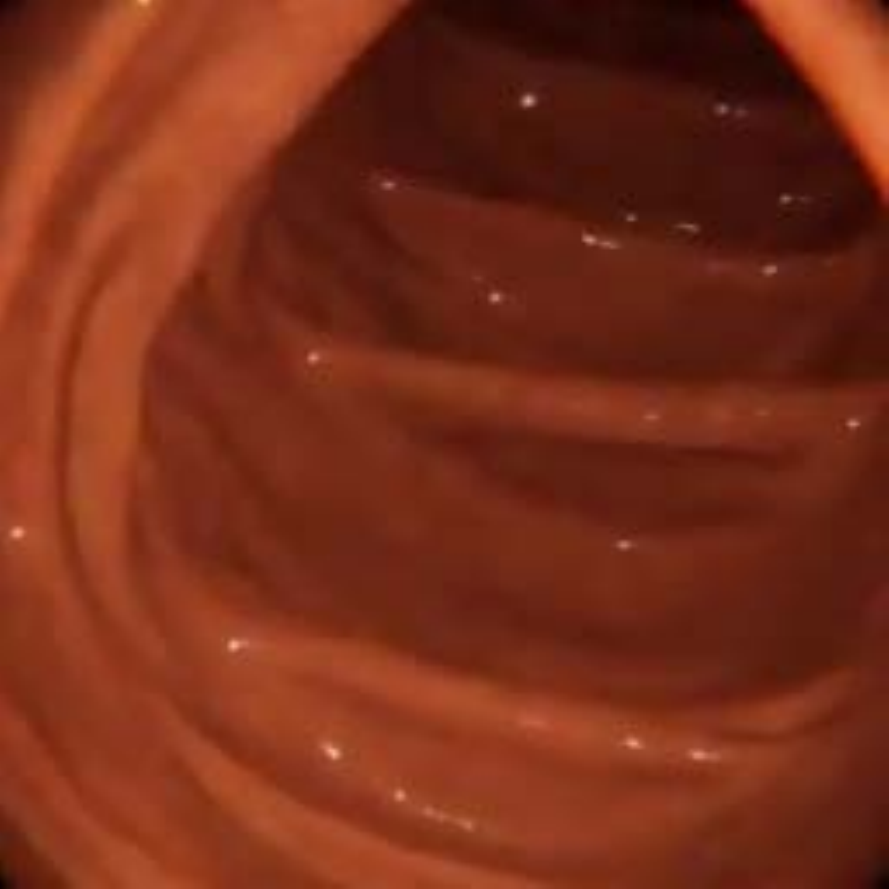}&
\includegraphics[width=0.115\textwidth]{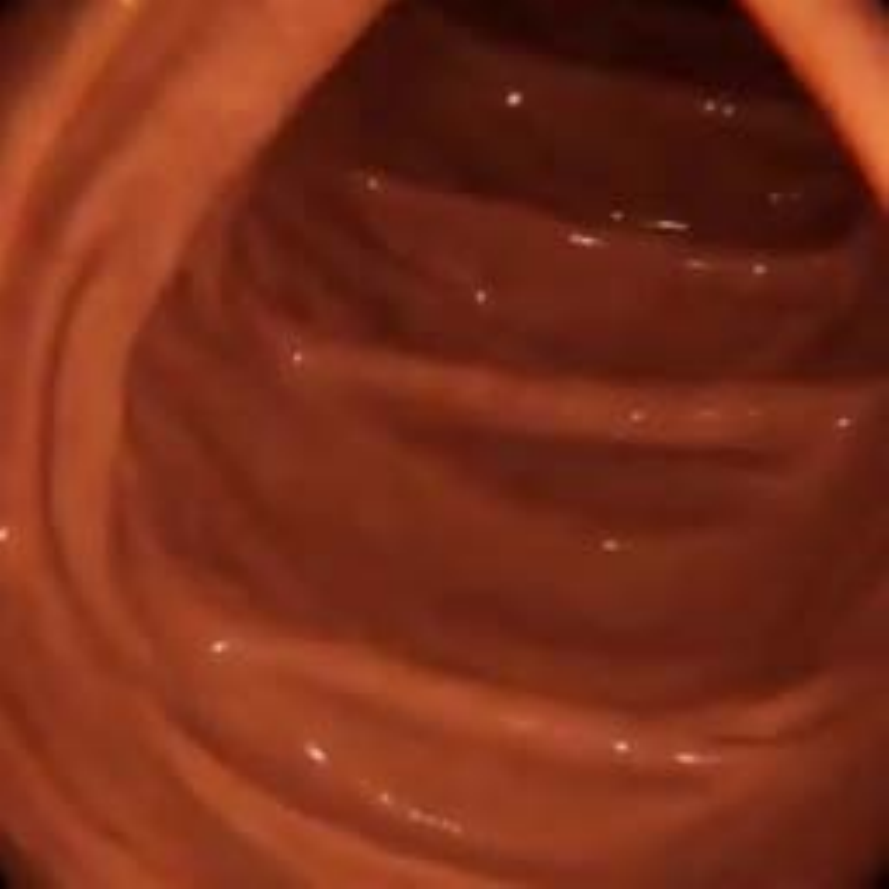}&
\includegraphics[width=0.115\textwidth]{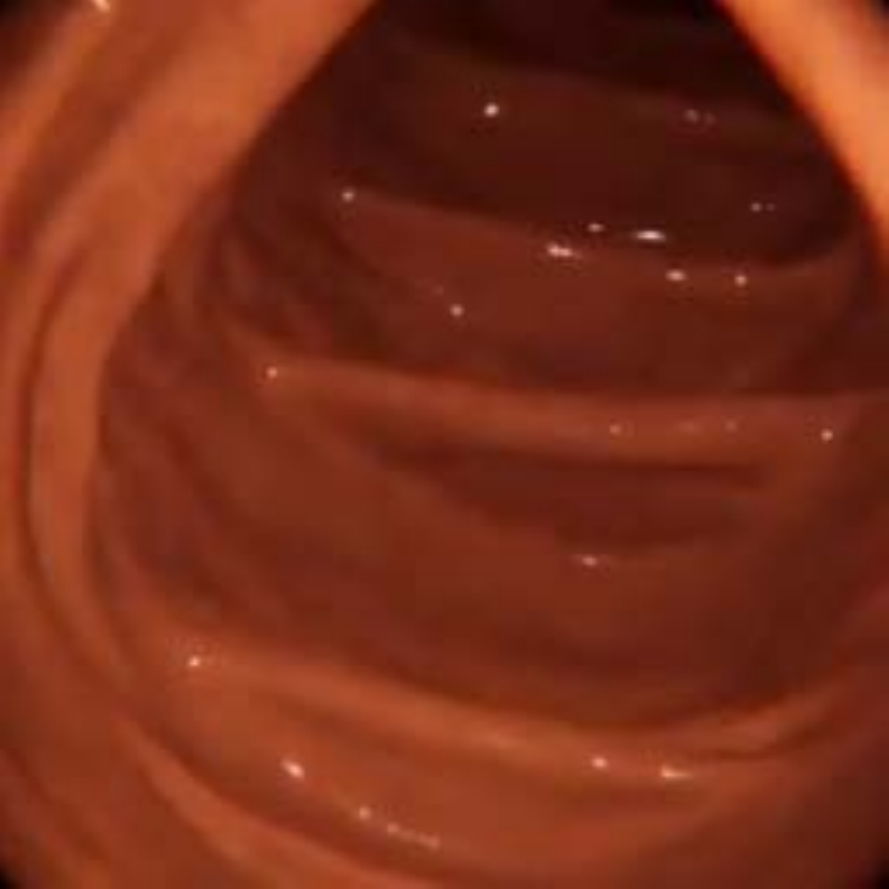}&

\includegraphics[width=0.115\textwidth]{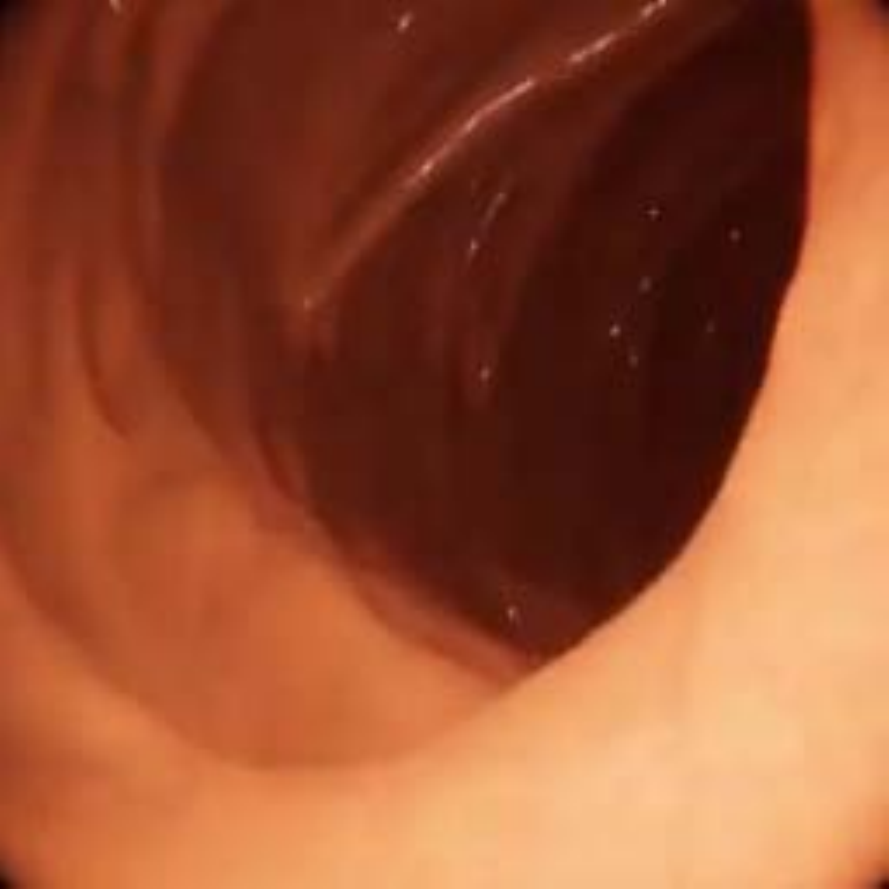}&
\includegraphics[width=0.115\textwidth]{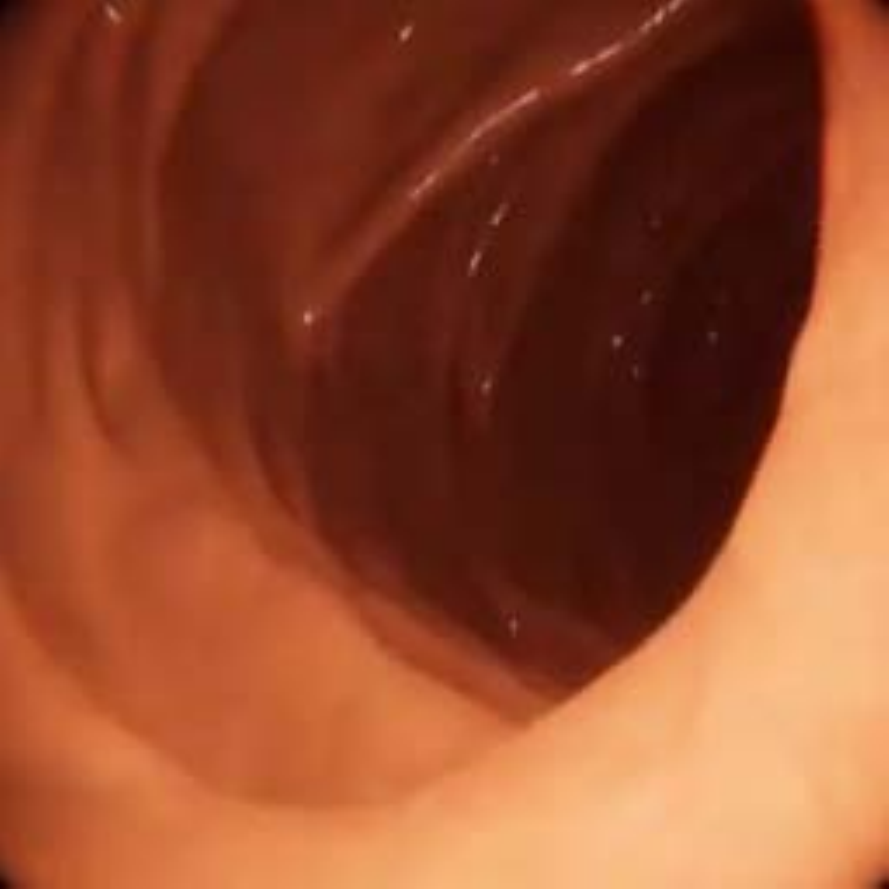}&
\includegraphics[width=0.115\textwidth]{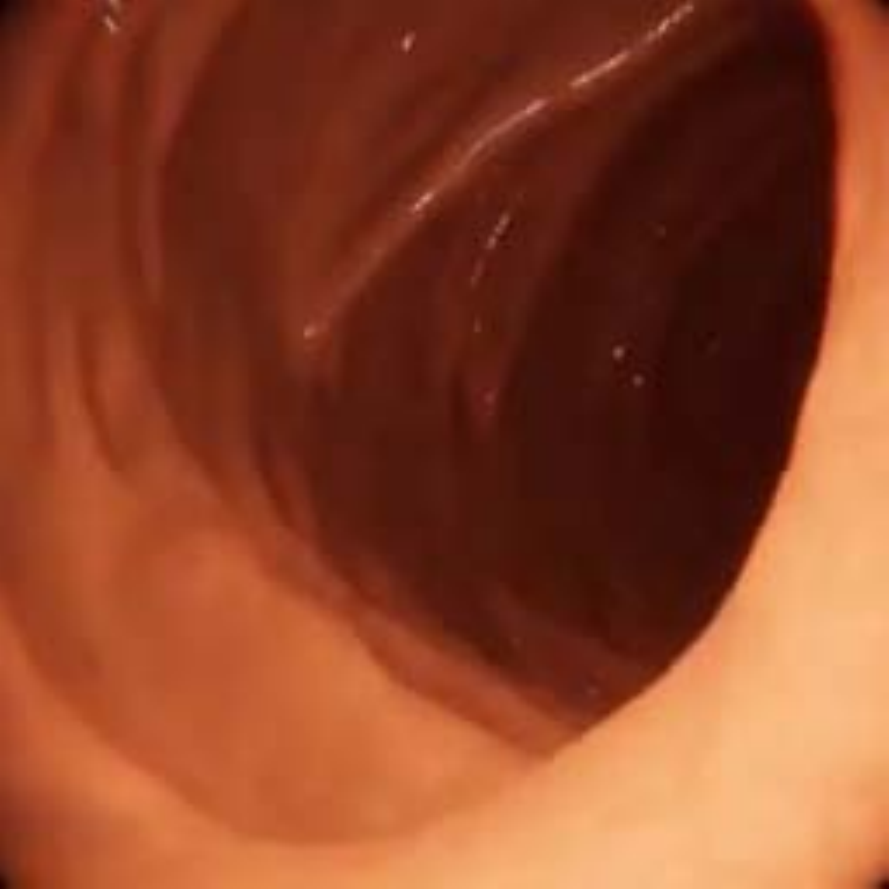}&
\includegraphics[width=0.115\textwidth]{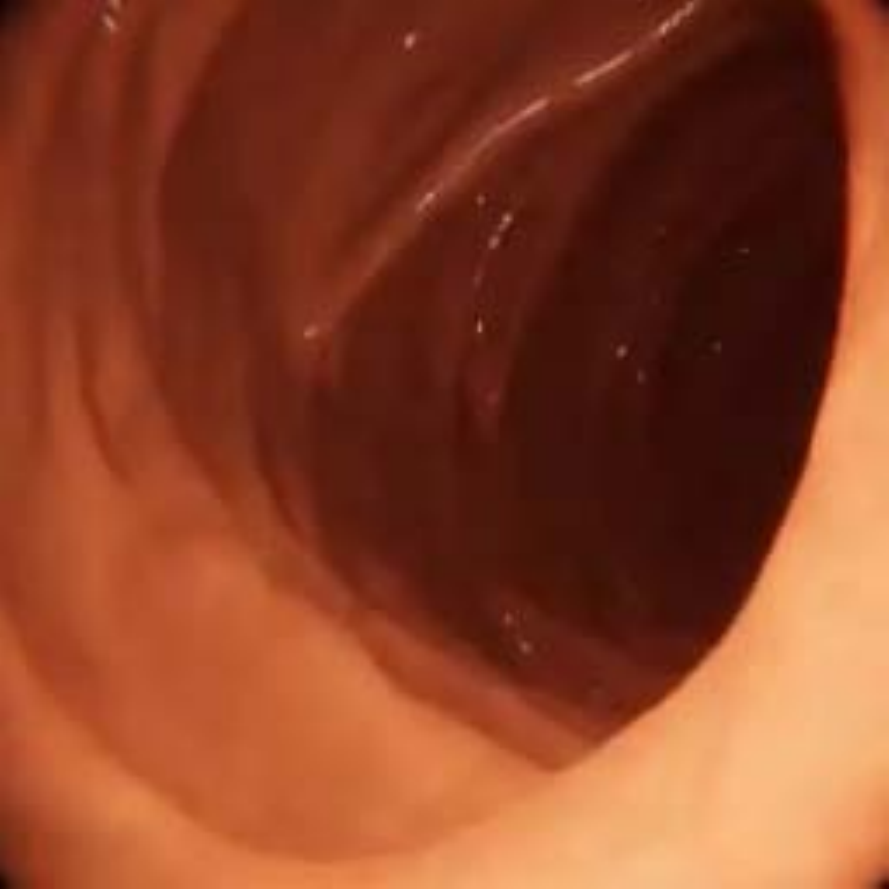}\\

\rotatebox{90}{~~~~\textbf{Ours}}&
\includegraphics[width=0.115\textwidth]{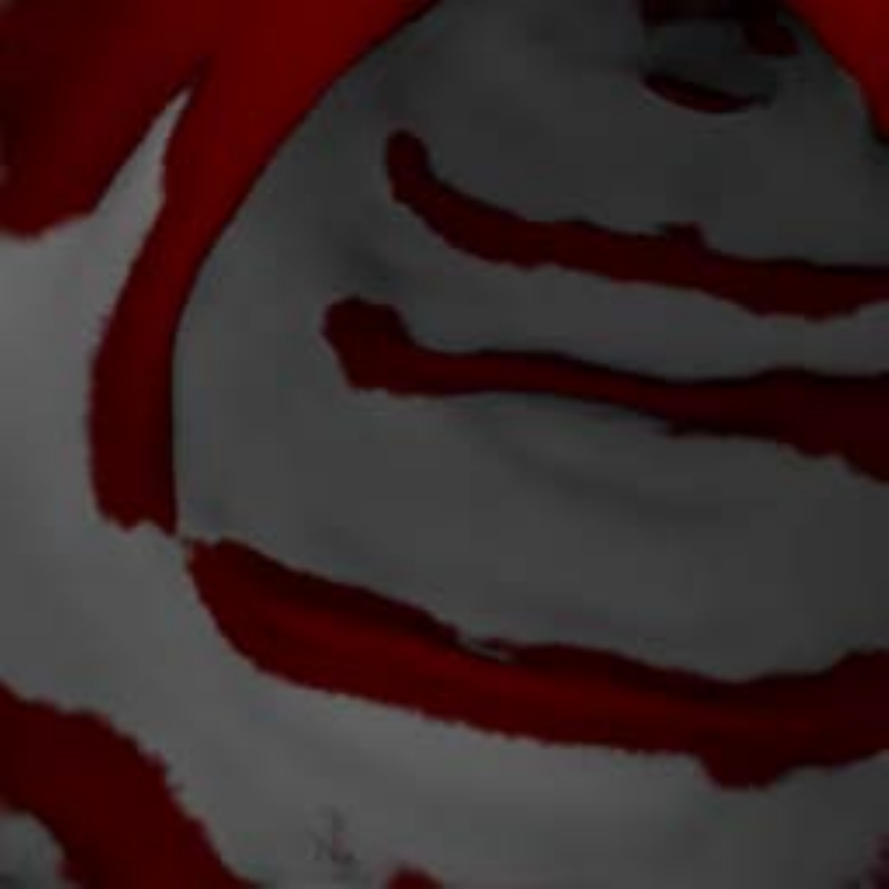}&
\includegraphics[width=0.115\textwidth]{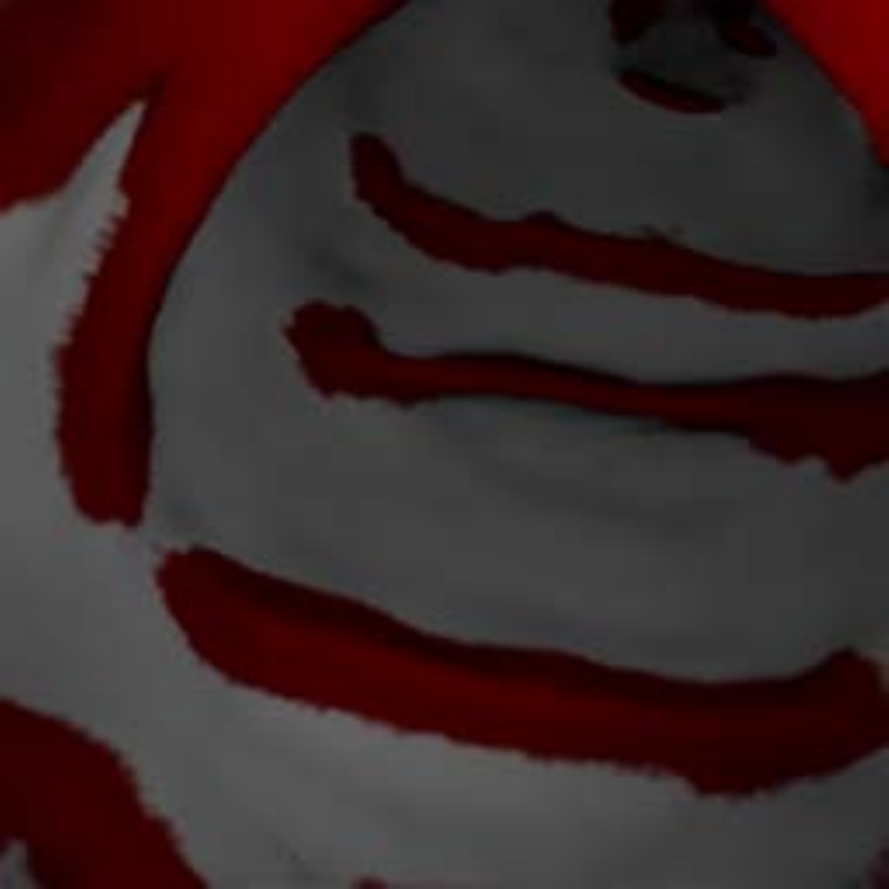}&
\includegraphics[width=0.115\textwidth]{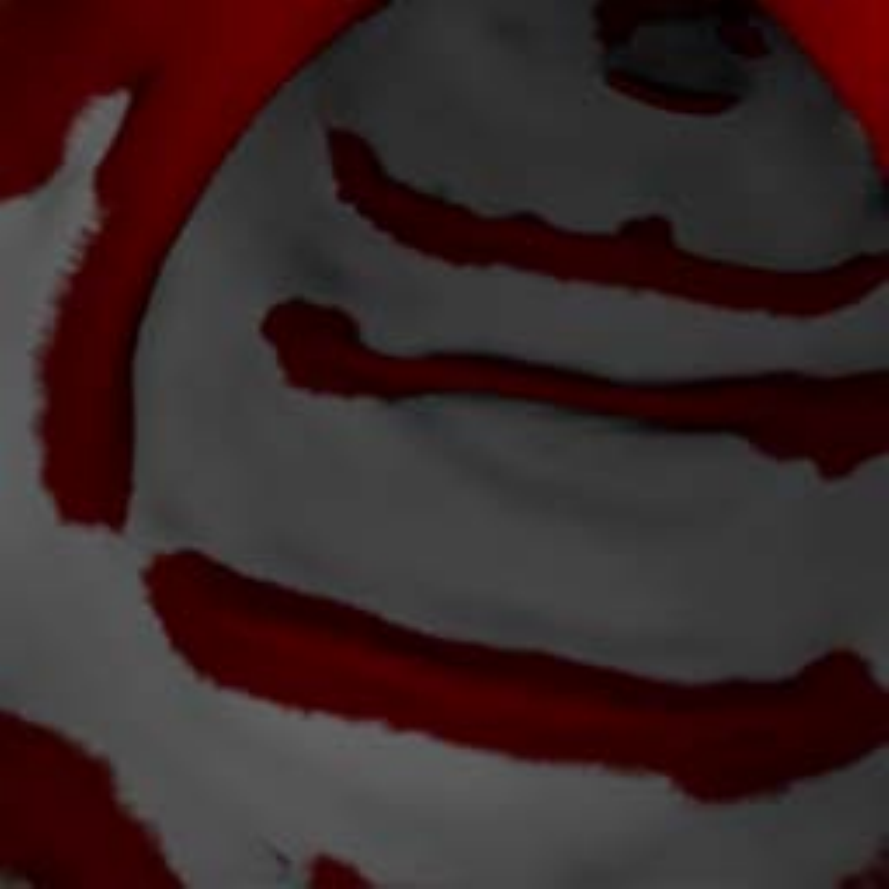}&
\includegraphics[width=0.115\textwidth]{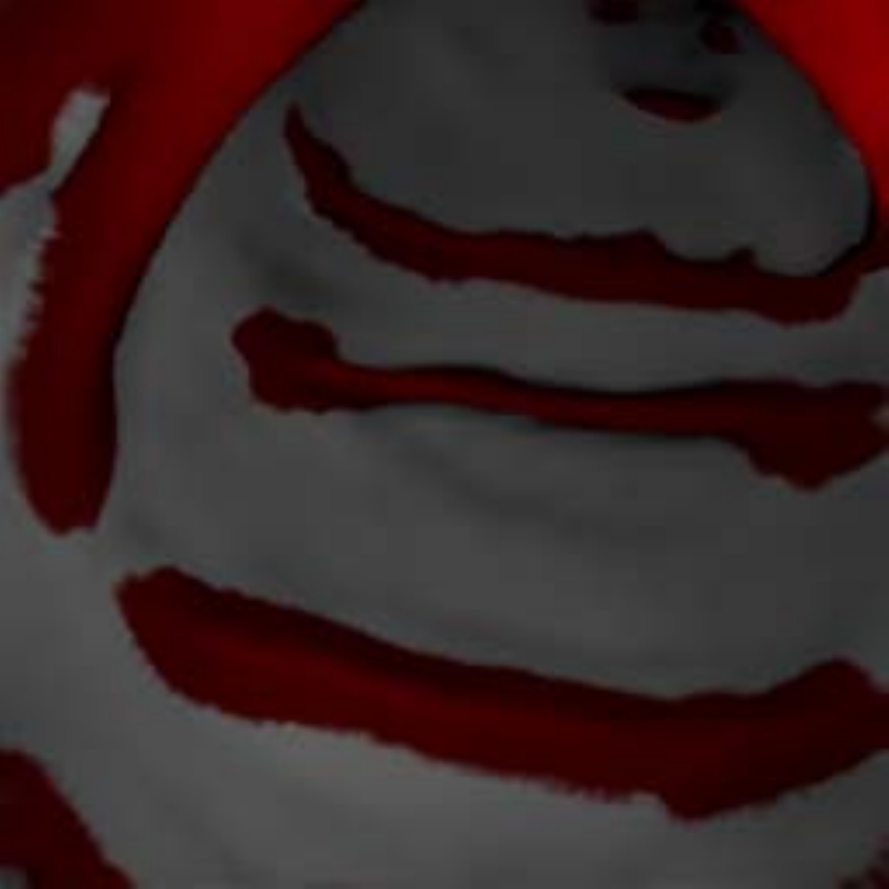}&

\includegraphics[width=0.115\textwidth]{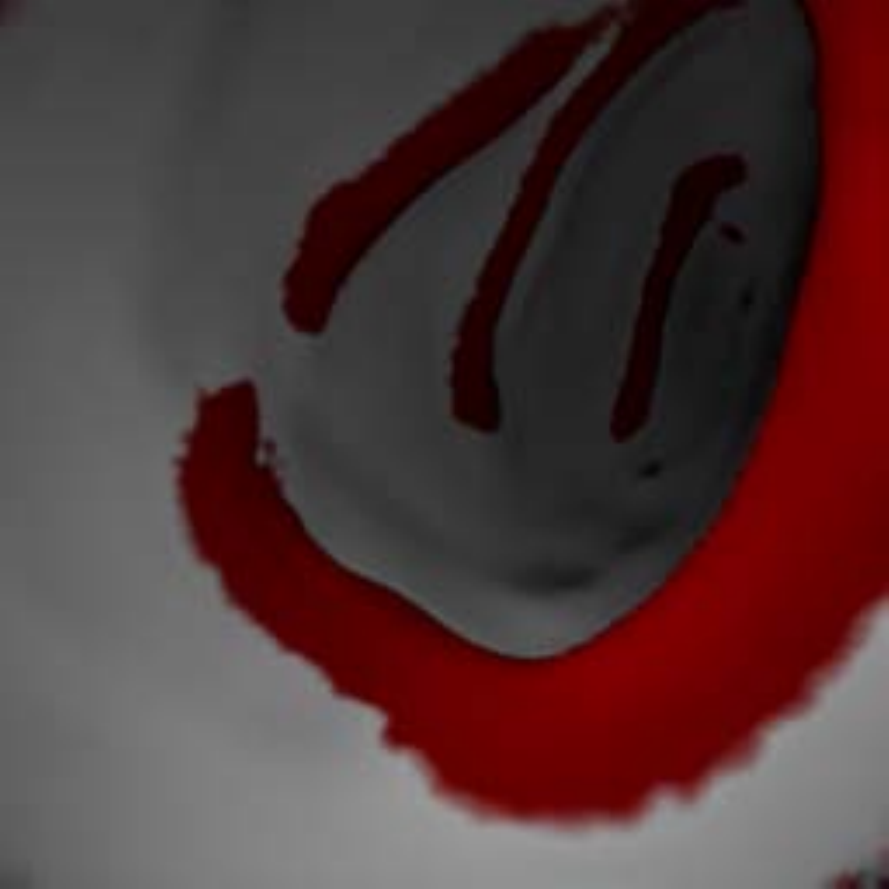}&
\includegraphics[width=0.115\textwidth]{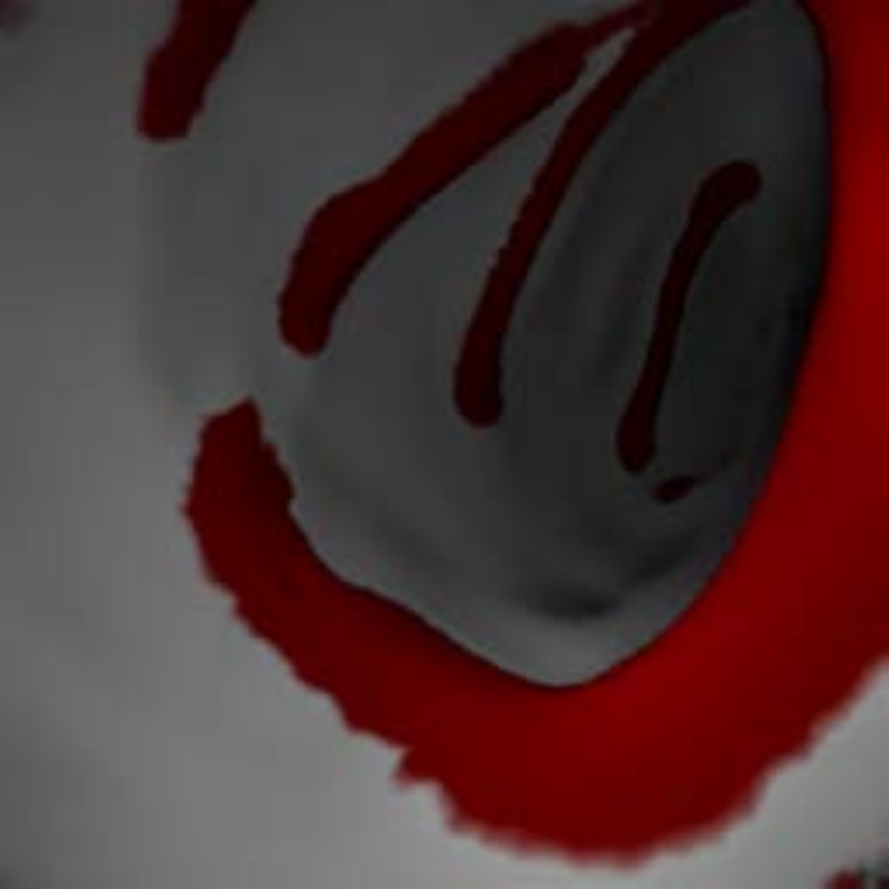}&
\includegraphics[width=0.115\textwidth]{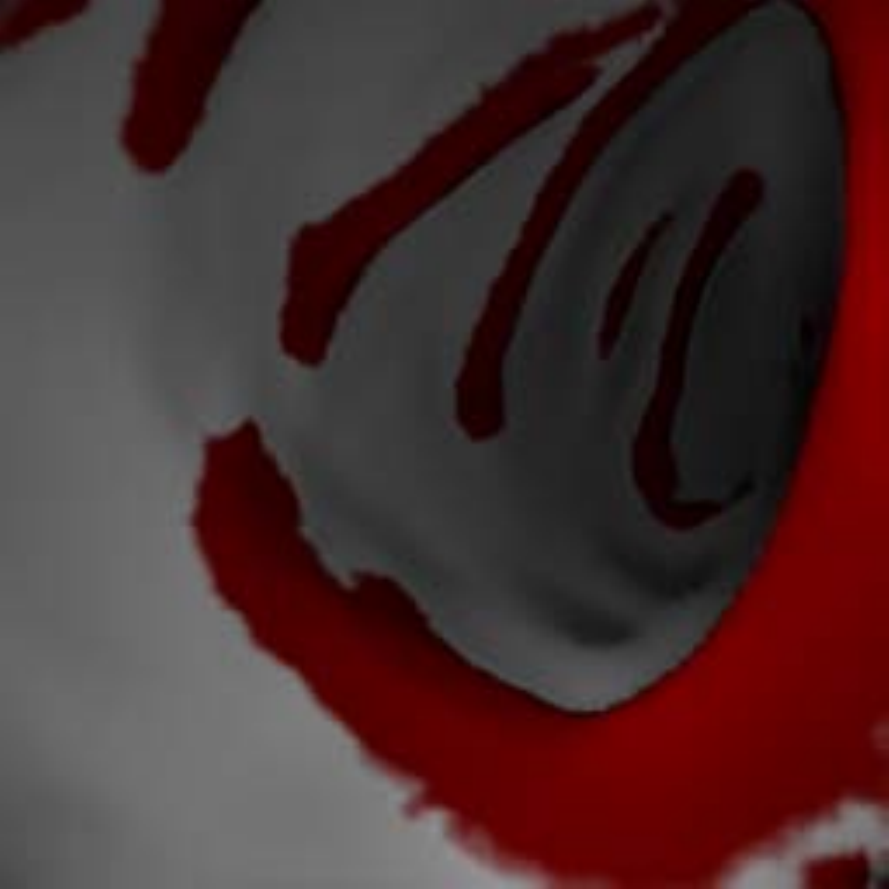}&
\includegraphics[width=0.115\textwidth]{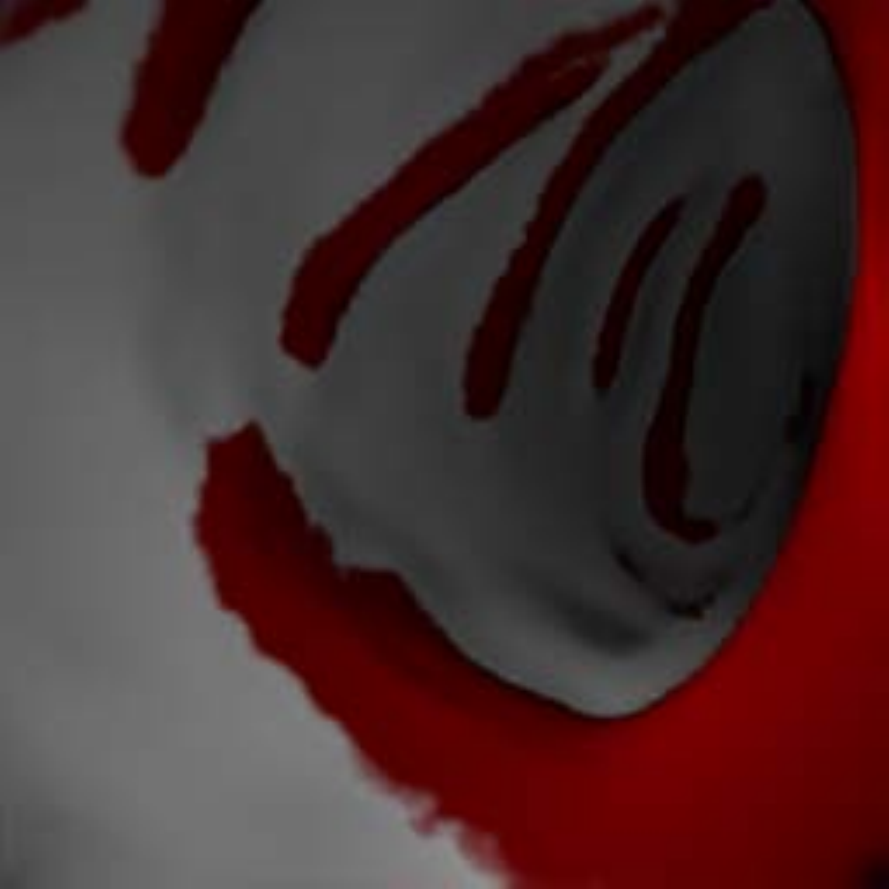}\\

\rotatebox{90}{Mathew\cite{mathew2020augmenting}}&
\includegraphics[width=0.115\textwidth]{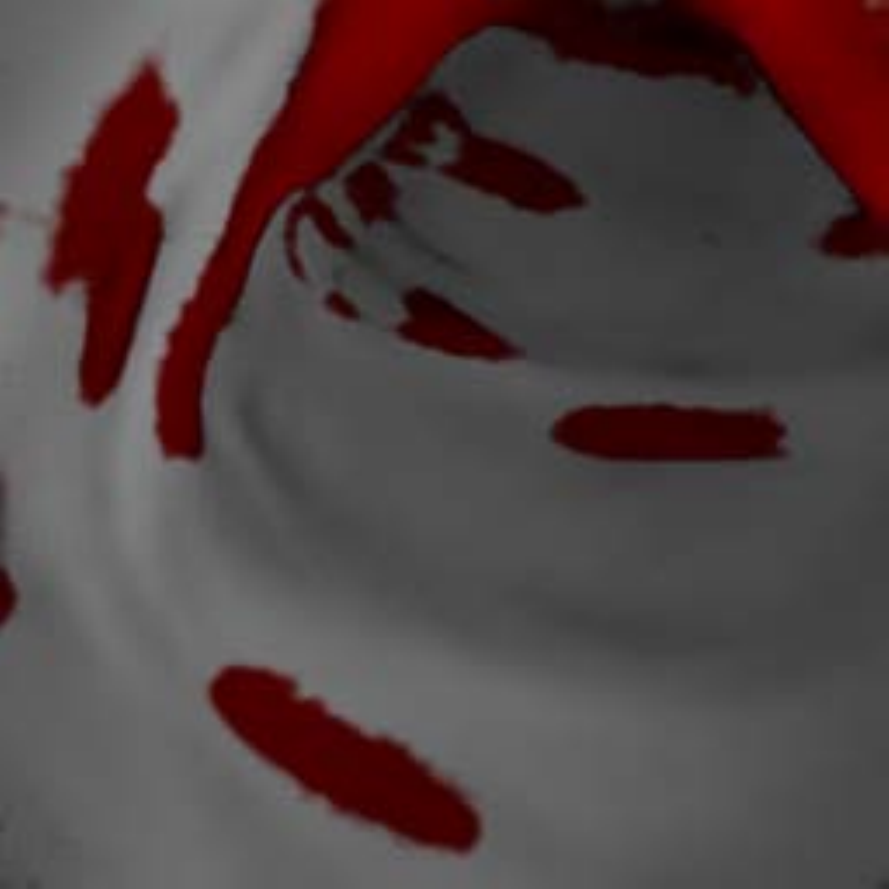}&
\includegraphics[width=0.115\textwidth]{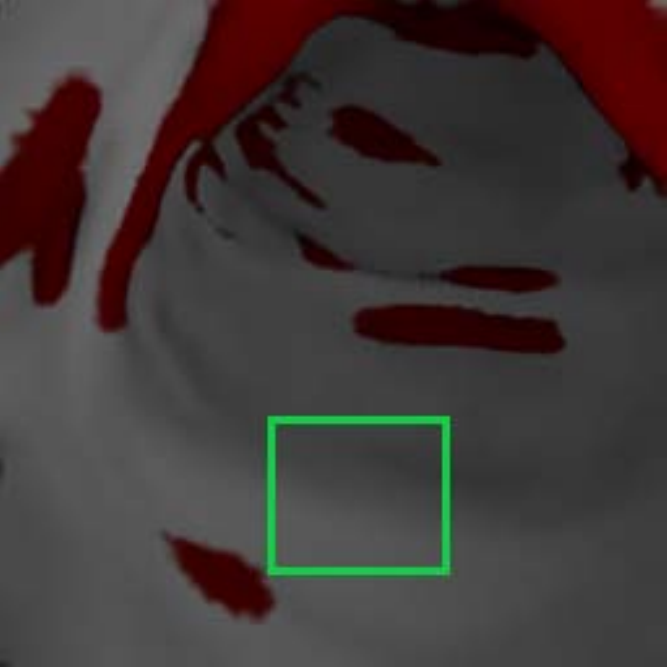}&
\includegraphics[width=0.115\textwidth]{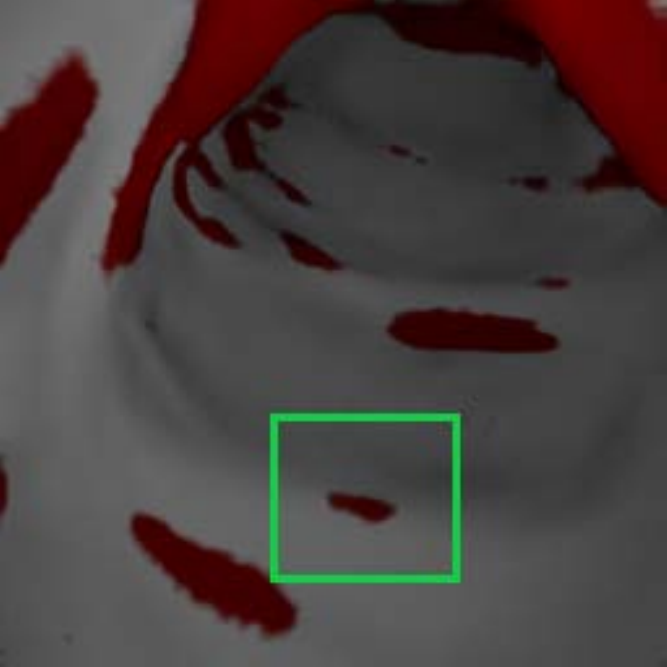}&
\includegraphics[width=0.115\textwidth]{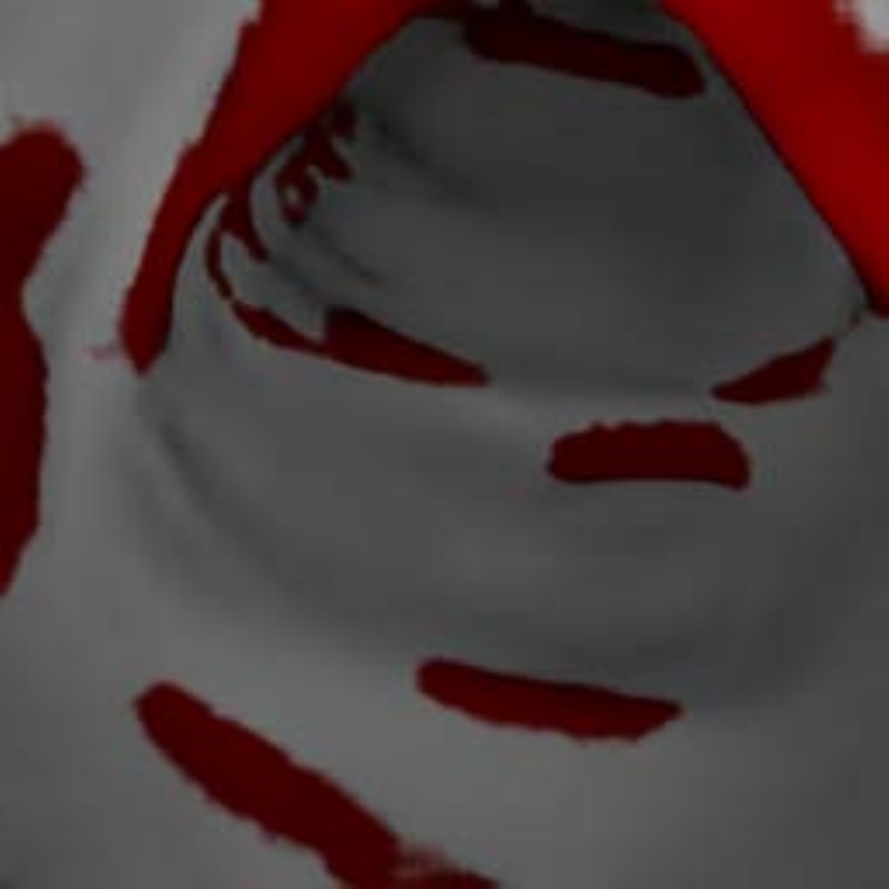}&

\includegraphics[width=0.115\textwidth]{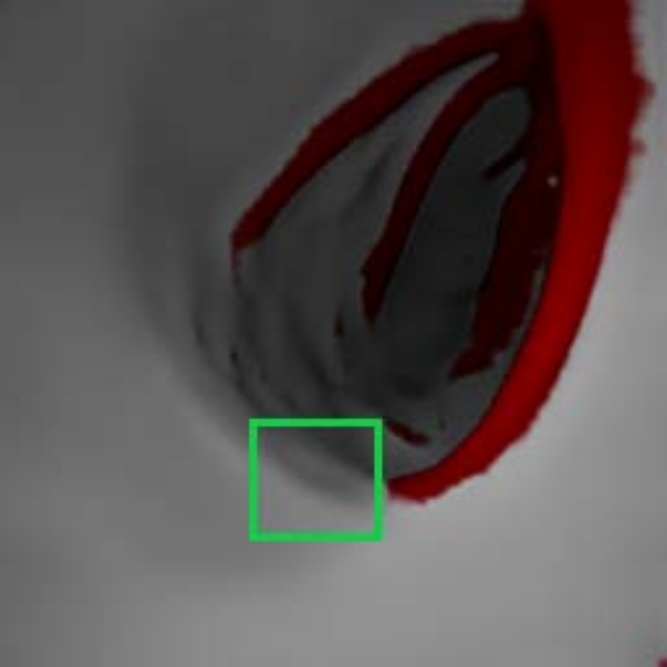}&
\includegraphics[width=0.115\textwidth]{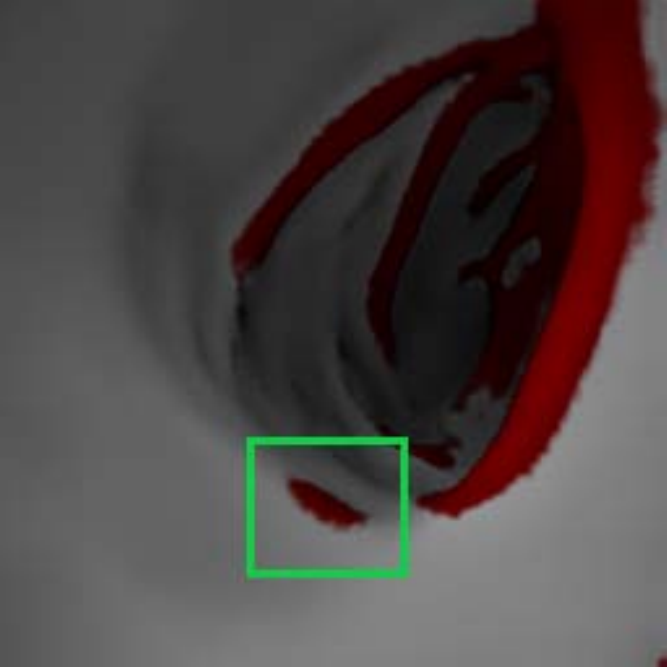}&
\includegraphics[width=0.115\textwidth]{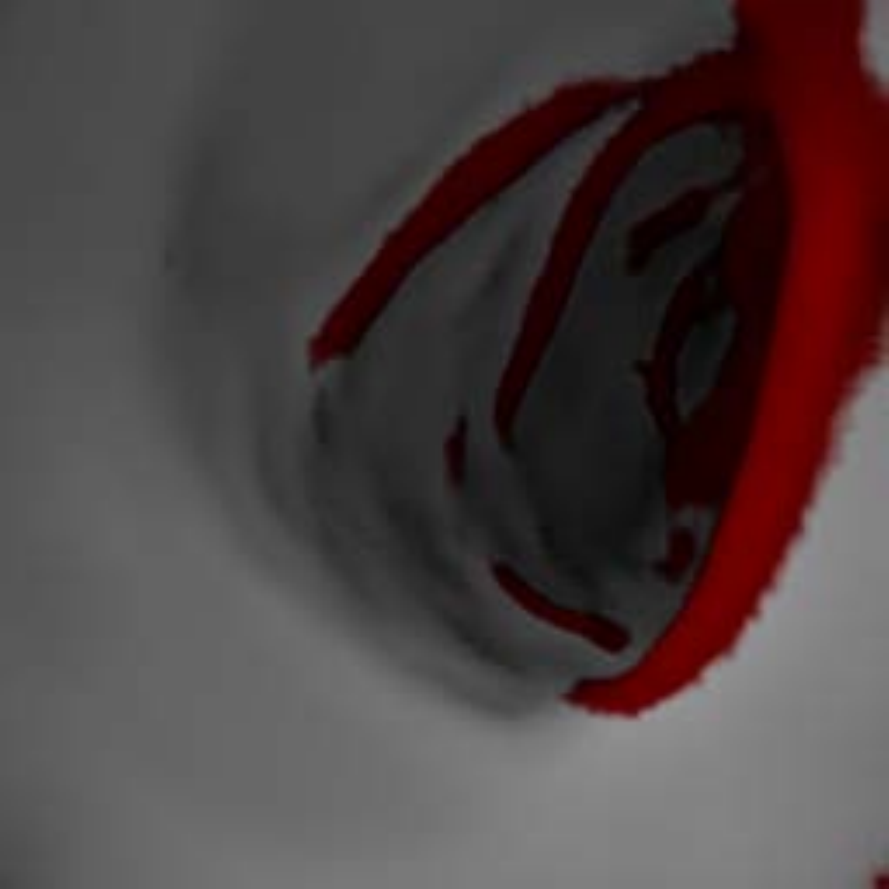}&
\includegraphics[width=0.115\textwidth]{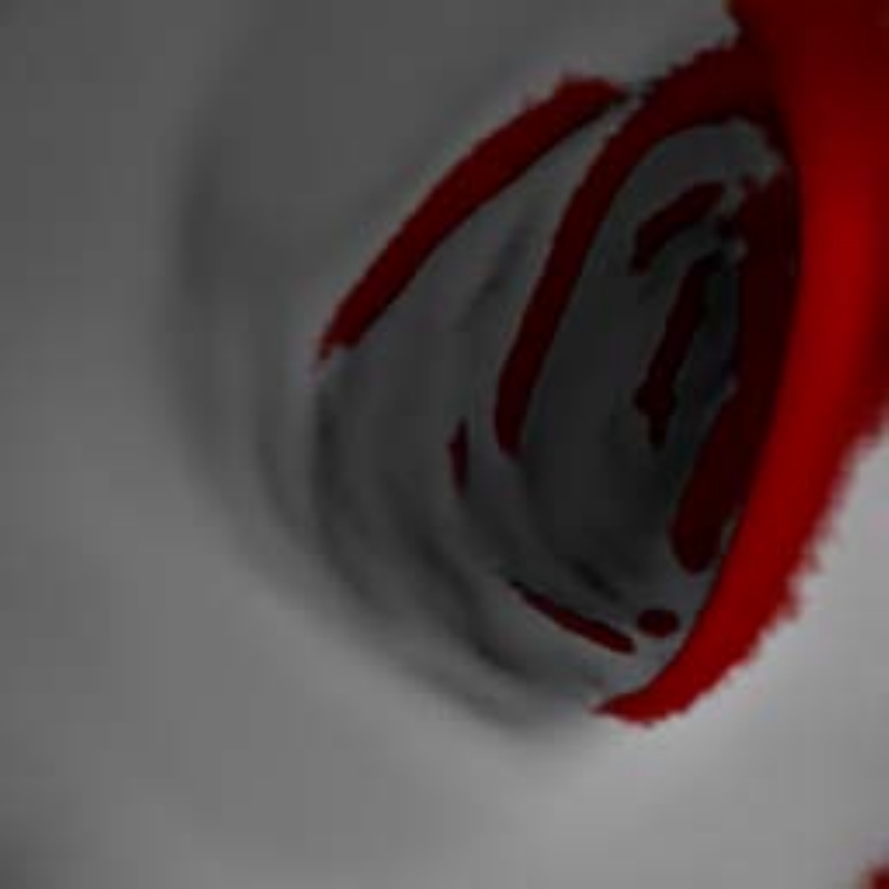}\\

\rotatebox{90}{~~~~\textbf{Ours}}&
\includegraphics[width=0.115\textwidth]{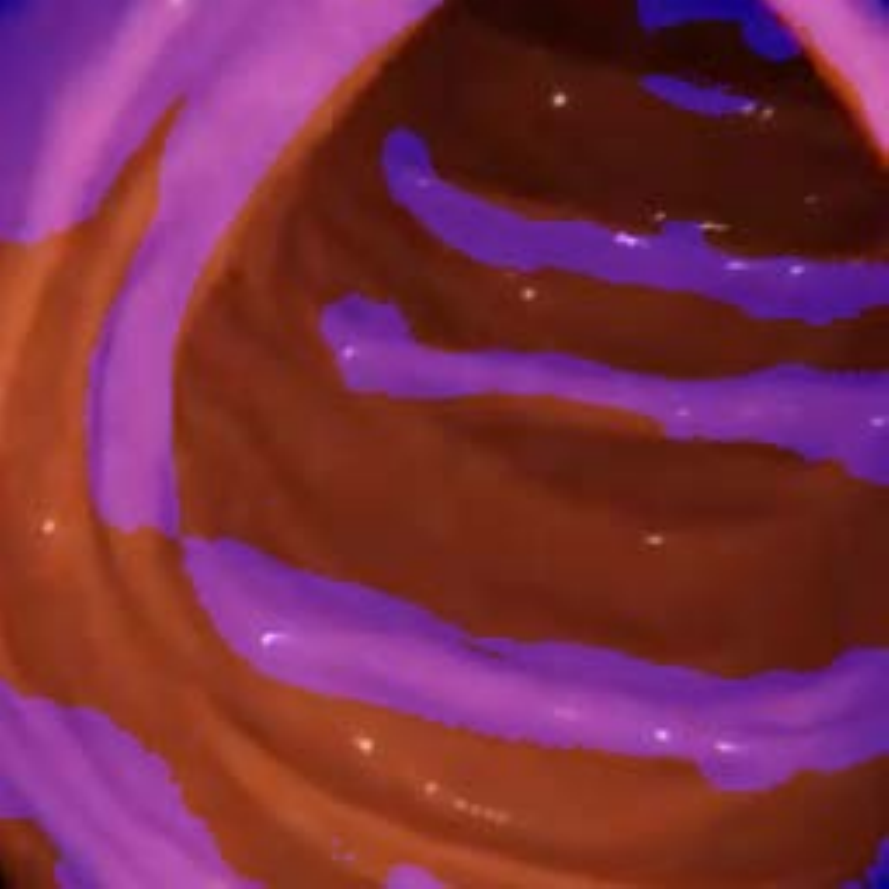}&
\includegraphics[width=0.115\textwidth]{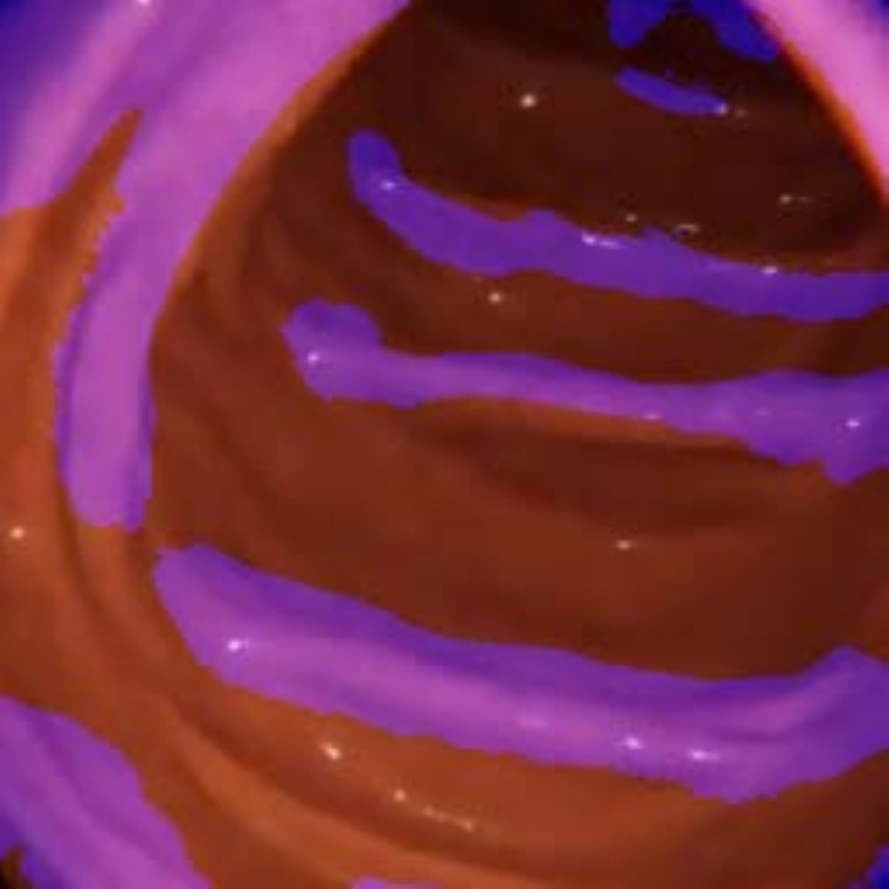}&
\includegraphics[width=0.115\textwidth]{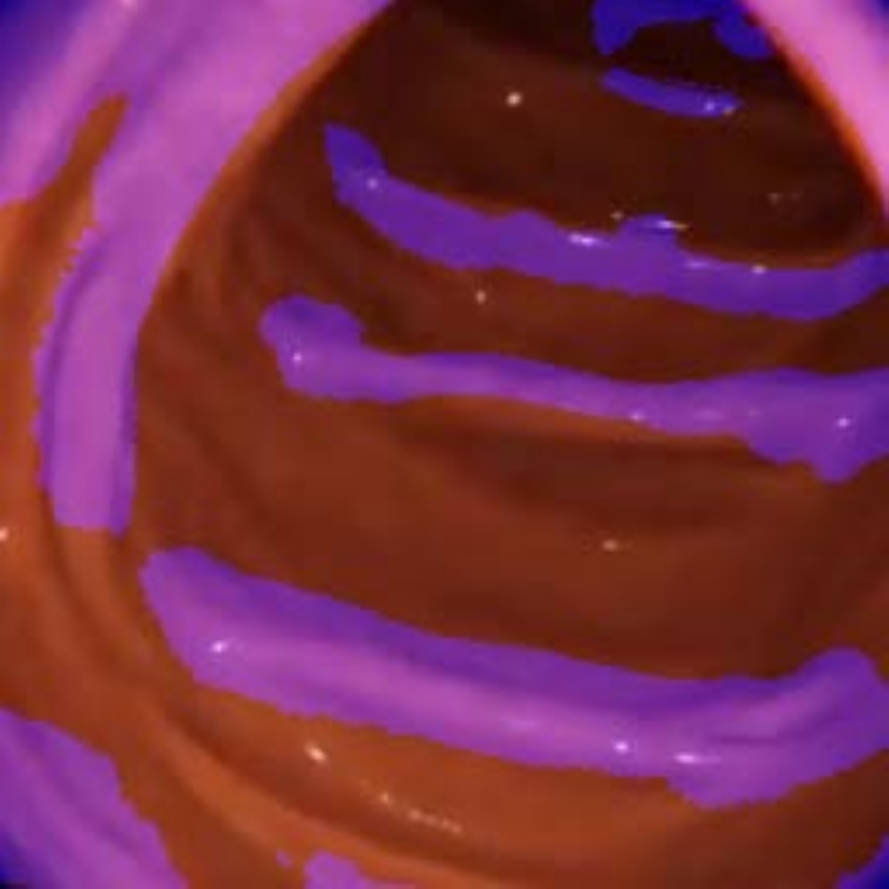}&
\includegraphics[width=0.115\textwidth]{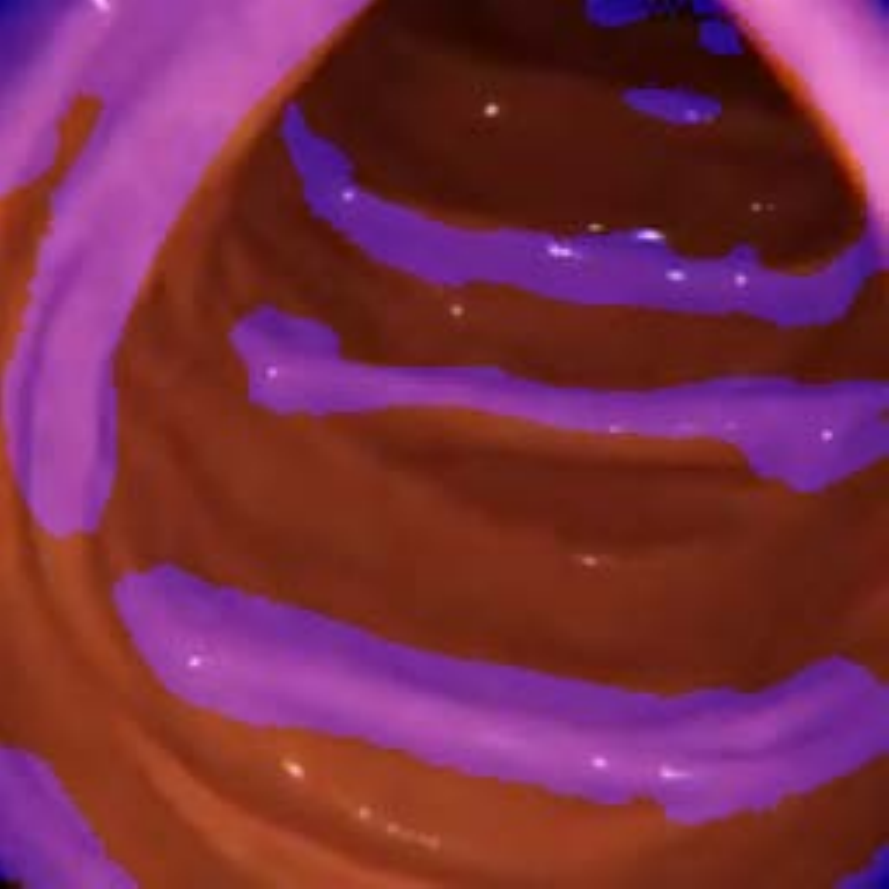}&

\includegraphics[width=0.115\textwidth]{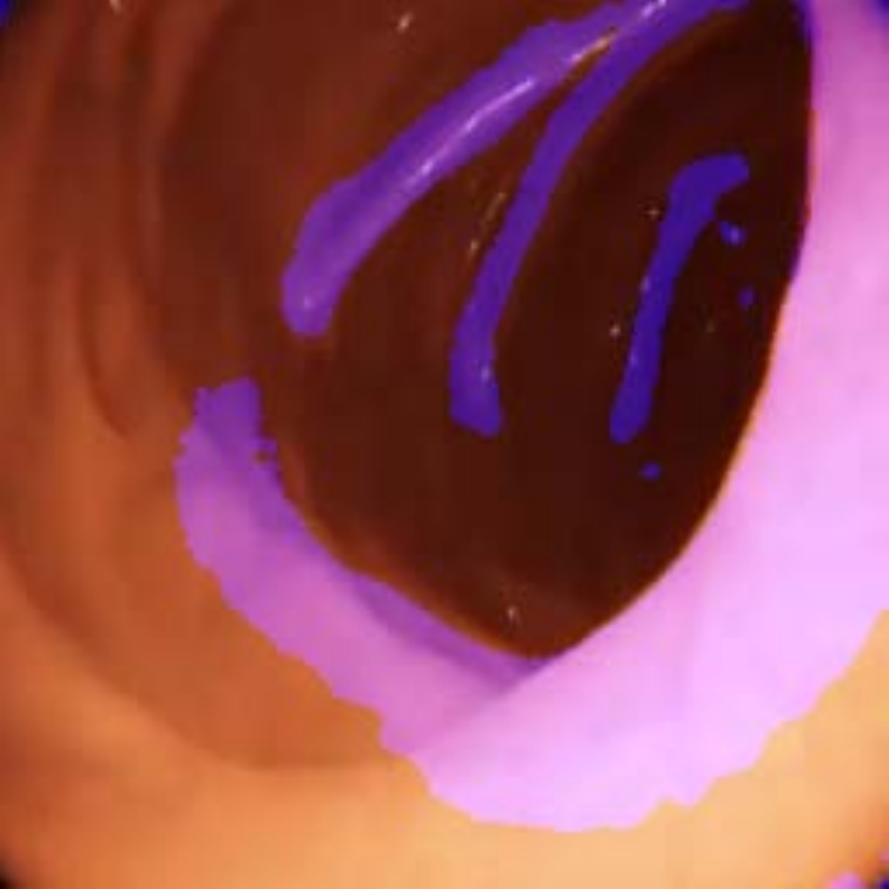}&
\includegraphics[width=0.115\textwidth]{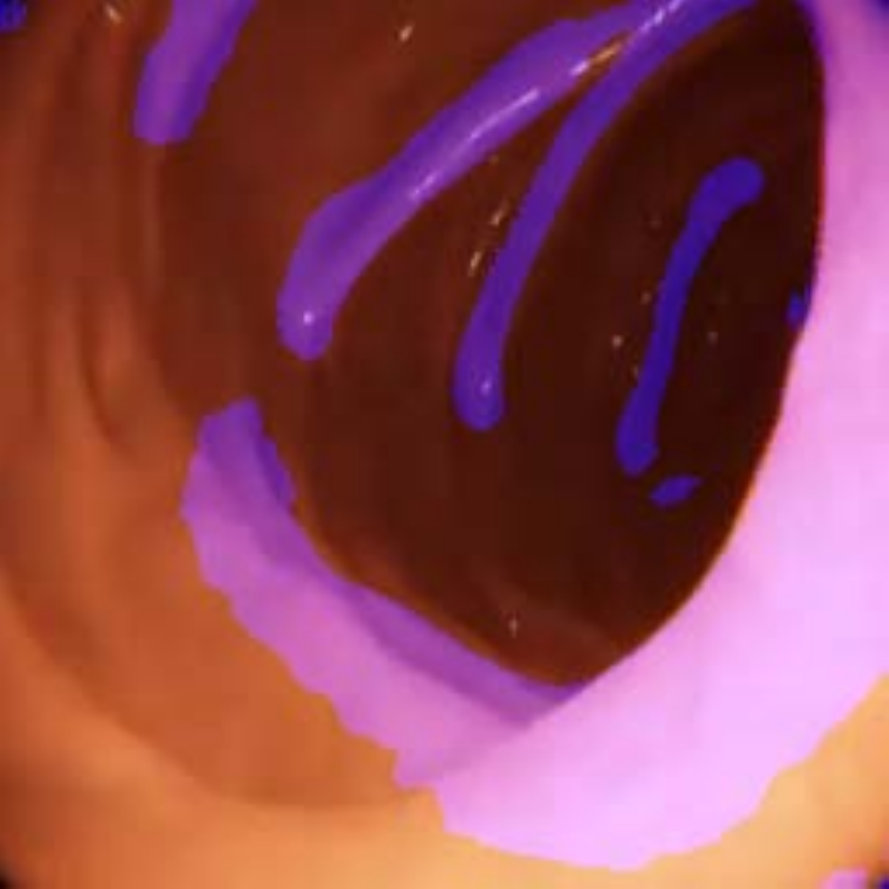}&
\includegraphics[width=0.115\textwidth]{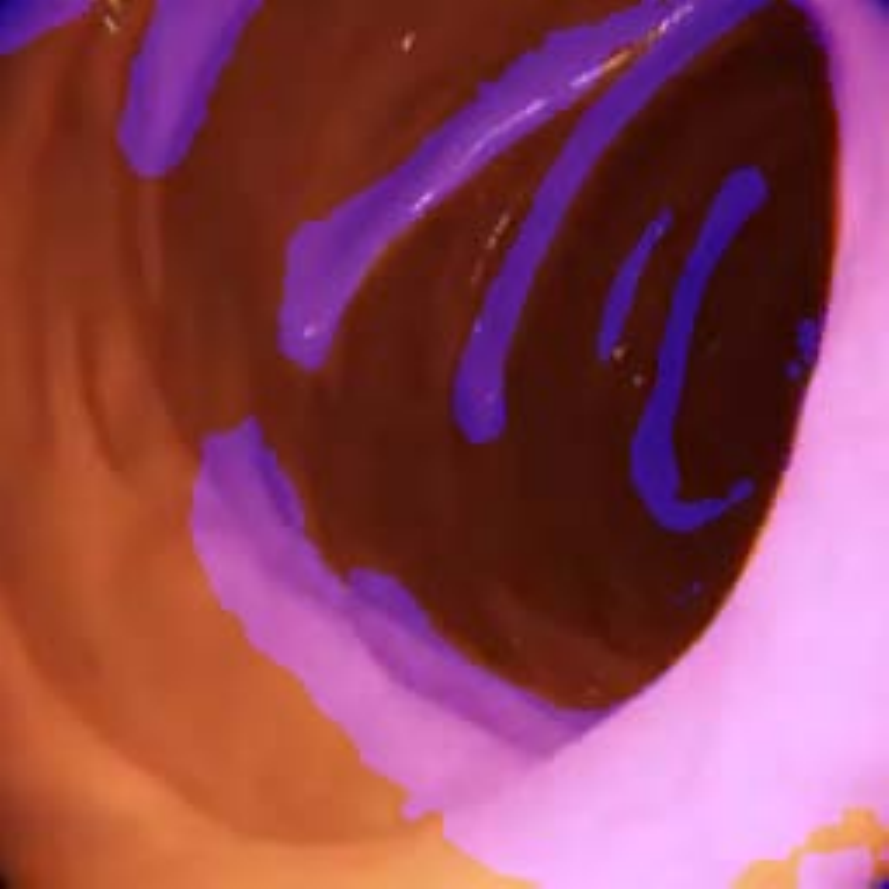}&
\includegraphics[width=0.115\textwidth]{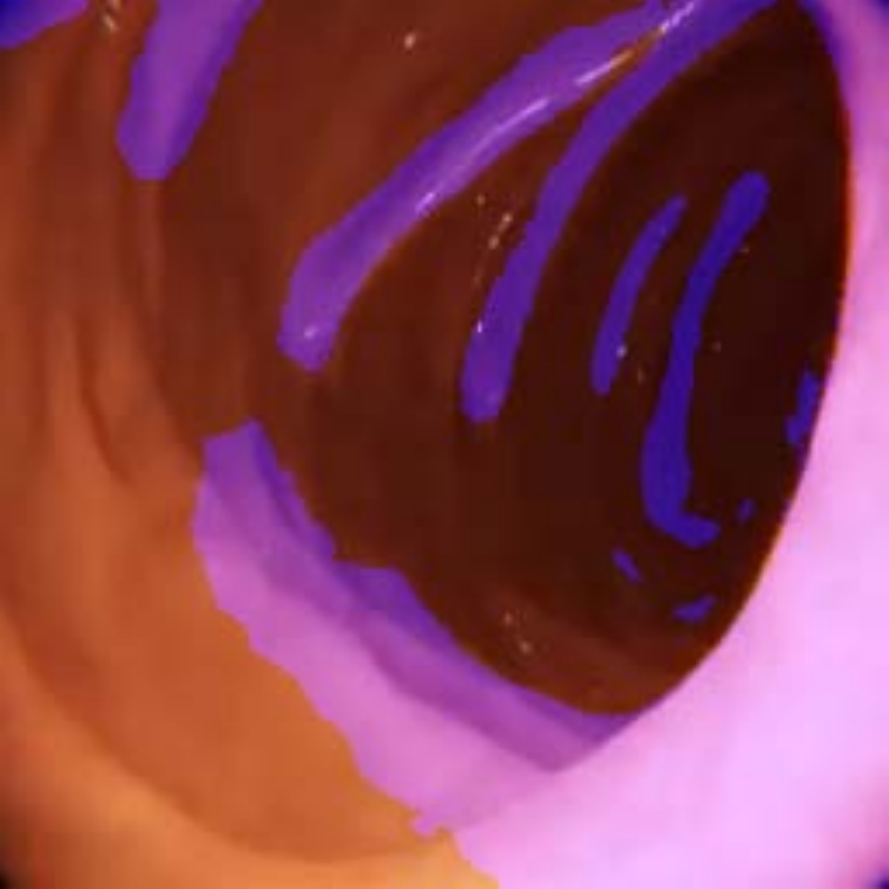}\\

\rotatebox{90}{Mathew\cite{mathew2020augmenting}}&
\includegraphics[width=0.115\textwidth]{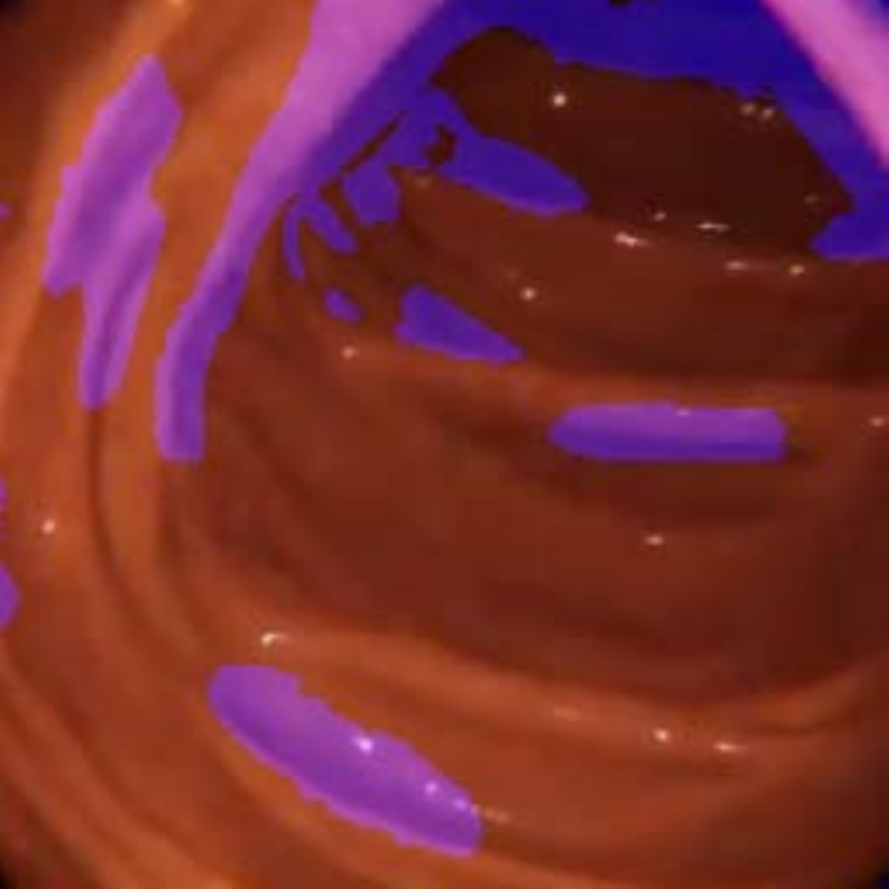}&
\includegraphics[width=0.115\textwidth]{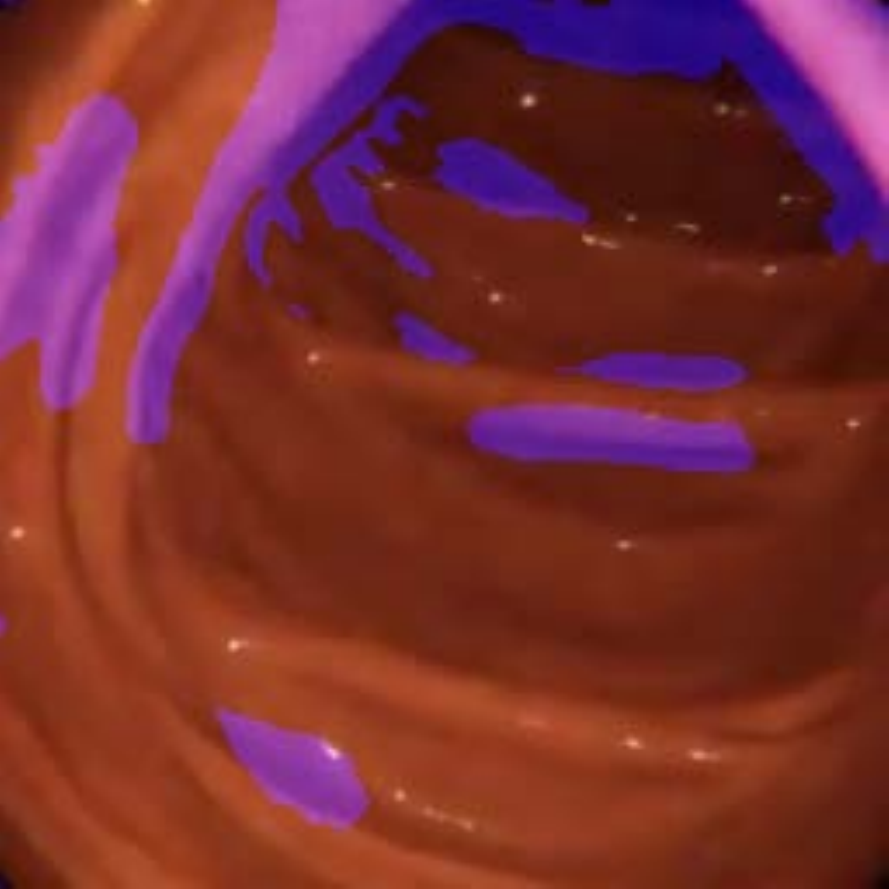}&
\includegraphics[width=0.115\textwidth]{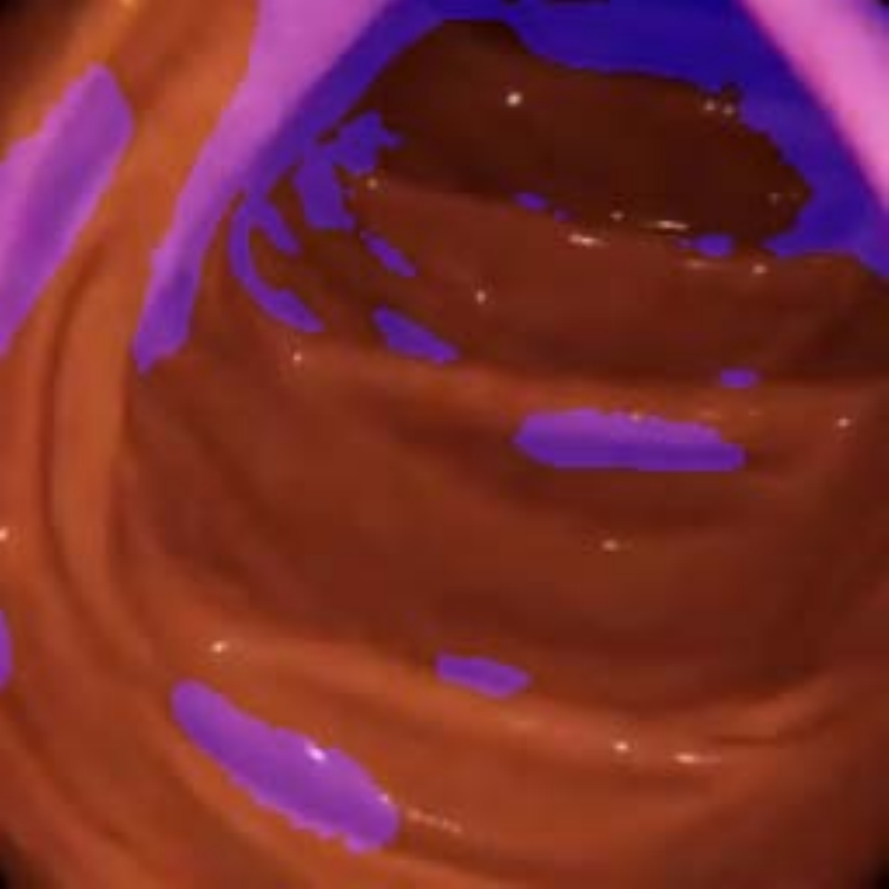}&
\includegraphics[width=0.115\textwidth]{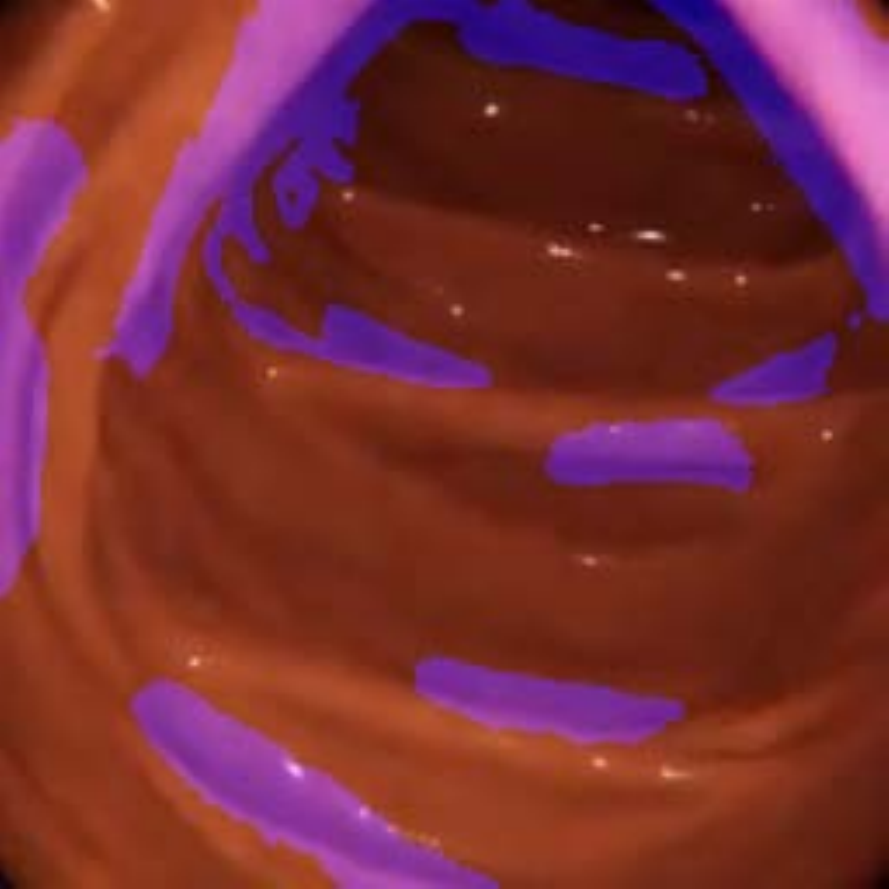}&

\includegraphics[width=0.115\textwidth]{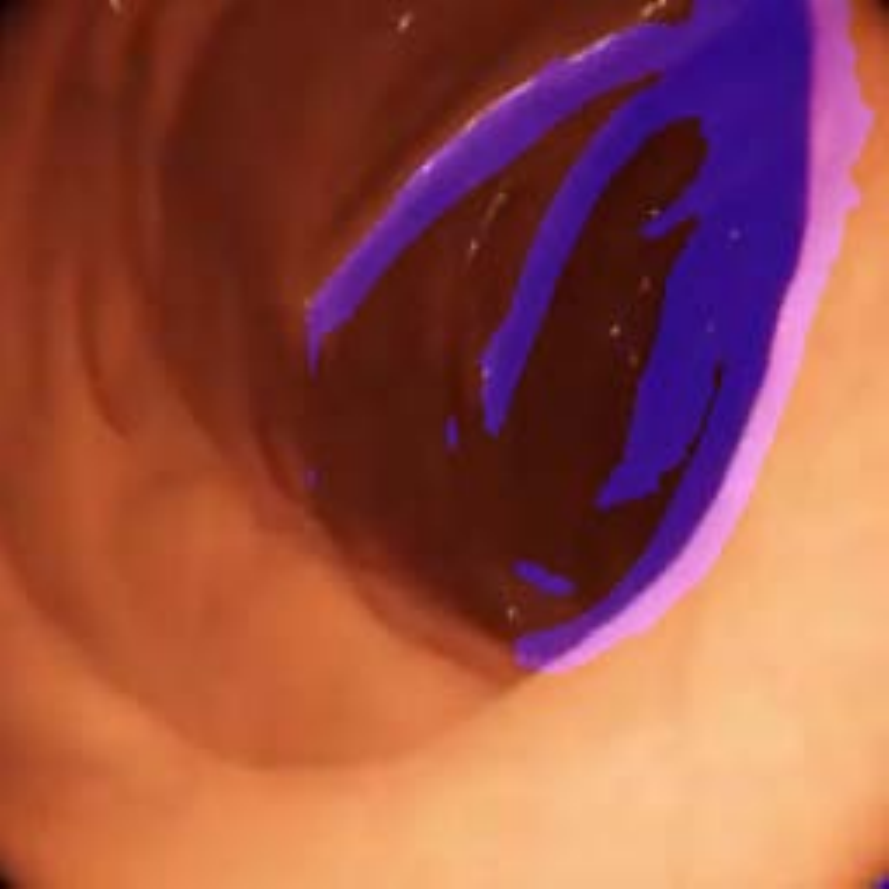}&
\includegraphics[width=0.115\textwidth]{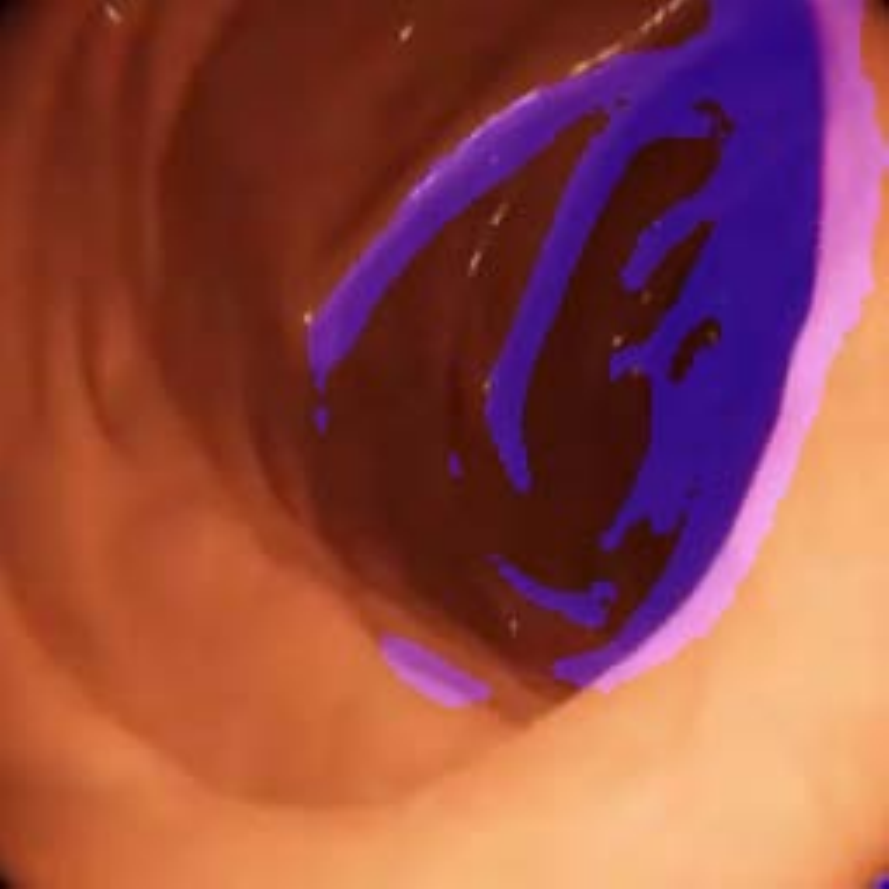}&
\includegraphics[width=0.115\textwidth]{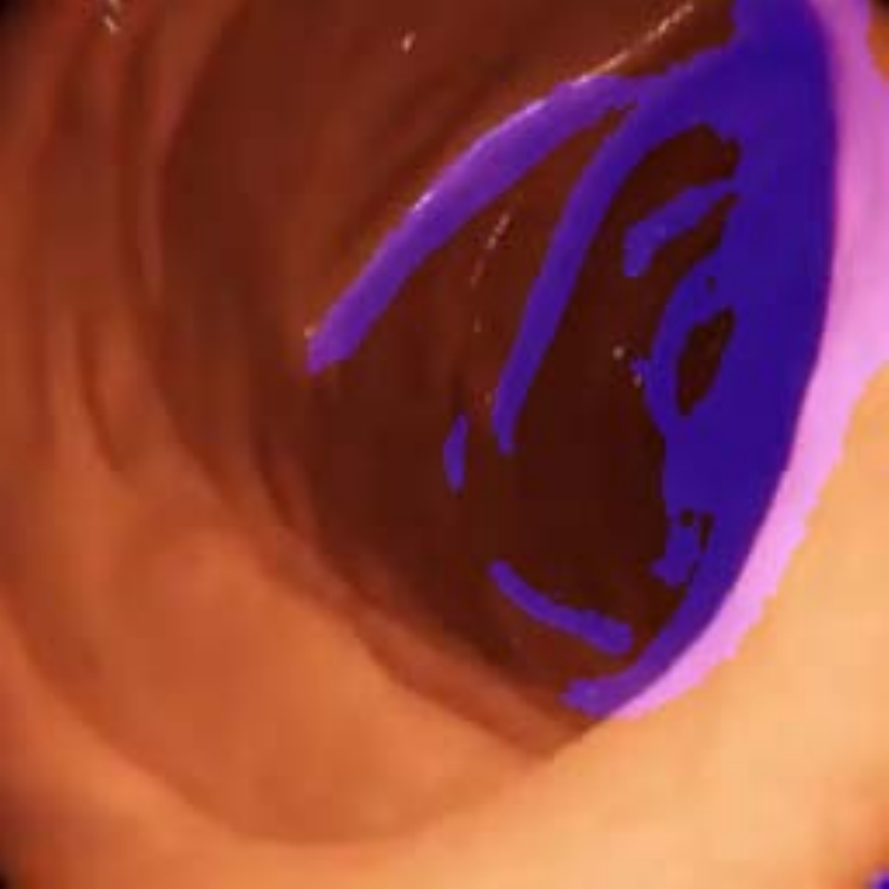}&
\includegraphics[width=0.115\textwidth]{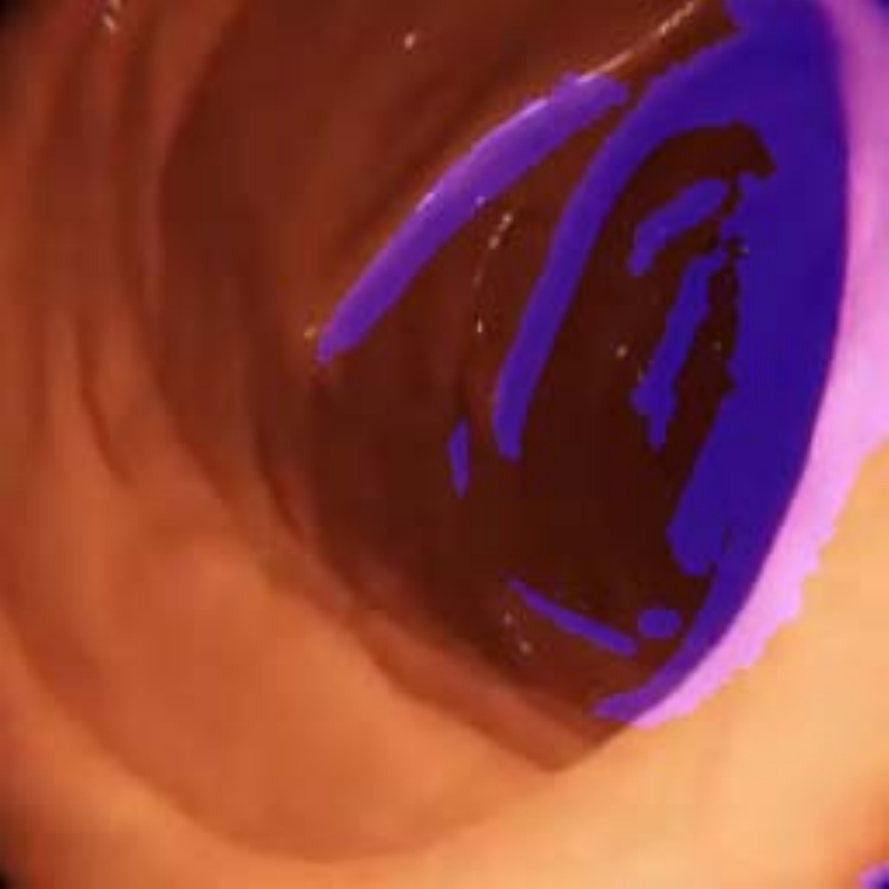}\\

\end{tabular}
\caption{Results of FoldIt (ours) and XDCycleGAN model \cite{mathew2020augmenting} are shown on video sequences from Ma et al. \cite{ma2019real} and recently released VR-CAPS colon simulator \cite{incetan2021vr}. Complete video sequences are provided in the \textbf{supplementary video}. The OC frames are followed by FoldIt and XDCycleGAN output and the fold segmentations overlaid on the OC images. Green bounding boxes indicate examples of XDCycleGAN losing haustral folds information across frames whereas our FoldIt preserves these features.}
\label{fig:unc}
\end{center}
\end{figure}

We also show results in Fig. \ref{fig:unc} for 11 real OC video sequences from Ma et al. \cite{ma2019real} and 2 from recently-released VR-CAPS simulator \cite{incetan2021vr}. For both, FoldIt and XDCycleGAN, we show an overlay view, where the fold segmentation is extracted from the network output and superimposed in blue on the OC input. While XDCycleGAN is capable of doing scale-consistent depth maps for these video sequences, it struggles to retain haustral fold features across frames, as shown by the green bounding boxes and results in flickering in the final output (as shown in the \textbf{supplementary video}\footnote{Supplementary Video: \url{https://youtu.be/_iWBJnDMXjo}}). FoldIt delineates the folds more accurately and is more consistent in preserving fold features across frames (and does not result in the flickering effect seen with XDCycleGAN). 

\begin{figure}[t!]
\begin{center}
\setlength{\tabcolsep}{1.5pt}
\begin{tabular}{cccc}

\includegraphics[width=0.2\textwidth]{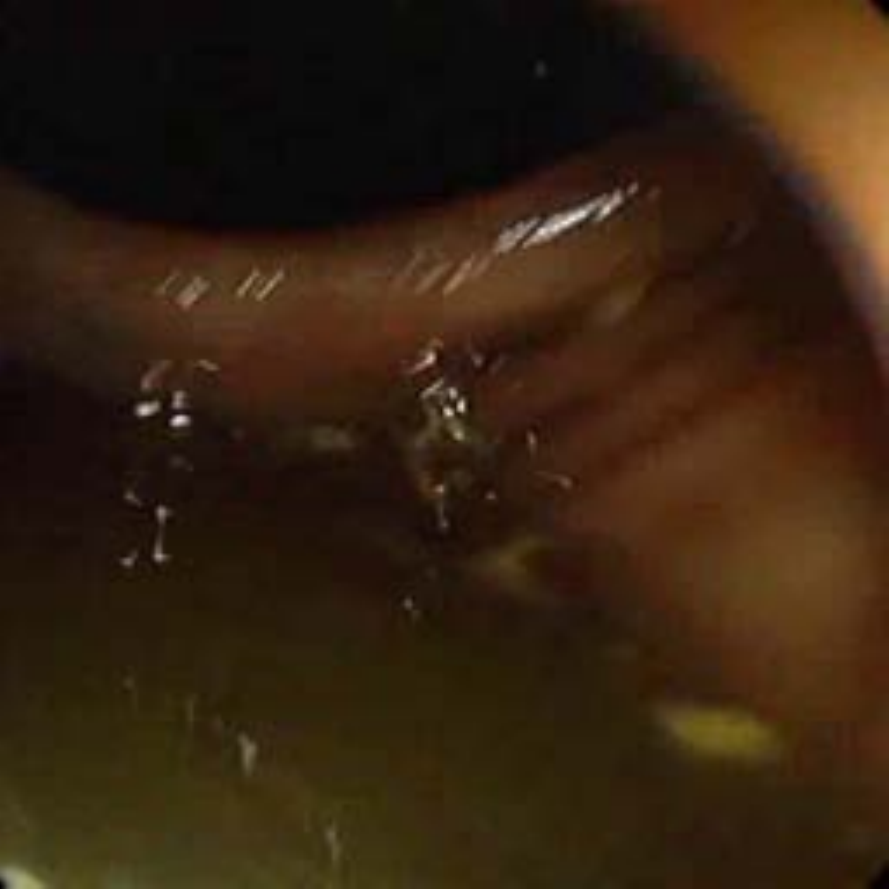}&
\includegraphics[width=0.2\textwidth]{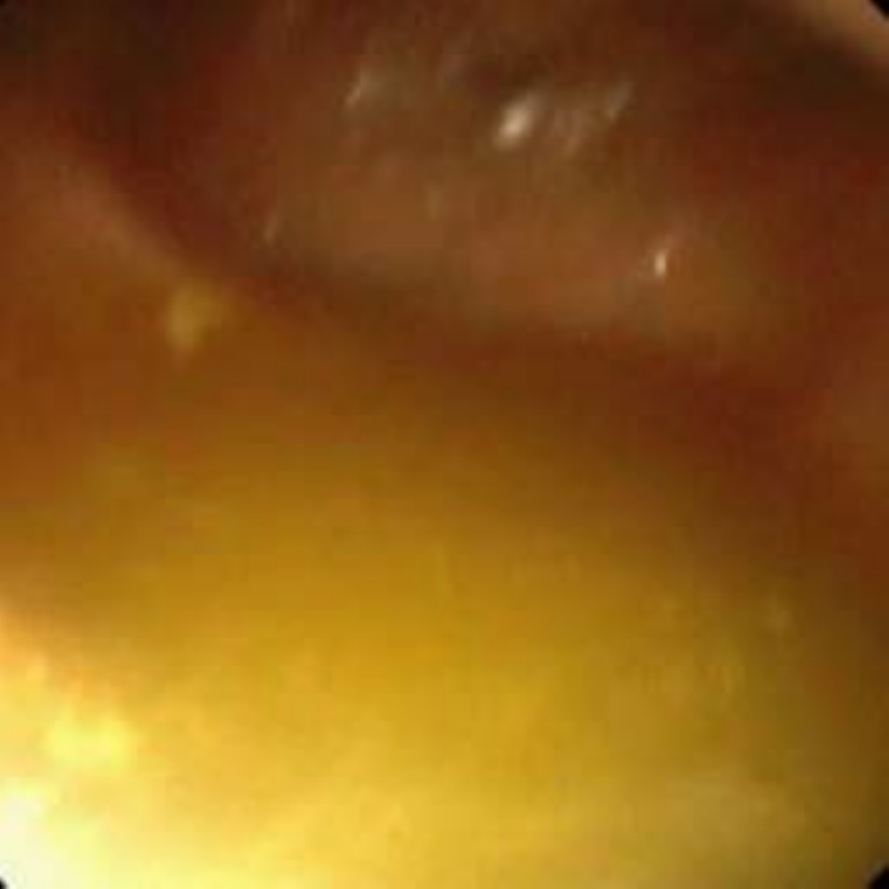}&
\includegraphics[width=0.2\textwidth]{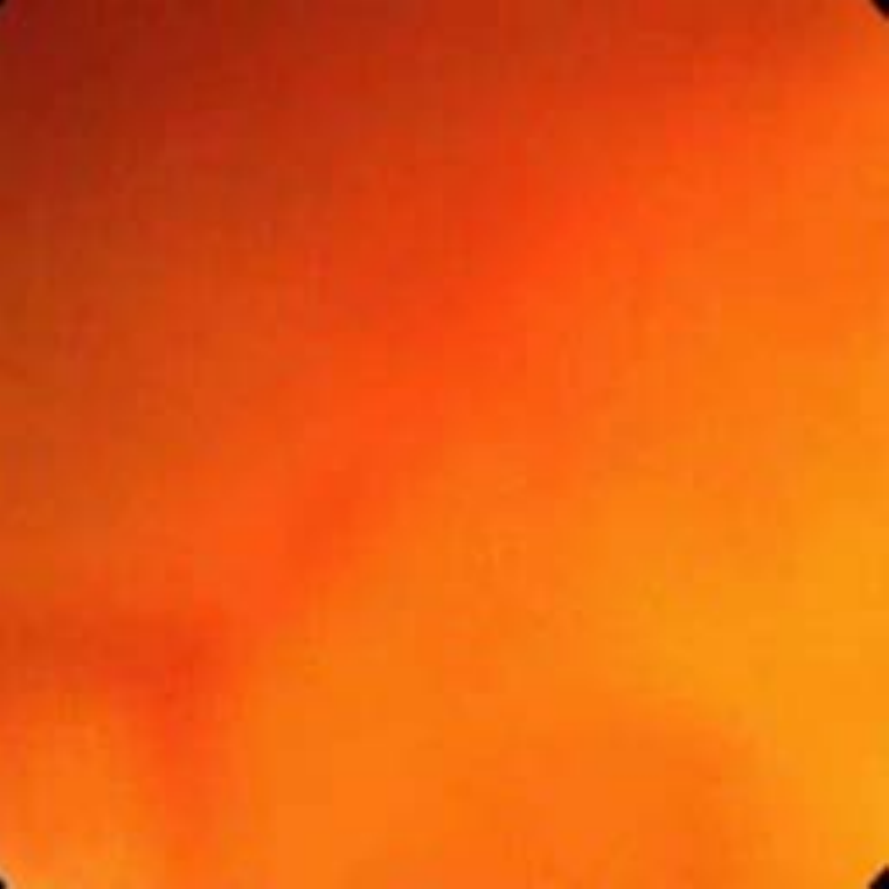}&
\includegraphics[width=0.2\textwidth]{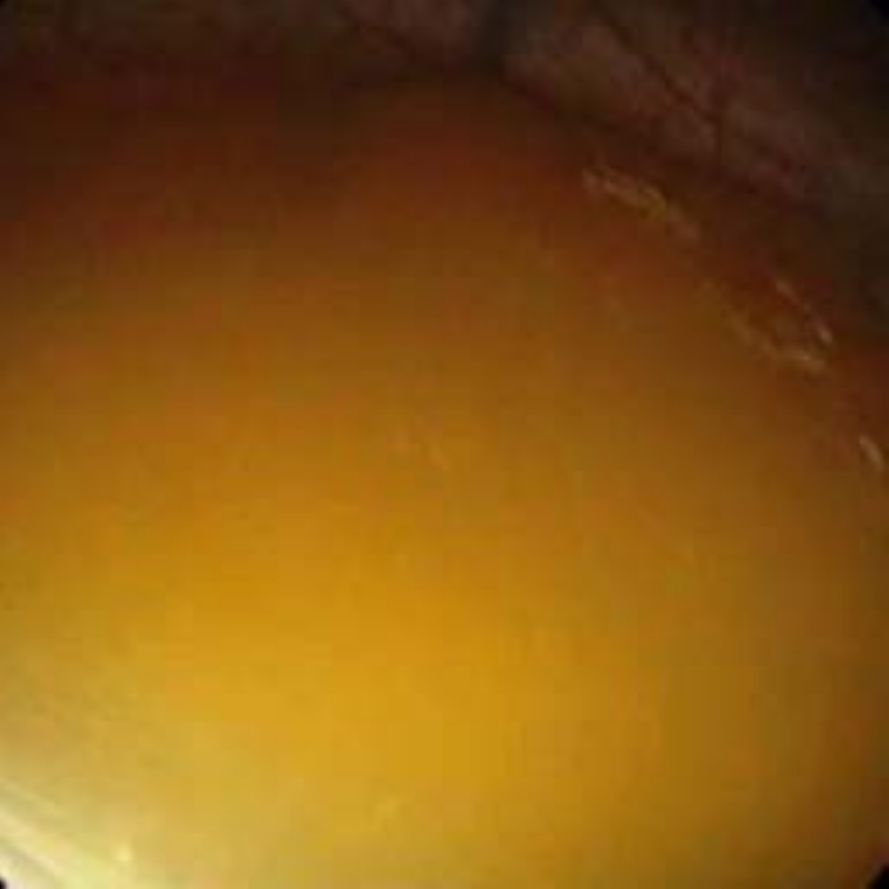}\\

\includegraphics[width=0.2\textwidth]{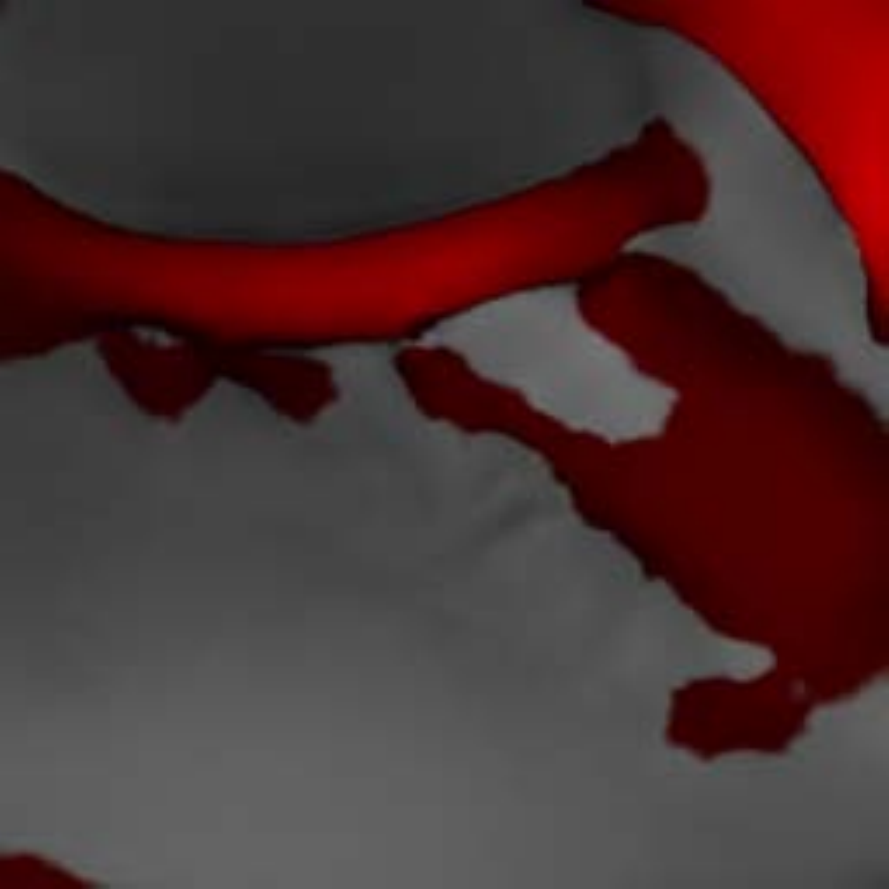}&
\includegraphics[width=0.2\textwidth]{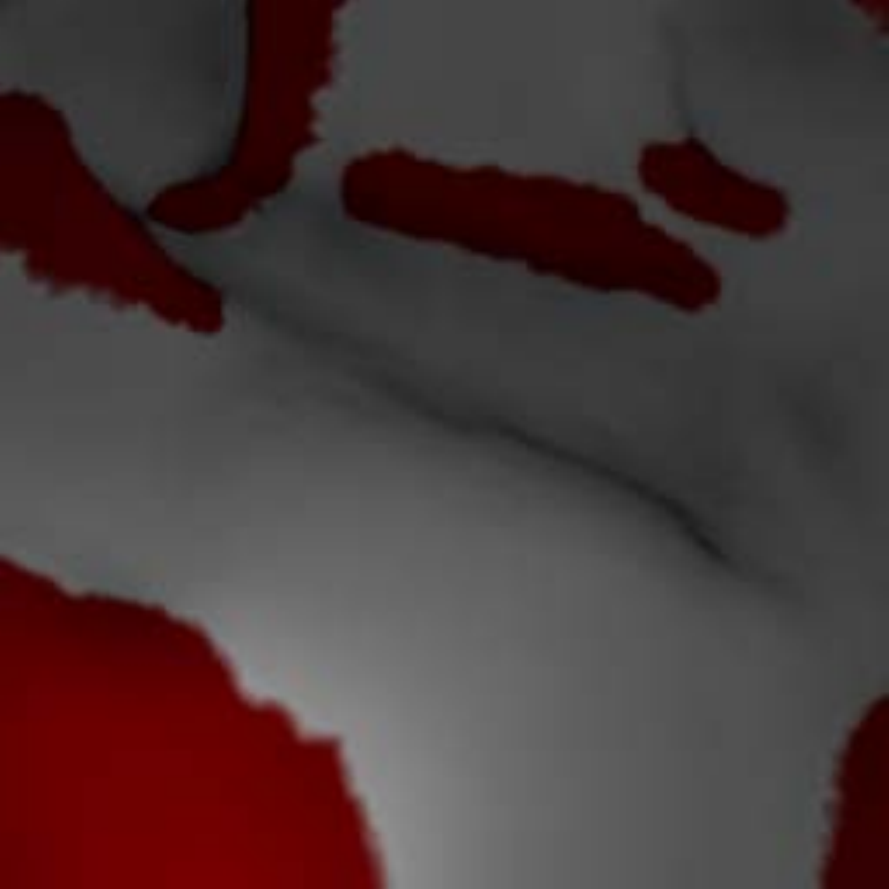}&
\includegraphics[width=0.2\textwidth]{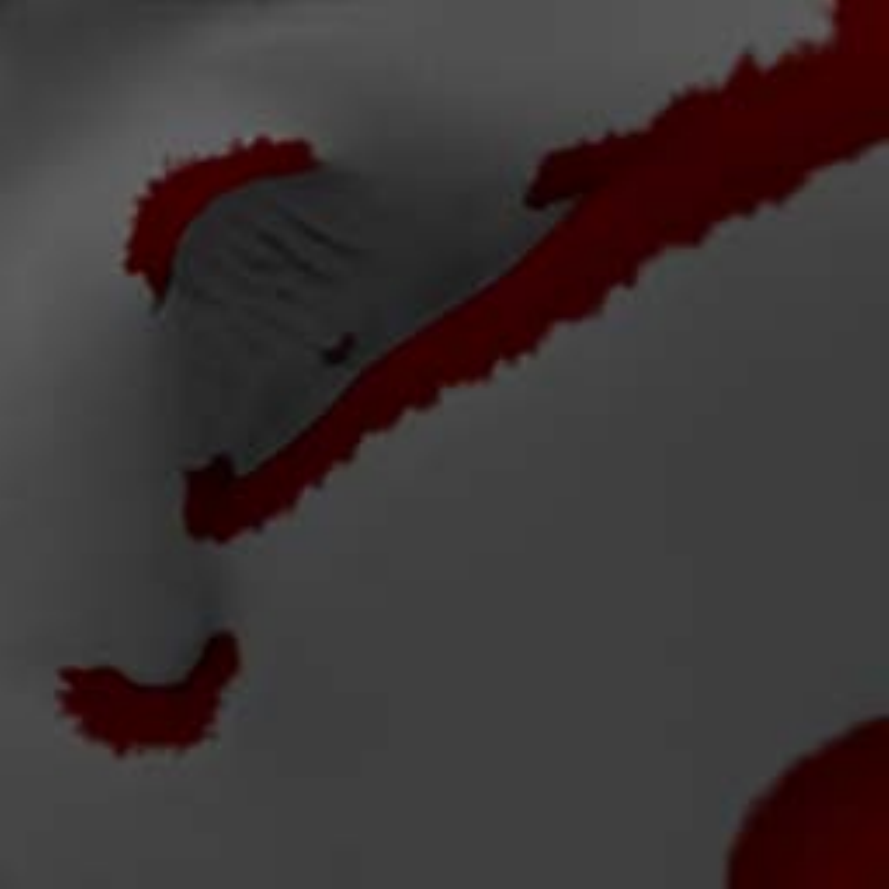}&
\includegraphics[width=0.2\textwidth]{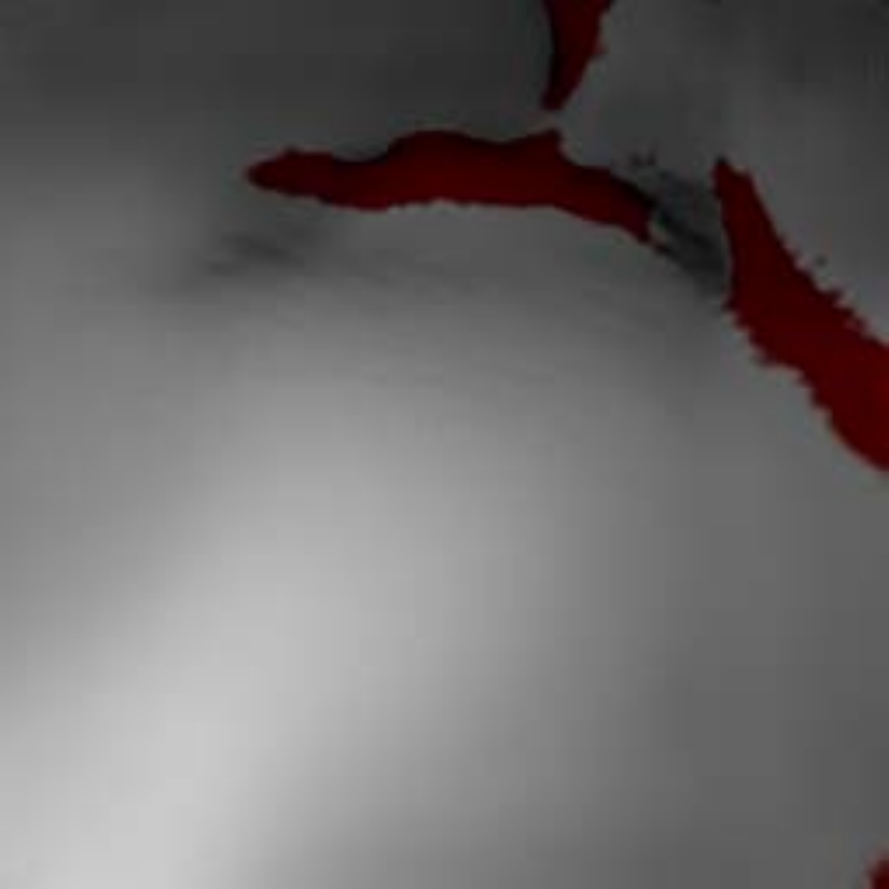}\\

\end{tabular}
\caption{Examples of our network failure in blurry and fluid-motion frames.}
\vspace{-5mm}
\label{fig:failure}
\end{center}
\end{figure}

\section{Limitations and Future Work}
In this work, we introduced FoldIt, a new model to detect and segment haustral folds in colonoscopy videos. Our model has some limitations. It fails to handle cases where there are large amounts of fluid or blurriness, as shown in Fig. \ref{fig:failure}. This is due to the fact that our network is trained on clean VC images where there is no fluid or blur. In effect, these types of OC frames do not have valid corresponding frames in our VC training data. In the future, fluid and blur can be introduced in our VC renderings to add a valid correspondence for these failed OC frames. Moreover, even though our current model gives consistent results across video frames, we have not incorporated any explicit temporal constraint. In the future, we will incorporate temporal consistency in our model to track haustral folds across frames; the tracked folds will be used to register OC and VC videos. Finally, we will use additional anatomical features such as taenia coli (three longitudinal muscles that run across colon) to further enrich our model.  

\section*{Acknowledgements}
This project was supported by MSK Cancer Center Support Grant/Core Grant (P30 CA008748), and NSF grants CNS1650499, OAC1919752, and ICER1940302.


\end{document}